%% file: paper.tex
\documentclass[a4paper,11pt]{article}
\pdfoutput=1   
\usepackage{jheppub}  
\usepackage{braket}
\usepackage{relsize}
\title{\boldmath Two-loop QCD Correction to Massive Spin-2 Resonance $ \to q ~ 
\bar{q} ~  g $}
\author[a]{Taushif Ahmed,}
\author[b]{Goutam Das,}
\author[b]{Prakash Mathews,}
\author[a]{Narayan Rana}
\author[a]{and V. Ravindran}
\affiliation[a]{The Institute of Mathematical Sciences,\\
                IV Cross Road, CIT Campus, Chennai 600 113, India}
\affiliation[b]{Theory Division, Saha Institute of Nuclear Physics,\\
                1/AF, Bidhannagar, Kolkata 700 064, India}
\emailAdd{taushif@imsc.res.in}
\emailAdd{goutam.das@saha.ac.in}
\emailAdd{prakash.mathews@saha.ac.in}
\emailAdd{rana@imsc.res.in}
\emailAdd{ravindra@imsc.res.in}
\abstract{Two-loop QCD correction to massive spin-2 Graviton decaying to
$q ~ + ~ \bar{q}~ + ~g$ is presented considering a generic universal spin-2 coupling
to the SM through the conserved energy-momentum tensor. Such a massive spin-2
particle can arise in extra-dimensional models. The ultraviolet and infrared structure of the 
QCD amplitudes are studied. In dimensional regularisation, the infrared pole structure is in 
agreement with Catani's proposal, confirming the universal factorization property of QCD 
amplitudes, even with the spin-2 tensorial coupling. 
This computation now completes the full two-loop QCD corrections for the production of a spin-2 
in association with a jet.
}
\newcommand{\del}{\partial}
\newcommand{\sig}{\sigma}
\newcommand{\omg}{\omega}
\newcommand{\eps}{\epsilon}

\newcommand{\qqqquad}{\qquad \qquad \qquad}
\newcommand{\rabar}[1]{\overset{\rightarrow}{#1}}

\begin{document} 
\keywords{Perturbative QCD, Spin-2}
\maketitle
\section{Introduction}
After the discovery of the Higgs boson one of the main motivation of Large Hadron Collider (LHC) is
to search for new physics beyond the Standard Model (BSM). It is well understood that the Standard Model
(SM) is not complete on many accounts. One of the main motivation to look for BSM physics is the large 
scale hierarchy between the Electroweak scale and the Planck scale, which is commonly known as the
hierarchy problem. There are a plethora of models to address this issue. The models with spin-2 particles
in brane world scenarios are particularly interesting. Recently there has been a renewed interest in
search of spin-2 particle in the context of the Higgs boson discovery. Of these models, theories with
large extra dimension like  Arkani-Hamed-Dimopoulos-Dvali (ADD)
\cite{ArkaniHamed:1998rs, Antoniadis:1998ig,ArkaniHamed:1998nn} and warped extra dimension like 
Randall-Sundrum (RS) \cite{Randall:1999ee} have gained a lot of attention as their signature can be 
tested at LHC energies. In fact, dedicated groups at the LHC are studying the search potentials of
such exotic particles. In the simplest of these models gravity is allowed to explore the full space-time
dimensions. The effect of gravity on the 3-brane where the SM resides is realized by a tower of spin-2
Kaluza-Klein (KK) excitations. When coupled to the SM, these KK modes may provide important signatures
about these models to be ultimately tested by the LHC. The channels like $\ell^{+}\ell^{-}, \gamma\gamma,
ZZ, W^{+}W^{-}$ and di-jet are the main channels to look for any deviations from the SM as a result of
the virtual spin-2 exchange. At hadron colliders, QCD corrections to these processes are necessary
to make precise quantitative predictions and to have control over the theoretical uncertainties. 
To next-to-leading-order (NLO) accuracy, these processes have been studied in the context of ADD
\cite{Mathews:2004xp,Kumar:2006id,Mathews:2005zs,Kumar:2008pk,Agarwal:2009xr,Agarwal:2010sp} and RS  
\cite{Mathews:2005bw,Kumar:2009nn,Agarwal:2009zg,Agarwal:2010sn,Li:2014awa}. 
The effect of parton shower (PS) has also been considered 
\cite{Frederix:2012dp,Frederix:2013lga,Das:2014tva}.
Moreover the signature can also be found in tri-gauge boson productions
\cite{Kumar:2011jq,XiaoZhou:2012cg,Xiao-Zhou:2013uqa,Chong:2014cwa,Das:2015bda}.
Recently all the important processes are fully automatized \cite{Das:2016pbk} in {\sc MadGraph5aMC}
framework \cite{Alwall:2014hca} with NLO+PS accuracy for a generic spin-2 particle. Nevertheless 
the NLO corrections are often very large, resulting in large K-factors and large scale uncertainties.
To have precise predictions the next stage is to compute full next-to-next-to-leading-order (NNLO) correction.
Higher order correction involves pure virtual contributions, pure real corrections and 
appropriate interference terms at each order. For the spin-2 case, the quark and gluon form-factors have
been calculated at two-loop level \cite{deFlorian:2013sza} and at three-loop level \cite{Ahmed:2015qia}
in peturbative QCD (pQCD). An attempt to consider the soft and collinear contribution at NNLO accuracy
was made \cite{deFlorian:2013wpa} in the context of spin-2. This reduces the unphysical scale
uncertainties, thereby improving the predictions. Recently the full NNLO correction has been studied
\cite{Ahmed:2016qhu} for Drell-Yan process with spin-2 mediators in the context of ADD. 

The processes with missing energy associated with a SM particle can also give significant 
information about new physics. Particularly, jet + missing energy is one important process studied at the
LHC to look for BSM signature like  dark matter in simplified models (see for example
\cite{Abercrombie:2015wmb} and references therein). A massive spin-2 particle which goes undetected could
be an important  dark matter candidate. The NLO correction for this process in large extra dimension has
been considered in \cite{Karg:2009xk} and it is shown that the QCD corrections for this process could be
as large as 50\% and suffer from large scale uncertainties. Therefore a full NNLO correction is important
for such process to have accurate prediction of cross-section and distributions and also to have the scale
uncertainties under control. An attempt has been made towards this direction in \cite{Ahmed:2014gla},
where the virtual NNLO QCD correction has been studied for massive spin-2 $\to g+g+g$. Although the $gg$
initiated sub-process is the dominant contribution at the LHC, as the perturbative order increases the 
other sub-processes like $q\bar{q}, q(\bar{q}) g $ begin to contribute significantly. In fact for full NNLO
correction one needs to have the virtual contributions from all the sub-processes, the real-virtual piece
and the pure real corrections. In this article we have considered the two loop virtual QCD correction to the
process massive spin-2 $\to ~  q + \bar{q} + g$. After appropriate analytical continuation of the
kinematical variables to the respective regions \cite{Gehrmann:2002zr}, the result in this paper can be
used for other scattering sub-processes \textit{viz.}
$q + \bar{q} \to G + g$ and $ ~ q(\bar{q})+g \to G +q (\bar{q})$, where $G$ denotes the spin-2 field.
This computation along with the $G\to g+g+g$ \cite{Ahmed:2014gla} to two-loop now completes the full
two-loop QCD corrections to the production of spin-2+jet at a hadron collider.

Our main  motivation for this work is two-fold; first to probe the structure of Quantum Field Theory
in the presence of a spin-2 field, to check the universality of infrared (IR) pole structure in QCD
\cite{Catani:1998bh}. The correct IR pole structure has been already realized in the case of 
spin-2$\to g+g+g$  \cite{Ahmed:2014gla}. Here we demonstrate the same for spin-2$\to q+\bar{q}+g$.
Secondly, we present one of the important ingredients for full two-loop QCD correction for real
graviton production associated with a jet. We consider a minimal universal coupling of spin-2 to the 
SM through energy-momentum tensor. Due to the conserved spin-2 current, no additional ultraviolet (UV)
renormalization is needed other than the QCD one. We encounter more than 800 Feynman diagrams with
higher tensorial integrals. The rank-2 nature of spin-2 increases the complexity of our calculation.
Using Integration-By-Parts (IBP) identities \cite{Tkachov:1981wb,Chetyrkin:1981qh}  and Lorentz 
Invariance (LI)  identities \cite{Gehrmann:1999as}, we are able to reduce all the scalar integrals to a
fewer set of Master Integrals (MI). These MIs are available in
\cite{Gehrmann:1999as,Binoth:2000ps,Smirnov:2000vy,Smirnov:2000ie,Gehrmann:2000zt,Gehrmann:2001ck}. 
Finally we observe the universality of infrared factorization of QCD amplitudes as predicted by
Catani \cite{Catani:1998bh} (see also \cite{Sterman:2002qn}).

The paper is organized as follows: in section \ref{sec:2}, we discuss the theoretical framework for our
work, particularly the spin-2 action of interaction with the SM, the notation used, the procedure of UV 
renormalization and infrared factorization. In section \ref{sec:3}, we discuss our approach to the 
computation. Section \ref{sec:4} and appendix are devoted to present the results. We conclude in section 
\ref{sec:5}.
\section{Theoretical Framework} \label{sec:2}
We consider a generic spin-2 particle minimally coupled to the SM fields through the conserved SM energy-momentum
tensor. The effective action which describes a spin-2 particle ($G^{\mu\nu}(x)$) interacting with colored
particles is given by 
\cite{ArkaniHamed:1998rs,Randall:1999ee,Antoniadis:1998ig,ArkaniHamed:1998nn,Han:1998sg,Giudice:1998ck}
\begin{equation}
\mathcal{S}_{int} = -\frac{\kappa}{2} ~ \int d^{4}x ~ T_{\mu\nu}^{QCD}(x) ~ 
              G^{\mu\nu}(x),
\end{equation}
where $\kappa$ is a dimensionful universal coupling which determines the strength of graviton coupling to
the SM. $T_{\mu\nu}^{QCD}$ is given by
\begin{equation}\label{lag}
\begin{split}
T_{\mu\nu}^{QCD} = 
& -g_{\mu\nu} \mathcal{L}_{QCD} - F_{\mu\rho}^{a} F^{a\rho}_{\nu}  
- \frac{1}{\xi}g_{\mu\nu}\del^{\rho}(A_{\rho}^{a}\del^{\sig}A_{\sig}^a) 
+ \frac{1}{\xi} (A_{\nu}^{a}\del_{\mu}(\del^{\sig} A_{\sig}^{a}) + 
 A_{\mu}^{a}\del_{\nu}(\del^{\sig} A_{\sig}^{a})) \\ 
& +\frac{i}{4}
\Big[\bar{\psi}\gamma_{\mu}(\del_{\nu} -i g_{s} T^{a} A_{\nu}^{a})\psi
-\bar{\psi}(\del_{\nu} + i g_{s} T^{a} A_{\nu}^{a})\gamma_{\mu}\psi 
+ (\mu \leftrightarrow \nu)\Big] \\
& + \Big[\del_{\mu}\bar{\omg}^{a}(\del_{\nu}\omg^{a} 
- g_{s} f^{abc} A_{\nu}^{c}\omg^{b}) + (\mu \leftrightarrow \nu)\Big],
\end{split}
\end{equation}
where $g_{s}$ is the strong coupling constant and $\xi$ is gauge fixing parameter. Here $\omg^{a}$ is the 
ghost field introduced in order to cancel the unphysical degrees of freedom associated with the gluon 
fileds $(A_{\mu}^{a})$. $T^{a}$ and $f^{abc}$ represent the generator and structure constants of $SU(N)$ 
gauge group, respectively. Throughout the computation, we consider $SU(N)$ as our gauge group and the QCD
corresponds to $N=3$. 

The decay process considered is
\begin{equation}
G(Q) \rightarrow q(p_{1}) + ~ \bar{q}(p_{2}) + ~ g(p_{3}).
\end{equation}
The corresponding Mandelstam variables for this process are defined as
\begin{equation}  
s \equiv (p_{1}+p_{2})^2, \qquad\qquad t \equiv (p_{2}+p_{3})^2, 
                          \qquad\qquad u \equiv (p_{3}+p_{1})^2.
\end{equation}  
They satisfy the following relation
\begin{equation} \label{stu}
 s + t + u = M_{G}^{2} \equiv Q^{2}.
\end{equation}
Here $M_{G}$ is the mass of the spin-2 field. The following dimensionless invariants which appear in the 
argument of harmonic polylogarithms (HPL) \cite{Remiddi:1999ew} and two-dimensional HPLs 
\cite{Gehrmann:2000zt} are also defined:
\begin{equation}
 x \equiv s/Q^2, \qquad \qquad y \equiv u/Q^2, \qquad\qquad z \equiv t/Q^2.
\end{equation}
Accordingly, Eq.\ (\ref{stu}) becomes
\begin{equation}
 x+y+z = 1.
\end{equation}
\subsection{Ultraviolet Renormalization}\label{uv}
Beyond leading order in perturbation theory, the on-shell QCD amplitudes develop both ultraviolet 
and infrared divergences. The spin-2 coupling to the SM particles $\kappa$ is free from such ultraviolet 
renormalization, which is due to the fact that spin-2 couples universally to the SM through conserved 
current. So the only UV renormalization required  is for the strong coupling constant $\hat{g}_{s}$.
Before performing the renormalization, we need to regularize  the theory in order to identify the true 
nature of the divergences. We regularize the theory under dimensional regularization where the space-time 
dimension is chosen to be $d=4+\eps$. Expanding the scattering amplitude in powers of 
$\hat{a}_{s}=\hat{g}_{s}^{2}/16\pi^2$, the matrix element (ME) is given by:
\begin{equation} 
\ket{\mathcal{M}} = \Big( \frac{\hat{a}_{s}}{\mu_{0}^{\eps}} S_{\eps}\Big)^{\frac{1}{2}}
                     \bigg(
                          \ket{\hat{\mathcal{M}}^{(0)}} + 
                     \Big( \frac{\hat{a}_{s}}{\mu_{0}^{\eps}} S_{\eps}\Big) \ket{\hat{\mathcal{M}}^{(1)}} +
                     \Big( \frac{\hat{a}_{s}}{\mu_{0}^{\eps}} S_{\eps}\Big)^{2} \ket{\hat{\mathcal{M}}^{(2)}} +
                     \mathcal{O}(\hat{a}_{s}^{3})
                     \bigg)
\end{equation}
where $\hat{g}_{s}$ is the unrenormalized strong coupling constant,
$S_{\eps} = \text{exp}[~\frac{\eps}{2} ( \gamma_{E} - \text{ln}~4\pi )]$ and 
$\gamma_{E} = 0.5772 ~.~.~.$ is the Euler constant. $\ket{\hat{\mathcal{M}}^{(i)}}$ is the unrenormalized 
color-space vector representing the $i^{th}$ loop-amplitude. $\mu_{0}$ is a mass scale introduced
to make the strong coupling constant $(\hat{g}_{s})$ dimensionless in $d$-dimension. We work within the 
$\overline{\text{MS}}$ scheme for performing the UV renormalization, in which the renormalized coupling 
constant $a_{s}\equiv a_{s}(\mu_R^2)$ is defined at the renormalization scale $\mu_R$ and is related to 
the unrenormalized $\hat{a}_{s}$ by
\begin{equation}
\begin{aligned}
\frac{\hat{a}_{s}}{\mu_{0}^{\eps}} ~ S_{\eps} &= \frac{a_{s}}{\mu_{R}^{\eps}} ~ Z(\mu_{R}^{2}),\\
                                              &= \frac{a_{s}}{\mu_{R}^{\eps}} \Big[ 1 + a_{s} \frac{2 \beta_{0}}{\eps} 
                  + a_{s}^{2} ~\Big( \frac{4 \beta_{0}^{2}}{\eps} + \frac{\beta_{1}}{\eps}\Big)  + \mathcal{O}(a_{s}^{3})   \Big],
\end{aligned}
\end{equation}
where 
\begin{equation}
\beta_0 = \Big(  \frac{11}{3} C_{A} - \frac{4}{3} T_{F}~ n_{f}  \Big) , ~~~~ 
\beta_1 = \Big(  \frac{34}{3} C_{A}^{2} - \frac{20}{3} C_{A} T_{F}~ n_{f} - 4 C_{F} T_{F}~ n_{f}  \Big).
\end{equation}
Here $C_{A} = N$  and $C_{F} = (N^{2} -1)/2N$ are the quadratic Casimir of the SU(N) group.
$ ~  T_{F} = 1/2 $ and $n_{f}$ is the number of light active quark flavors. 

The matrix element can also be expressed as a power series of renormalized strong 
coupling constant with UV finite matrix elements  $\ket{\mathcal{M}^{(i)}}$,
\begin{equation}
\ket{\mathcal{M}} = (a_{s})^{\frac{1}{2}} ~ \Big( \ket{\mathcal{M}^{(0)}} + 
                                           a_{s}  \ket{\mathcal{M}^{(1)}} + 
                                          a_{s}^2 \ket{\mathcal{M}^{(2)}} + 
                                                       \mathcal{O}(a_{s}^3) \Big)
\end{equation}
where 
\begin{equation}
\begin{aligned}
\ket{\mathcal{M}^{(0)}} &= \Big( \frac{1}{\mu_{R}^{\eps}} \Big)^{\frac{1}{2}} ~ \ket{\hat{\mathcal{M}}^{(0)}} , \\ 
\ket{\mathcal{M}^{(1)}} &= \Big( \frac{1}{\mu_{R}^{\eps}} \Big)^{\frac{3}{2}} ~ \Big[ \ket{\hat{\mathcal{M}}^{(1)}}
                                                      + \mu_{R}^{\eps} \frac{r_{_{1}}}{2} ~ \ket{\hat{\mathcal{M}}^{(0)}} \Big], \\
\ket{\mathcal{M}^{(2)}} &= \Big( \frac{1}{\mu_{R}^{\eps}} \Big)^{\frac{5}{2}} ~ \Big[ \ket{\hat{\mathcal{M}}^{(2)}}
                                                      + \mu_{R}^{\eps} \frac{3 r_{_{1}}}{2} ~ \ket{\hat{\mathcal{M}}^{(1)}}
                      + \mu_{R}^{2\eps} ~ \Big( \frac{r_{_{2}}}{2} - \frac{r_{_{1}}^2}{8} \Big)  ~ \ket{\hat{\mathcal{M}}^{(0)}} \Big]
\end{aligned}
\end{equation}
with 
\begin{equation}
 r_{_{1}} = \frac{2\beta_0}{\eps} , ~  \qquad r_{_{2}} = \Big(  \frac{4\beta_0^2}{\eps^2}  + \frac{\beta_1}{\eps} \Big)
\end{equation}
\subsection{Infrared Factorization}\label{ir}
In higher order calculation, the UV renormalized matrix elements contain singularities of infrared
origin. Generally two kinds of singularities arise -- the soft and collinear while working with 
massless QCD. According to the KLN theorem, these singularities get canceled against the similar 
contribution from real emission Feynman diagrams, resulting in infrared safe observables. 
The IR divergences have a universal structure in dimensional regularization, which was predicted in 
\cite{Catani:1998bh} to two loop, except the two loop single pole in $\eps$. In 
\cite{Sterman:2002qn,Aybat:2006wq,Aybat:2006mz} the infrared structure of scattering amplitudes are studied 
and connection of the single pole to soft anomalous dimension matrix is predicted. The factorization of the
single pole in quark and gluon form factors in terms of soft and collinear anomalous 
dimensions was demonstrated to two-loop level \cite{Ravindran:2004mb} whose validity at three-loop was later 
established in \cite{Moch:2005tm}. The proposal by Catani was generalized beyond two loops in 
\cite{Becher:2009cu, Gardi:2009qi}.

According to Catani's prediction \cite{Catani:1998bh}, the renormalized amplitude factorizes in 
dimensional regularization. The ME at a given order $\ket{\mathcal{M}^{(i)}}$ can be 
expressed as the sum of the lower order MEs times appropriate insertion operators 
($\mathbf{I}_{q}^{(i)}(\eps)$) and a finite piece $\ket{\mathcal{M}^{(i) fin}}$. These insertion operators 
contain the infrared pole structure which are universal. For the present case, we have two external 
massless quarks and a gluon, for which the one-loop and the two-loop ME can be written in the following form,
\begin{equation}
\begin{aligned}
\ket{\mathcal{M}^{(1)}} &=  2 ~\mathbf{I}_{q}^{(1)}(\eps) ~\ket{\mathcal{M}^{(0)}} ~ + ~ \ket{\mathcal{M}^{(1) fin}}, \\
\ket{\mathcal{M}^{(2)}} &=  2 ~\mathbf{I}_{q}^{(1)}(\eps) ~\ket{\mathcal{M}^{(1)}} 
                        ~+~ 4 ~\mathbf{I}_{q}^{(2)}(\eps) ~\ket{\mathcal{M}^{(0)}} ~ + ~ \ket{\mathcal{M}^{(2) fin}}
\end{aligned}
\end{equation}
where the one-loop and two-loop insertion operators are given by
\begin{equation}
\begin{split}
\mathbf{I}_{q}^{(1)}(\eps) &= \frac{1}{2} ~ \frac{e^{-\frac{\eps}{2} \gamma_{E}}}{\Gamma(1+\frac{\eps}{2})} 
                              \Bigg\{
                              \Big( \frac{4}{\eps^2} - \frac{3}{\eps} \Big)(C_A - 2 C_F) 
                              \bigg[ \Big( -\frac{s}{\mu_{R}^2} \Big)^{\frac{\eps}{2}} \bigg]\\
                          & + \Big(-\frac{4 C_A}{\eps^2} + \frac{3 C_A}{2\eps} + \frac{\beta_0}{2\eps}\Big)
                              \bigg[ \Big( -\frac{t}{\mu_{R}^2} \Big)^{\frac{\eps}{2}} +  \Big( -\frac{u}{\mu_{R}^2} \Big)^{\frac{\eps}{2}} \bigg]
                              \Bigg\},\\
\mathbf{I}_{q}^{(2)}(\eps) &= \frac{1}{2} ~ \mathbf{I}_{q}^{(1)}(\eps) ~
                              \bigg[ \mathbf{I}_{q}^{(1)}(\eps) - \frac{2\beta_0}{\eps} \bigg]
                            + \frac{e^{\frac{\eps}{2} \gamma_{E}} ~ \Gamma(1+\eps) }{\Gamma(1+\frac{\eps}{2})}
                              \bigg[ -\frac{\beta_0}{\eps}  + K \bigg]~ \mathbf{I}_{q}^{(1)}(2\eps)\\
                          & + \Big( 2 \mathbf{H}_{q}^{(2)}(\eps) + \mathbf{H}_{g}^{(2)}(\eps) \Big)
\end{split}
\end{equation}
where 
\begin{equation}
 K = \bigg( \frac{67}{18} ~-~ \frac{\pi^2}{6} \bigg) C_A ~-~ \frac{10}{9}T_{F}~n_{f}.
\end{equation}
The functions $\mathbf{H}_{q}^{(2)}(\eps), \mathbf{H}_{g}^{(2)}(\eps)$ are dependent on the 
renormalization scheme. In the $\overline{\text{MS}}$ scheme these are given by 
\begin{equation}
\begin{split}
\mathbf{H}_{q}^{(2)}(\eps) &=  \frac{1}{\eps}
                              \Bigg\{ 
                              C_{A}C_{F} \Big( -\frac{245}{432} + \frac{23}{16}\zeta_{2} - \frac{13}{4}\zeta_{3}  \Big)
                            + C_{F}^{2}\Big( \frac{3}{16} - \frac{3}{2}\zeta_{2} + 3\zeta_{3}\Big)\\
                          & + C_{F}~n_{f} \Big( \frac{25}{216} - \frac{1}{8}\zeta_{2}\Big)
                              \Bigg\},\\
\mathbf{H}_{g}^{(2)}(\eps) &=  \frac{1}{\eps}
                              \Bigg\{ 
                              C_{A}^{2} \Big( -\frac{5}{24} - \frac{11}{48}\zeta_{2} - \frac{1}{4}\zeta_{3}  \Big)
                            + C_{A}~n_{f}\Big( \frac{29}{54} + \frac{1}{24}\zeta_{2} \Big)
                            + \frac{1}{4} C_{F}~n_{f} - \frac{5}{54} n_{f}^{2}
                              \Bigg\}.
\end{split}
\end{equation}
Here $\zeta_{i}$  is the Riemann Zeta function.
\section{Calculation of Amplitudes} \label{sec:3}
In this section we discuss the calculational details of the amplitudes 
$\ket{\hat{\mathcal{M}}^{(i)}}$ for the process $G \rightarrow q ~+~ \bar{q} ~+~ g$ 
to two-loop level in pQCD. Particularly we calculate the squared matrix elements 
$\braket{\mathcal{M}^{(0)} | \mathcal{M}^{(1)}}$ and $\braket{\mathcal{M}^{(0)} | \mathcal{M}^{(2)}}$.
Due to the tensorial coupling of spin-2 with the SM, the computational procedure becomes tedious.
Starting from the generation of Feynman amplitudes, we systematically automatize the calculational
procedure using  in-house codes based on FORM \cite{Vermaseren:2000nd}, Mathematica,
Reduze 2 \cite{vonManteuffel:2012np} and LiteRed \cite{Lee:2012cn,Lee:2013mka}.
\subsection{Generation of Feynman Diagrams and Simplification}
QGRAF \cite{Nogueira:1991ex} is used to generate all the Feynman amplitudes in terms of symbolic expressions. 
For the process under consideration, we have 4 Feynman diagrams in the Born, 43 in the one-loop and 847 in 
the two-loop level; where we have excluded all the tadpoles and self-energy corrections to the external legs. 
The raw QGRAF output is then manipulated using in-house FORM routines to incorporate the Feynman rules 
\cite{Han:1998sg,Giudice:1998ck}  and to take care of the color and Dirac matrix ordering.
For the internal gluons Feynman gauge is used and the ghost-graviton interactions \cite{Mathews:2004pi}
are introduced in the Lagrangian (see Eq.\ (\ref{lag})) as is necessary for higher order computations.
For the external gluon the physical polarisations are summed using
\begin{equation}
\sum_{\lambda=\pm 1} \eps^{\mu}(p_{3},\lambda)\eps^{\nu*}(p_{3},\lambda) =
        -g^{\mu\nu} ~+~ \frac{p_{3}^{\mu}n^{\nu} + n^{\mu}p_{3}^{\nu}}{p_{3}.n}.
\end{equation}
Here $\lambda$ is the helicity and $p_3$ is the momentum of the external gluon. $n$ is an arbitrary 
light-like 4-vector. We choose $n = p_1$, one of the external fermion momenta without loss of generality. 
The spin-2 polarisation sum in d-dimension is given by \cite{Han:1998sg,Mathews:2004pi}:
\begin{equation}
\begin{aligned}
 B^{\mu\nu;\rho\sig}(q) ~=~ &\bigg( g^{\mu\rho} - \frac{q^{\mu}q^{\sig} }{q.q} \bigg)
                             \bigg( g^{\nu\sig} - \frac{q^{\nu}q^{\sig} }{q.q} \bigg)
                         +   \bigg( g^{\mu\sig} - \frac{q^{\mu}q^{\sig} }{q.q} \bigg)
                             \bigg( g^{\nu\rho} - \frac{q^{\nu}q^{\rho} }{q.q} \bigg)\\
           &-\frac{2}{d-1}    \bigg( g^{\mu\nu} - \frac{q^{\mu}q^{\nu} }{q.q} \bigg)  
                             \bigg( g^{\rho\sig} - \frac{q^{\rho}q^{\sig} }{q.q} \bigg).
\end{aligned}
\end{equation}
where the metric $g_{\mu\nu} = Diag(1,-1,-1,-1)$. The squared matrix elements are further 
processed using in-house codes based on LiteRed and Mathematica.
\subsection{Reduction of Tensor Integrals}
Two loop calculation involves a large number of higher rank Feynman integrals, particularly in the present 
case it contains tensorial integrals. The conventional approach is to convert the different Feynman 
integrals into scalar integrals. This generates thousands of scalar integrals which need to be properly
classified. The idea is to connect all the scalar integrals to belong to a particular basis. The basis is 
chosen keeping in mind that any scalar products of loop momenta and external momenta can be expressed only 
in terms of linear combinations of the propagators. In the case of one-loop there are four different scalar 
products which are written in terms of four propagators. The choice of basis for one-loop and two-loop case 
is given in \cite{Ahmed:2014gla}, for completeness we present those here. It is straight forward to choose 
the following set as the basis for the one-loop case,
\begin{equation} \label{class1}
\begin{aligned}
B_{11} &= \{ \mathcal{D}_1,\mathcal{D}_{1;1},\mathcal{D}_{1;12},\mathcal{D}_{1;123} \}, \\
B_{12} &= \{ \mathcal{D}_1,\mathcal{D}_{1;2},\mathcal{D}_{1;23},\mathcal{D}_{1;123} \}, \\
B_{13} &= \{ \mathcal{D}_1,\mathcal{D}_{1;3},\mathcal{D}_{1;31},\mathcal{D}_{1;123} \}
\end{aligned}
\end{equation}
where 
\begin{equation}
\mathcal{D}_{1}   = k_{1}^2, ~
\mathcal{D}_{1;i}   = (k_1 - p_i)^2, ~
\mathcal{D}_{1;ij}  = (k_1 - p_i- p_j)^2, ~
\mathcal{D}_{1;ijk} = (k_1 - p_i- p_j - p_k)^2
\end{equation}
with $i,j,k = 1,2,3$.
At two-loop one has 9 independent scalar products of loop momenta and external momenta, \textit{viz.} 
$\{ (k_{\alpha} \cdot k_{\beta}), (k_{\alpha} \cdot p_{i}) \}$; $\alpha,\beta = 1,2$ and $i=1,2,3$. 
The physical diagrams contain at most 7 different propagators. Hence we need to increase the number of
propagators to 9. With the help of Reduze 2, all the  two-loop diagrams are classified into six different
auxiliary topologies presented below:
 \begin{equation} \label{class2}
 \begin{aligned}
B_{21} &=  \{ \mathcal{D}_{0},~\mathcal{D}_{1},~\mathcal{D}_{2},
    ~\mathcal{D}_{1;1},~\mathcal{D}_{2;1},~\mathcal{D}_{1;12},
    ~\mathcal{D}_{2;12},~\mathcal{D}_{1;123},~\mathcal{D}_{2;123} \}, \\
B_{22} &=  \{ \mathcal{D}_{0},~\mathcal{D}_{1},~\mathcal{D}_{2},
    ~\mathcal{D}_{1;2},~\mathcal{D}_{2;2},~\mathcal{D}_{1;23},
    ~\mathcal{D}_{2;23},~\mathcal{D}_{1;123},~\mathcal{D}_{2;123} \}, \\
B_{23} &=  \{ \mathcal{D}_{0},~\mathcal{D}_{1},~\mathcal{D}_{2},
    ~\mathcal{D}_{1;3},~\mathcal{D}_{2;3},~\mathcal{D}_{1;31},
    ~\mathcal{D}_{2;31},~\mathcal{D}_{1;123},~\mathcal{D}_{2;123} \}, \\
B_{24} &=  \{ \mathcal{D}_{0},~\mathcal{D}_{1},~\mathcal{D}_{2},
    ~\mathcal{D}_{1;1},~\mathcal{D}_{2;1},~\mathcal{D}_{0;3},
    ~\mathcal{D}_{1;12},~\mathcal{D}_{2;12},~\mathcal{D}_{1;123} \}, \\
B_{25} &=  \{ \mathcal{D}_{0},~\mathcal{D}_{1},~\mathcal{D}_{2},
    ~\mathcal{D}_{1;2},~\mathcal{D}_{2;2},~\mathcal{D}_{0;1},
    ~\mathcal{D}_{1;23},~\mathcal{D}_{2;23},~\mathcal{D}_{1;123} \}, \\
B_{26} &=  \{ \mathcal{D}_{0},~\mathcal{D}_{1},~\mathcal{D}_{2},
    ~\mathcal{D}_{1;3},~\mathcal{D}_{2;3},~\mathcal{D}_{0;2},
    ~\mathcal{D}_{1;31},~\mathcal{D}_{2;31},~\mathcal{D}_{1;123} \}
 \end{aligned}
 \end{equation}
where 
\begin{equation}
\begin{aligned}
&\mathcal{D}_{0}        = (k_{1} -k_{2})^2, ~
\mathcal{D}_{\alpha}   = k_{\alpha}^2, ~
\mathcal{D}_{\alpha;i}   = (k_{\alpha} - p_{i} )^2, ~
\mathcal{D}_{\alpha;ij}   = (k_{\alpha} - p_{i} - p_{j} )^2, \\
&\mathcal{D}_{0;i}   = (k_{1} - k_{2} - p_{i} )^2, ~
\mathcal{D}_{\alpha;ijk}   = (k_{\alpha} - p_{i} - p_{j} - p_{k} )^2.
\end{aligned}
\end{equation}

Although properly classified, these large number of scalar integrals are not all independent. In fact 
they are related by the IBP identities \cite{Tkachov:1981wb,Chetyrkin:1981qh} and LI identities 
\cite{Gehrmann:1999as} which follow from the Poincare invariance. At a fixed order, they result in a 
large linear system of equations for the integrals. The inclusion of LI identities accelerates the 
solution of the system of equations, although they are not independent from the IBP identities 
\cite{Lee:2008tj}. We generate the IBP relations and LI identities using Laporta algorithm 
\cite{Laporta:2001dd} as implemented in LiteRed. Using LiteRed along with Mint 
\cite{Nason:2007vt,Lee:2013hzt} we reduce all 
different scalar integrals to a fewer set of irreducible scalar integrals \textit{i.e.} the MIs.

At one-loop, two kinds of MIs appear, \textit{viz.} the \textit{Bubble}- two-propagator MI and 
\textit{Box}- four-propagator MI. In case of two-loop we find a total of 24 topologies, out of 
which 8 are non-planar topologies and 16 planar topologies. All the two-loop MIs in our calculation 
can be related to the MI computed in \cite{Gehrmann:2000zt,Gehrmann:2001ck}. At this point we would 
like to note that some of the MIs in our case do not appear as given in 
\cite{Gehrmann:2000zt,Gehrmann:2001ck}. The reason behind this is the different convention in the basis 
in LiteRed and in \cite{Gehrmann:2000zt,Gehrmann:2001ck}. Thus we found topologies containing higher 
power of propagators instead of the irreducible numerator in \cite{Gehrmann:2000zt,Gehrmann:2001ck}. 
Nevertheless those can be related by properly using the IBP and LI identities. In this way we reduce 
all the scalar integrals to the known set of MIs. We also found two extra topologies for the MI 
\cite{Ahmed:2014gla}, \textit{viz.} \textit{Kite} and \textit{GlassS} which are basically product of 
two one-loop MIs. \textit{GlassS} is found to be the product of two \textit{Bubbles}. \textit{Kite}  
is the product of one \textit{Bubble} and one \textit{Box} one-loop MIs. Using all the MIs we finally 
find the unrenormalized one-loop $\braket{\hat{\mathcal{M}}^{(0)} | \hat{\mathcal{M}}^{(1)}}$ and
two-loop $\braket{\hat{\mathcal{M}}^{(0)} | \hat{\mathcal{M}}^{(2)}}$ matrix elements which are 
presented in the next section.
\section{Results} \label{sec:4}
We have checked that the amplitudes are gauge invariant, which serves a crucial check on our computation.
Following the renormalization prescription in section (\ref{uv}), we compute the 
UV renormalized matrix elements
$\braket{\mathcal{M}^{(0)} | \mathcal{M}^{(1)}}$ and $\braket{\mathcal{M}^{(0)} | \mathcal{M}^{(2)}}$
in terms of the unrenormalized ones,
\begin{equation} \label{res:eq1}
\begin{aligned}
\braket{\mathcal{M}^{(0)} | \mathcal{M}^{(1)}} =
         \bigg( \frac{1}{\mu_{R}^{\eps}} \bigg)^{2} ~ 
         &\bigg[ 
               \braket{\hat{\mathcal{M}}^{(0)} | \hat{\mathcal{M}}^{(1)}}   
 ~+~\mu_{R}^{\eps} \frac{r_{_{1}}}{2} 
               \braket{\hat{\mathcal{M}}^{(0)} | \hat{\mathcal{M}}^{(0)}}
         \bigg]  \\
\braket{\mathcal{M}^{(0)} | \mathcal{M}^{(2)}} =
         \bigg( \frac{1}{\mu_{R}^{\eps}} \bigg)^{3} ~ 
         &\bigg[ 
               \braket{\hat{\mathcal{M}}^{(0)} | \hat{\mathcal{M}}^{(2)}}   
 ~+~\mu_{R}^{\eps}  \frac{3r_{_{1}}}{2}
               \braket{\hat{\mathcal{M}}^{(0)} | \hat{\mathcal{M}}^{(1)}}\\   
&~+~\mu_{R}^{2\eps} \bigg( \frac{r_{_{2}}}{2} - \frac{r_{_{1}}^2}{8} \bigg) 
               \braket{\hat{\mathcal{M}}^{(0)} | \hat{\mathcal{M}}^{(0)}}
         \bigg].
\end{aligned}
\end{equation}
Similarly the renomalized matrix elements according to Catani's prescription can be written as 
\begin{equation} \label{res:eq2}
\begin{aligned}
\braket{\mathcal{M}^{(0)} | \mathcal{M}^{(1)}} &=
         2~\mathbf{I}_{q}^{(1)}(\eps) ~ \braket{\mathcal{M}^{(0)} | \mathcal{M}^{(0)}}
                                    ~+~ \braket{\mathcal{M}^{(0)} | \mathcal{M}^{(1)fin}} \\
\braket{\mathcal{M}^{(0)} | \mathcal{M}^{(2)}} &=
         2~\mathbf{I}_{q}^{(1)}(\eps) ~ \braket{\mathcal{M}^{(0)} | \mathcal{M}^{(1)}}
       + 4~\mathbf{I}_{q}^{(2)}(\eps) ~ \braket{\mathcal{M}^{(0)} | \mathcal{M}^{(0)}}
                                    ~+~ \braket{\mathcal{M}^{(0)} | \mathcal{M}^{(2)fin}}.
\end{aligned}
\end{equation}
From Eq.\ (\ref{res:eq1}) we extract the coefficients of different poles \textit{viz.}\ the 
$1/\eps^4, 1/\eps^3, 1/\eps^2, 1/\eps$ and we find that they exactly agree with the respective poles 
coming from Eq.\ (\ref{res:eq2}). By comparing the $\mathcal{O}(\eps^0)$ terms from these two sets of 
equations (Eq.\ (\ref{res:eq1})) and (Eq.\ (\ref{res:eq2})), we obtain the unknown pieces 
$\braket{\mathcal{M}^{(0)} | \mathcal{M}^{(1)fin}}$ and 
$\braket{\mathcal{M}^{(0)} | \mathcal{M}^{(2)fin}}$.\\
The final result is written in the following form:
\begin{equation}
\begin{aligned}
\braket{\mathcal{M}^{(0)} | \mathcal{M}^{(0)}} &=
         \mathcal{F}_{b}~ \mathcal{A}^{(0)},                        \\ 
\braket{\mathcal{M}^{(0)} | \mathcal{M}^{(1)fin}} &=
         \mathcal{F}_{b} 
         \Bigg\{ 
       ~ \mathcal{A}_{0}^{(1)} ~ \text{ln}\bigg( -\frac{Q^2}{\mu_{R}^2} \bigg)
       +  \Big(\mathcal{A}_{1}^{(1)} \zeta_{2} +  \mathcal{A}_{2}^{(1)}    \Big)
         \Bigg\},                                                   \\
\braket{\mathcal{M}^{(0)} | \mathcal{M}^{(2)fin}} &=
         \mathcal{F}_{b} 
         \Bigg\{
               ~ \mathcal{A}_{0}^{(2)}   ~\text{ln}^{2}\bigg( -\frac{Q^2}{\mu_{R}^2} \bigg) \\
      &+  \Big(~ \mathcal{A}_{1}^{(2)} \zeta_{3} 
             + ~  \mathcal{A}_{2}^{(2)} \zeta_{2} 
               + \mathcal{A}_{3}^{(2)} 
          \Big) ~\text{ln}\bigg( -\frac{Q^2}{\mu_{R}^2} \bigg)                  \\
      &+  \Big(  \mathcal{A}_{4}^{(2)} \zeta_{2}^{2} 
               + \mathcal{A}_{5}^{(2)} \zeta_{3} 
               + \mathcal{A}_{6}^{(2)} \zeta_{2} 
               + \mathcal{A}_{7}^{(2)}
          \Big)
         \Bigg\}
\end{aligned}
\end{equation}
where 
\begin{equation}
\begin{aligned}
\mathcal{F}_{b} &= 16 \pi^{2} \kappa^{2} \big(N^{2} -1\big)~Q^{2},
\end{aligned}
\end{equation}
\begin{equation}
\begin{aligned}
\mathcal{A}^{(0)} =  \frac{ \bigg( 2 + y^2\big(3 - 9z\big) - 4z + 3z^2 - z^3 + y^3\big(-1 + 4z\big) 
                                              + y\big(-4 + 12z - 9z^2 + 4z^3\big) \bigg)}{4yz(1 - y - z)},
\end{aligned}
\end{equation}
\begin{equation}
\begin{aligned}
\mathcal{A}_{i}^{(1)} &= \mathcal{A}_{i;C_A}^{(1)} ~ C_A + \mathcal{A}_{i;C_F}^{(1)} ~ C_F + \mathcal{A}_{i;n_f}^{(1)} ~ n_f,  \\
\mathcal{A}_{i}^{(2)} &= \mathcal{A}_{i;C_{A}^{2}}^{(2)}  C_{A}^{2} 
                       + \mathcal{A}_{i;C_{F}^{2}}^{(2)}  C_{F}^{2} 
                       + \mathcal{A}_{i;n_{f}^{2}}^{(2)}  n_{f}^{2} 
                       + \mathcal{A}_{i;C_{A}C_{F}}^{(2)}  C_{A}C_{F}
                       + \mathcal{A}_{i;C_{A}n_{f}}^{(2)}  C_{A}n_{f}
                       + \mathcal{A}_{i;C_{F}n_{f}}^{(2)}  C_{F}n_{f}.
\end{aligned}
\end{equation}
We notice that unlike the Higgs decay i.e. $H \rightarrow b + \bar{b} + g$  \cite{Ahmed:2014pka},
there is no $C_{F}$ term in the $\mathcal{A}_{0}^{(1)}$, which is due to the absence of Yukawa-like term 
in spin-2 case. All the one-loop and two-loop coefficients are presented in the Appendix \ref{app:1lcof} 
and \ref{app:2lcof} respectively except the $A_{7}^{(2)}$ term which is provided as the ancillary file in
the arXiv. 
\section{Conclusions} \label{sec:5}
In this article, we present the two-loop virtual QCD correction to massive $G \to q +\bar{q}+ g$ considering
the minimal and universal coupling between spin-2 and the SM particles. We confine ourselves within the
framework of massless QCD where only the light quark degrees of freedom are taken into account. We employ 
the Feynman diagrammatic approach to achieve our goal. As expected, the computation becomes very tedious 
not only due to the presence of a large number of Feynman diagrams but also due to the involvement of a 
tensorial coupling. In-house codes and state-of-the-art techniques, in particular, IBP and LI identities, 
are employed extensively to execute the computation successfully. The bare matrix elements contain UV as 
well as IR divergences. The strong coupling constant renormalization is sufficient to make it UV finite. 
No extra UV renormalization is required for the spin-2 coupling as a consequence of the conserved SM 
energy-momentum tensor through which it couples universally to the SM fields. The UV finite matrix elements
exhibit poles of infrared origin in dimensional regularization. The resulting infrared pole structures are 
in exact agreement with the Catani's prescription which ensures the universal factorization property of QCD 
amplitudes even in the presence of spin-2 field. This serves a crucial check on the correctness of our 
computation. The result presented here is an important piece, which now completes the full two-loop 
calculation of real graviton production associated with a jet. The full NNLO computation to this process 
needs additional inputs that we reserve for future study.

\begin{acknowledgments}
We sincerely thank Thomas Gehrmann for providing the necessary MIs which are used in this calculation. 
GD would like to thank R. N. Lee for help with LiteRed. GD also thanks Kuntal Nayek for help
with Mathematica. GD is supported by funding from Department of Atomic Energy, India.
\end{acknowledgments}
\appendix 
\section{Harmonic Polylogarithms} \label{app:hpl}
All our results are presented in terms of HPLs which are generalisation of Neilson's polylogarithms.
Here we briefly discuss some important properties of HPLs. For more details one can see 
 \cite{Remiddi:1999ew,Gehrmann:2000zt}.
The HPLs are represented by $H(\rabar{m}_{\omg};y) $; $\rabar{m}_{\omg}$ being a $\omg$-dimensional 
vector which belongs to the set \{-1,0,1\} through which we define the following rational functions
\begin{equation}
f(1;y) \equiv \frac{1}{1-y}, \qquad
f(0;y) \equiv \frac{1}{y}, \qquad 
f(-1;y) \equiv \frac{1}{1+y}.
\end{equation}
The weight-1 ($\omg = 1$) HPLs are then defined as
\begin{equation}
H(1,y)  ~\equiv~ -\text{ln}(1-y), \qquad
H(0,y)  ~\equiv~  \text{ln}(y), \qquad 
H(-1,y) ~\equiv~  \text{ln}(1+y).
\end{equation}
For $\omg > ~1$ HPLs are defined as
\begin{equation} \label{eq:a3}
H(m,\rabar{m}_{\omg};y) ~\equiv~ \int_{0}^{y} ~ dx ~ f(m,x) 
                                                ~ H(\rabar{m}_{\omg};x),
                                              \qqqquad m \in 0, \pm{1}.
\end{equation}
Moreover the higher dimensional HPLs are defined following the Eq.\ (\ref{eq:a3}) for the new elements 
{2,3} in $\rabar{m}_{\omg}$, representing a new class of rational functions
\begin{equation}
f(2;y) ~\equiv~ f(1-z;y) ~\equiv~ \frac{1}{1-y-z}, \qquad
f(3;y) ~\equiv~ f(z;y)   ~\equiv~ \frac{1}{y+z} ;
\end{equation}
correspondingly the weight-1 ($\omg = 1$) two-dimensional HPLs are given by
\begin{equation}
H(2,y)  ~\equiv~ -\text{ln}\Big(1-\frac{y}{1-z} \Big), \qquad
H(3,y)  ~\equiv~  \text{ln}\Big(\frac{y+z}{z}\Big).
\end{equation}
\subsection*{Properties}
 HPLs follow some important properties\\
\textit{\underline{Shuffle algebra}: }
Product of two HPLs with weights $\omg_1$ and $\omg_2$ of same argument $y$ is a HPL of weight 
($\omg_1 + \omg_1$) and argument $y$. This way all possible permutations of the elements of  
$\rabar{m}_{\omg_1}$ and $\rabar{m}_{\omg_2}$ are considered preserving the relative orders of 
the element of $\rabar{m}_{\omg_1}$ and $\rabar{m}_{\omg_2}$,
\begin{equation} \label{B:6}
   H(\rabar{m}_{\omg_1};y)~H(\rabar{m}_{\omg_2};y) ~~ =
    \sum_{ \scriptscriptstyle{ \mathsmaller{ \rabar{m}_{\omg} = \rabar{m}_{\omg_1} \biguplus \rabar{m}_{\omg_2} }  } } 
    H(\rabar{m}_{\omg};y)
\end{equation}
\textit{\underline{Integration-by-parts identities}: }
Using the integration-by-parts identities the ordering of the elements of $\rabar{m}_{\omg}$ inside 
the HPL can be reversed and in this process some products of two HPLs can be generated.
\begin{equation}\label{B:7}
\begin{aligned}
H(\rabar{m}_{\omg};y) ~\equiv~ H(m_{1},m_{2},... m_{\omg} ; y)
&=    H(m_{1} ; y)~H(m_{2},... m_{\omg} ; y) \\
&-   H(m_{2},m_{1} ; y)~H(m_{3},... m_{\omg} ; y)\\
&+ ... + (-1)^{\omg + 1}~H(m_{\omg},..., m_{2}, m_{1} ; y).
\end{aligned}
\end{equation}
Equations \ref{B:6} and \ref{B:7} are very useful in writing the higher-weight HPLs in terms of 
lower-weight HPLs or products of HPLs, which have been used in the checks of the different poles in 
section \ref{sec:5}. Below we provide the necessary relations of HPLs used in our computation.
\begin{equation}
\begin{aligned}
H(0,2,0,y) &=  H(0,y) H(0,2,y) - 2 H(0,0,2,y),\\
H(1,0,2,y) &= H(1,y) H(0,2,y) - H(0,1,2,y) - H(0,2,1,y),\\
H(1,2,0,y) &= H(1,y) H(2,0,y) - H(2,1,y) H(0,y) + H(0,2,1,y),\\
H(2,0,0,y) &=  H(2,y) H(0,0,y) - H(0,2,y) H(0,y) + H(0,0,2,y),\\
H(2,0,2,y) &=  H(2,y) H(0,2,y) - 2 H(0,2,2,y),\\
H(2,2,2,y) &=  \frac{1}{3} H(2,y) H(2,2,y),\\
H(2,2,0,y) &=  H(2,y) H(2,0,y) - H(2,2,y) H(0,y) + H(0,2,2,y),\\
H(3,0,2,y) &=  H(3,y) H(0,2,y) - H(0,3,2,y) - H(0,2,3,y),\\
H(3,2,0,y) &=  H(3,y) H(2,0,y) - H(2,3,y) H(0,y) + H(0,2,3,y),\\
H(3,3,2,y) &=  H(3,y) H(3,2,y) - H(3,3,y) H(2,y) + H(2,3,3,y),\\
H(2,1,0,y) &=  H(2,y) H(1,0,y) - H(1,2,y) H(0,y) + H(0,1,2,y),\\
H(2,3,2,y) &=  H(2,y) H(3,2,y) - 2 H(3,2,2,y)
\end{aligned}
\end{equation}
\section{One-loop Coefficients} \label{app:1lcof}
\input{oneloop}
\section{Two-loop Coefficients} \label{app:2lcof}
\input{twoloop}
\bibliographystyle{JHEP}
\bibliography{goutam}
\end{document}

%% file: oneloop.tex
\tiny{
\begin{equation*}
\begin{aligned}
\mathcal{A}_{0}^{(1)}     &=    -\frac{\beta_{0}}{2} ~ \mathcal{A}_{0};  \\
\mathcal{A}_{1;C_{A}}^{(1)} &=  - {C_A} ~ \mathcal{A}_{0}; \qquad 
\mathcal{A}_{1;C_{F}}^{(1)} = 0; \qquad  
\mathcal{A}_{1;n_{f}}^{(1)} = 0;\\
\mathcal{A}_{2;C_{A}}^{(1)} &=
\Bigg\{-6 (-1 + y) y^4 (-1 + z) (y + z) + 6 (-1 + y) y^5 (-1 + z) (y + z) +  84 (-1 + y) y^3 (-1 + z) z (y + z) - 54 (-1 + y) y^4 (-1\\
  &+ z) z (y + z) +  44 (-1 + y) y^5 (-1 + z) z (y + z) + 132 (-1 + y) y^2 (-1 + z) z^2  (y + z) - 152 (-1 + y) y^3 (-1 + z) z^2 (y\\
  &+ z) +  48 (-1 + y) y^5 (-1 + z) z^2 (y + z) + 84 (-1 + y) y (-1 + z) z^3 (y + z) -  152 (-1 + y) y^2 (-1 + z) z^3 (y + z) - 88 (-1\\
  &+ y) y^3 (-1 + z) z^3  (y + z) + 144 (-1 + y) y^4 (-1 + z) z^3 (y + z) -  6 (-1 + y) (-1 + z) z^4 (y + z) - 54 (-1 + y) y (-1\\
  &+ z) z^4 (y + z) +  144 (-1 + y) y^3 (-1 + z) z^4 (y + z) + 6 (-1 + y) (-1 + z) z^5 (y + z) +  44 (-1 + y) y (-1 + z) z^5 (y + z)\\
  &+ 48 (-1 + y) y^2 (-1 + z) z^5 (y + z) +  3 (-1 + y) y (-1 + z) (y + z)^4 (16 y^3 + y^2 (-41 + 12 z) -  2 (7 - 6 z + 3 z^2)\\
  &+ 3 y (13 - 9 z + 4 z^2)) H(0, y)^2 +  3 (-1 + y) (-1 + z) z (y + z)^4 (-14 + 39 z - 41 z^2 + 16 z^3 +  6 y^2 (-1 + 2 z) + 3 y (4\\
  &- 9 z + 4 z^2)) H(0, z)^2 +  3 (-1 + y) (-1 + z) (y + z)^4 (4 + 16 y^4 - 22 z + 45 z^2 - 43 z^3 +  16 z^4 + y^3 (-43 + 20 z)\\
  &+ 3 y^2 (15 - 17 z + 8 z^2) +  y (-22 + 48 z - 51 z^2 + 20 z^3)) H(1, z)^2 +  2 (-1 + z) (y + z)^4 H(0, y) (-10 + 30 y - 35 y^2\\
  &+ 20 y^3 - 5 y^4 + 20 z -  80 y z + 132 y^2 z - 116 y^3 z + 44 y^4 z - 15 z^2 + 60 y z^2 -  72 y^2 z^2 + 24 y^3 z^2 + 5 z^3\\
  &- 25 y z^3 + 20 y^2 z^3 +  3 (-1 + y) (2 + y^2 (3 - 9 z) - 4 z + 3 z^2 - z^3 + y^3 (-1 + 4 z) +  y (-4 + 12 z - 9 z^2\\
  &+ 4 z^3)) H(0, z) +  3 (-1 + y) (2 + 16 y^4 - 4 z + 3 z^2 - z^3 + 2 y^3 (-21 + 8 z) +  6 y^2 (7 - 6 z + 2 z^2) + y (-18 + 24 z\\
  &- 15 z^2 + 4 z^3)) H(1, z)) +  (-1 + y) (-1 + z) (y^7 (-9 + 84 z) + 3 y^6 (9 - 83 z + 144 z^2) -  9 z^4 (-2 + 4 z - 3 z^2 + z^3)\\
  &+ 3 y^5 (-12 + 92 z - 373 z^2 + 324 z^3) +  y z^3 (-36 - 136 z + 276 z^2 - 249 z^3 + 84 z^4) +  y^2 z^2 (-60 - 292 z + 1053 z^2\\
  &- 1119 z^3 + 432 z^4) +  y^3 z (-36 - 292 z + 1608 z^2 - 2175 z^3 + 972 z^4) +  y^4 (18 - 136 z + 1053 z^2 - 2175 z^3\\
  &+ 1248 z^4)) H(2, y) +  3 (-1 + y) (-1 + z) (y + z)^4 (4 + 16 y^4 - 22 z + 45 z^2 - 43 z^3 +  16 z^4 + y^3 (-43 + 20 z) + 3 y^2 (15\\
  &- 17 z + 8 z^2) +  y (-22 + 48 z - 51 z^2 + 20 z^3)) H(2, y)^2 +  2 (-1 + y) (y + z)^4 H(0, z) (-5 (-1 + z)^2 (2 - 2 z + z^2)\\
  &+  5 y^3 (1 - 5 z + 4 z^2) + 3 y^2 (-5 + 20 z - 24 z^2 + 8 z^3) +  4 y (5 - 20 z + 33 z^2 - 29 z^3 + 11 z^4) +  3 (-1 + z) (y^3 (-1\\
  &+ 4 z) + 4 y (-1 + z)^2 (-1 + 4 z) +  3 y^2 (1 - 5 z + 4 z^2) + 2 (1 - 9 z + 21 z^2 - 21 z^3 + 8 z^4))  H(2, y)) + (-1 + y) (-1\\
  &+ z) H(1, z) (3 y^7 (-3 + 28 z) +  3 y^6 (9 - 83 z + 144 z^2) - 9 z^4 (-2 + 4 z - 3 z^2 + z^3) +  3 y^5 (-12 + 92 z - 373 z^2\\
  &+ 324 z^3) +  y z^3 (-36 - 136 z + 276 z^2 - 249 z^3 + 84 z^4) +  y^2 z^2 (-60 - 292 z + 1053 z^2 - 1119 z^3 + 432 z^4)\\
  &+  y^3 z (-36 - 292 z + 1608 z^2 - 2175 z^3 + 972 z^4) +  y^4 (18 - 136 z + 1053 z^2 - 2175 z^3 + 1248 z^4) -  6 (y + z)^4 (4\\
  &+ 16 y^4 - 22 z + 45 z^2 - 43 z^3 + 16 z^4 +  y^3 (-43 + 20 z) + 3 y^2 (15 - 17 z + 8 z^2) +  y (-22 + 48 z - 51 z^2\\
  &+ 20 z^3)) H(3, y)) -  6 (-1 + y) y (-1 + z) (y + z)^4 (16 y^3 + y^2 (-41 + 12 z) -  2 (7 - 6 z + 3 z^2) + 3 y (13 - 9 z\\
  &+ 4 z^2)) H(0, 0, y) -  6 (-1 + y) (-1 + z) z (y + z)^4 (-14 + 39 z - 41 z^2 + 16 z^3 +  6 y^2 (-1 + 2 z) + 3 y (4 - 9 z\\
  &+ 4 z^2)) H(0, 0, z) -  6 (-1 + y) (-1 + z) (y + z)^4 (2 + 16 y^4 - 4 z + 3 z^2 - z^3 +  2 y^3 (-21 + 8 z) + 6 y^2 (7 - 6 z\\
  &+ 2 z^2) +  y (-18 + 24 z - 15 z^2 + 4 z^3)) H(0, 1, z) +  6 (-1 + y) (-1 + z) (y + z)^4 (2 + 16 y^4 - 4 z + 3 z^2 - z^3\\
  &+  2 y^3 (-21 + 8 z) + 6 y^2 (7 - 6 z + 2 z^2) +  y (-18 + 24 z - 15 z^2 + 4 z^3)) H(0, 2, y) -  6 (-1 + y) y (-1 + z) (y\\
  &+ z)^4 (16 y^3 + y^2 (-41 + 12 z) -  2 (7 - 6 z + 3 z^2) + 3 y (13 - 9 z + 4 z^2)) H(1, 0, y) +  6 (-1 + y) (-1 + z) (y + z)^4 (2\\
  &+ y^2 (3 - 9 z) - 4 z + 3 z^2 - z^3 +  y^3 (-1 + 4 z) + y (-4 + 12 z - 9 z^2 + 4 z^3)) H(1, 0, z) -  6 (-1 + y) (-1 + z) (y\\
  &+ z)^4 (4 + 16 y^4 - 22 z + 45 z^2 - 43 z^3 +  16 z^4 + y^3 (-43 + 20 z) + 3 y^2 (15 - 17 z + 8 z^2) +  y (-22 + 48 z - 51 z^2\\
  &+ 20 z^3)) H(1, 1, z) +  6 (-1 + y) (-1 + z) (y + z)^4 (2 + 16 y^4 - 4 z + 3 z^2 - z^3 +  2 y^3 (-21 + 8 z) + 6 y^2 (7 - 6 z\\
  &+ 2 z^2) +  y (-18 + 24 z - 15 z^2 + 4 z^3)) H(2, 0, y) -  6 (-1 + y) (-1 + z) (y + z)^4 (4 + 16 y^4 - 22 z + 45 z^2 - 43 z^3\\
  &+  16 z^4 + y^3 (-43 + 20 z) + 3 y^2 (15 - 17 z + 8 z^2) +  y (-22 + 48 z - 51 z^2 + 20 z^3)) H(2, 2, y) -  6 (-1 + y) (-1 + z) (y\\
  &+ z)^4 (4 + 16 y^4 - 22 z + 45 z^2 - 43 z^3 +  16 z^4 + y^3 (-43 + 20 z) + 3 y^2 (15 - 17 z + 8 z^2) +  y (-22 + 48 z - 51 z^2\\
  &+ 20 z^3)) H(3, 2, y)\Bigg\}\Big/ \Big(24 (-1 + y) y (-1 + z) z (-1 + y + z) (y + z)^4\Big);\\[5pt]
\mathcal{A}_{2;C_{F}}^{(1)} &=
\Bigg\{20 (-1 + y) y (-1 + z) (y + z) - 59 (-1 + y) y^2 (-1 + z) (y + z) +  68 (-1 + y) y^3 (-1 + z) (y + z) - 39 (-1 + y) y^4 (-1\\
  &+ z) (y + z) +  10 (-1 + y) y^5 (-1 + z) (y + z) + 20 (-1 + y) (-1 + z) z (y + z) -  138 (-1 + y) y (-1 + z) z (y + z) + 316 (-1\\
  &+ y) y^2 (-1 + z) z (y + z) -  318 (-1 + y) y^3 (-1 + z) z (y + z) + 161 (-1 + y) y^4 (-1 + z) z (y + z) -  41 (-1 + y) y^5 (-1\\
  &+ z) z (y + z) - 59 (-1 + y) (-1 + z) z^2 (y + z) +  316 (-1 + y) y (-1 + z) z^2 (y + z) - 578 (-1 + y) y^2 (-1 + z) z^2  (y + z)\\
  &+ 435 (-1 + y) y^3 (-1 + z) z^2 (y + z) -  142 (-1 + y) y^4 (-1 + z) z^2 (y + z) + 28 (-1 + y) y^5 (-1 + z) z^2  (y + z) + 68 (-1\\
  &+ y) (-1 + z) z^3 (y + z) -  318 (-1 + y) y (-1 + z) z^3 (y + z) + 435 (-1 + y) y^2 (-1 + z) z^3  (y + z) - 202 (-1 + y) y^3 (-1\\
\end{aligned}
\end{equation*}
\begin{equation*}
\begin{aligned}
  &+ z) z^3 (y + z) +  20 (-1 + y) y^4 (-1 + z) z^3 (y + z) - 39 (-1 + y) (-1 + z) z^4 (y + z) +  161 (-1 + y) y (-1 + z) z^4 (y + z)\\
  &- 142 (-1 + y) y^2 (-1 + z) z^4  (y + z) + 20 (-1 + y) y^3 (-1 + z) z^4 (y + z) +  10 (-1 + y) (-1 + z) z^5 (y + z) - 41 (-1\\
  &+ y) y (-1 + z) z^5 (y + z) +  28 (-1 + y) y^2 (-1 + z) z^5 (y + z) - (-1 + y)^2 (-1 + 4 y) (-1 + z)^2  (y + z)^2 (-2 + 2 y^3 + 4 z\\
  &- 3 z^2 + z^3 + y^2 (-6 + 4 z) +  y (6 - 8 z + 3 z^2)) H(0, y)^2 + (1 - y) (-1 + y)  (y^3 + 3 y^2 (-1 + z) + 4 y (-1 + z)^2 + 2 (-1\\
  &+ z)^3) (-1 + z)^2  (y + z)^2 (-1 + 4 z) H(0, z)^2 + (1 - y) (-1 + y) (-1 + z)^2 (y + z)^2  (4 + 8 y^4 - 18 z + 33 z^2 - 27 z^3\\
  &+ 8 z^4 + y^3 (-27 + 20 z) +  3 y^2 (11 - 17 z + 8 z^2) + y (-18 + 48 z - 51 z^2 + 20 z^3)) H(1, z)^2 +  (-1 + z)^2 (-1 + y + z) (y\\
  &+ z)^2 H(0, y)  (-(y z (8 y^3 + y (30 - 27 z) + 12 (-1 + z) + 2 y^2 (-13 + 6 z))) -  2 (-1 + y)^2 (-1 + 4 y) (2 + 2 y^2 + 2 y (-2\\
  &+ z) - 2 z + z^2) H(1, z)) +  (1 - y) (-1 + y) (-1 + z)^2 (y^5 (-3 + 20 z) + y^4 (9 - 73 z + 80 z^2) -  3 z^2 (-2 + 4 z - 3 z^2\\
  &+ z^3) + 6 y^3 (-2 + 15 z - 34 z^2 + 20 z^3) +  y z (-8 - 32 z + 90 z^2 - 73 z^3 + 20 z^4) +  2 y^2 (3 - 16 z + 81 z^2 - 102 z^3\\
  &+ 40 z^4)) H(2, y) +  (1 - y) (-1 + y) (-1 + z)^2 (y + z)^2 (4 + 8 y^4 - 18 z + 33 z^2 - 27 z^3 +  8 z^4 + y^3 (-27 + 20 z)\\
  &+ 3 y^2 (11 - 17 z + 8 z^2) +  y (-18 + 48 z - 51 z^2 + 20 z^3)) H(2, y)^2 +  (1 - y) (-1 + y) (-1 + y + z) (y\\
  &+ z)^2 H(0, z)  (y z (-12 + 30 z - 26 z^2 + 8 z^3 + 3 y (4 - 9 z + 4 z^2)) +  2 (y^2 + 2 y (-1 + z) + 2 (-1 + z)^2) (-1 + z)^2 (-1\\
  &+ 4 z) H(2, y)) +  (-1 + y + z) (-1 + y + z - y z)^2 H(1, z) (y^4 (3 - 20 z) +  y^3 (-6 + 50 z - 60 z^2) + 3 z^2 (2 - 2 z + z^2)\\
  &+  y^2 (6 - 34 z + 94 z^2 - 60 z^3) - 2 y z (4 + 17 z - 25 z^2 + 10 z^3) +  2 (y + z)^2 (-4 + 8 y^3 + 14 z - 19 z^2 + 8 z^3\\
  &+ y^2 (-19 + 12 z) +  2 y (7 - 10 z + 6 z^2)) H(3, y)) + 2 (1 - y) (1 - 5 y + 4 y^2) (-y - z)  (-1 + z)^2 (2 y^4 + 6 y^3 (-1 + z)\\
  &+ y^2 (6 - 14 z + 7 z^2) +  z (-2 + 4 z - 3 z^2 + z^3) + y (-2 + 10 z - 11 z^2 + 4 z^3)) H(0, 0, y) +  2 (1 - y) (-1 + y) (-y\\
  &- z) (-1 + z) (1 - 5 z + 4 z^2)  (y^4 + 2 (-1 + z)^3 z + 2 y (-1 + z)^2 (-1 + 3 z) + y^3 (-3 + 4 z) +  y^2 (4 - 11 z\\
  &+ 7 z^2)) H(0, 0, z) + 2 (-1 + y)^2 (-1 + 4 y) (-1 + z)^2  (y + z)^2 (-2 + 2 y^3 + 4 z - 3 z^2 + z^3 + y^2 (-6 + 4 z) +  y (6 - 8 z\\
  &+ 3 z^2)) H(0, 1, z) - 2 (-1 + y)^2 (-1 + 4 y) (-1 + z)^2  (y + z)^2 (-2 + 2 y^3 + 4 z - 3 z^2 + z^3 + y^2 (-6 + 4 z) +  y (6 - 8 z\\
  &+ 3 z^2)) H(0, 2, y) + 2 (-1 + y)^2 (-1 + 4 y) (-1 + z)^2  (y + z)^2 (-2 + 2 y^3 + 4 z - 3 z^2 + z^3 + y^2 (-6 + 4 z) +  y (6 - 8 z\\
  &+ 3 z^2)) H(1, 0, y) + 2 (-1 + y)^2 (-1 + z)^2 (y + z)^2  (4 + 8 y^4 - 18 z + 33 z^2 - 27 z^3 + 8 z^4 + y^3 (-27 + 20 z)\\
  &+  3 y^2 (11 - 17 z + 8 z^2) + y (-18 + 48 z - 51 z^2 + 20 z^3))  H(1, 1, z) - 2 (-1 + y)^2 (-1 + 4 y) (-1 + z)^2 (y + z)^2  (-2\\
  &+ 2 y^3 + 4 z - 3 z^2 + z^3 + y^2 (-6 + 4 z) + y (6 - 8 z + 3 z^2))  H(2, 0, y) + 2 (-1 + y)^2 (-1 + z)^2 (y + z)^2  (4 + 8 y^4\\
  &- 18 z + 33 z^2 - 27 z^3 + 8 z^4 + y^3 (-27 + 20 z) +  3 y^2 (11 - 17 z + 8 z^2) + y (-18 + 48 z - 51 z^2 + 20 z^3))  H(2, 2, y)\\
  &+ 2 (-1 + y)^2 (-1 + z)^2 (y + z)^2  (4 + 8 y^4 - 18 z + 33 z^2 - 27 z^3 + 8 z^4 + y^3 (-27 + 20 z) +  3 y^2 (11 - 17 z + 8 z^2)\\
  &+ y (-18 + 48 z - 51 z^2 + 20 z^3)) H(3, 2, y)\Bigg\}\Big/ \Big(4 (-1 + y)^2 y (-1 + z)^2 z (-1 + y + z) (y + z)^2\Big);\\[5pt]
\mathcal{A}_{2;n_{f}}^{(1)} &=
\Bigg\{-24 y^4 z + 12 y^5 z - 8 y^6 z - 24 y^3 z^2 + 8 y^4 z^2 - 8 y^5 z^2 -  24 y^2 z^3 - 8 y^3 z^3 + 16 y^4 z^3 - 24 y z^4\\
  &+ 8 y^2 z^4 + 16 y^3 z^4 +  12 y z^5 - 8 y^2 z^5 - 8 y z^6 - (y + z)^4 (2 + y^2 (3 - 9 z) - 4 z +  3 z^2 - z^3 + y^3 (-1 + 4 z)\\
  &+ y (-4 + 12 z - 9 z^2 + 4 z^3)) H(0, y) -  (y + z)^4 (2 + y^2 (3 - 9 z) - 4 z + 3 z^2 - z^3 + y^3 (-1 + 4 z) +  y (-4 + 12 z\\
  &- 9 z^2 + 4 z^3)) H(0, z) +  4 y z (-5 y^3 + 3 y^4 + y z^2 + y^2 (6 + z - 6 z^2) +  z^2 (6 - 5 z + 3 z^2)) H(1, z) +  4 y z (-5 y^3\\
  &+ 3 y^4 + y z^2 + y^2 (6 + z - 6 z^2) +  z^2 (6 - 5 z + 3 z^2)) H(2, y)\Bigg\}\Big/\Big(24 y z (-1 + y + z) (y + z)^4\Big).\\
\end{aligned}
\end{equation*}
 }

%% file: twoloop.tex
\tiny{
\begin{equation*}
\begin{aligned}
\mathcal{A}_{0}^{(2)}     &=    -\frac{3 \beta_{0}^{2}}{8} ~ \mathcal{A}_{0}; \\[5pt]
\mathcal{A}_{1;C_{A}^{2}}^{(2)} &=  - ~ \mathcal{A}_{0}; \qquad
\mathcal{A}_{1;C_{F}^{2}}^{(2)} = 24 ~ \mathcal{A}_{0}; \qquad
\mathcal{A}_{1;n_{f}^{2}}^{(2)} = 0; \qquad
\mathcal{A}_{1;C_{A}C_{F}}^{(2)} = - 26 ~ \mathcal{A}_{0}; \qquad  
\mathcal{A}_{1;C_{A}n_{f}}^{(2)} = 0; \qquad
\mathcal{A}_{1;C_{F}n_{f}}^{(2)} = 0;  \\[5pt]
\mathcal{A}_{2;C_{A}^{2}}^{(2)} &=   \frac{55}{12}~ \mathcal{A}_{0}; \qquad
\mathcal{A}_{2;C_{F}^{2}}^{(2)} = -12 ~ \mathcal{A}_{0}; \qquad
\mathcal{A}_{2;n_{f}^{2}}^{(2)} = 0; \qquad
\mathcal{A}_{2;C_{A}C_{F}}^{(2)} = \frac{23}{2} ~ \mathcal{A}_{0}; \qquad
\mathcal{A}_{2;C_{A}n_{f}}^{(2)} = -\frac{5}{6}  \mathcal{A}_{0}; \qquad
\mathcal{A}_{2;C_{F}n_{f}}^{(2)} = -  ~ \mathcal{A}_{0};\\[5pt]
\mathcal{A}_{3;C_{A}^{2}}^{(2)} &=  
\Bigg\{8 (-1 + y)^2 y (-1 + z)^2 z (y + z) (39 y^6 (-1 + 4 z) +  y^5 (84 - 710 z + 204 z^2) + 3 z^3 (26 - 41 z + 28 z^2 - 13 z^3)\\
  &-  3 y^4 (41 - 372 z + 507 z^2 + 56 z^3) +  2 y z^2 (117 - 543 z + 558 z^2 - 355 z^3 + 78 z^4) -  2 y^3 (-39 + 543 z - 1354 z^2\\
  &+ 850 z^3 + 84 z^4) +  y^2 z (234 - 1662 z + 2708 z^2 - 1521 z^3 + 204 z^4)) -  8 (-1 + y)^2 y (-1 + z)^2 z (y + z)^4 (56 + 500 y^4\\
  &+ 34 z - 396 z^2 +  550 z^3 - 244 z^4 + y^3 (-1415 + 1348 z) +  3 y^2 (433 - 715 z + 180 z^2) - 4 y (110 - 219 z + 42 z^2\\
  &+ 38 z^3))  H(0, y)^3 + 8 (-1 + y)^2 y (-1 + z)^2 z (y + z)^4  (-56 + 244 y^4 + 440 z - 1299 z^2 + 1415 z^3 - 500 z^4 +  2 y^3 (-275\\
  &+ 76 z) + y^2 (396 + 168 z - 540 z^2) -  y (34 + 876 z - 2145 z^2 + 1348 z^3)) H(0, z)^3 -  32 (-1 + y)^2 y (-1 + z)^2 z (y + z)^4 (9\\
  &+ 364 y^4 - 203 z + 702 z^2 -  872 z^3 + 364 z^4 + y^3 (-872 + 564 z) + 6 y^2 (117 - 167 z + 92 z^2) +  y (-203 + 606 z - 1002 z^2\\
  &+ 564 z^3)) H(1, z)^3 +  (-1728 y^{12} (-1 + z)^2 + 216 (-1 + z)^5 z^6 (-1 + 4 z) +  24 y^{11} (-1 + z)^2 (375 + 718 z) + y^{10} (-19080\\
  &- 28289 z + 250122 z^2 -  338949 z^3 + 136304 z^4) - y (-1 + z)^2 z^5 (-8082 + 36428 z -  80967 z^2 + 91745 z^3 - 40992 z^4 + 1728 z^5)\\
  &+  y^9 (20808 + 63133 z - 748547 z^2 + 1624467 z^3 - 1359901 z^4 +  400256 z^5) + y^2 (-1 + z)^2 z^4 (25944 - 171016 z + 438304 z^2\\
  &-  585767 z^3 + 364650 z^4 - 73560 z^5 + 864 z^6) +  y^8 (-12024 - 69787 z + 1130000 z^2 - 3636494 z^3 + 4922656 z^4 -  3035579 z^5\\
  &+ 701120 z^6) + y^7 (3312 + 52501 z - 1005911 z^2 +  4577790 z^3 - 9135618 z^4 + 9138769 z^5 - 4491307 z^6 + 860032 z^7)\\
  &+  y^3 z^3 (32844 - 401168 z + 1810731 z^2 - 4338377 z^3 + 6108582 z^4 -  5103134 z^5 + 2392815 z^6 - 537669 z^7 + 35376 z^8)\\
  &+  y^4 z^2 (23424 - 374888 z + 2225864 z^2 - 6614805 z^3 + 11116384 z^4 -  10963698 z^5 + 6184528 z^6 - 1787581 z^7 + 190880 z^8)\\
  &+  y^5 z (6066 - 175600 z + 1559775 z^2 - 6176973 z^3 + 12917254 z^4 -  15269476 z^5 + 10196989 z^6 - 3546779 z^7 + 488960 z^8)\\
  &+  y^6 (-288 - 26312 z + 553704 z^2 - 3461777 z^3 + 9763000 z^4 -  14401588 z^5 + 11527644 z^6 - 4728907 z^7 + 774416 z^8)) H(2, y)^2\\
  &-  32 (-1 + y)^2 y (-1 + z)^2 z (y + z)^4 (9 + 364 y^4 - 203 z + 702 z^2 -  872 z^3 + 364 z^4 + y^3 (-872 + 564 z) + 6 y^2 (117 - 167 z\\
  &+ 92 z^2) +  y (-203 + 606 z - 1002 z^2 + 564 z^3)) H(2, y)^3 +  H(0, y)^2 ((y + z)^2 (1728 y^{10} (-1 + z)^2 + 216 (-1 + z)^5 z^3  (-1\\
  &+ 4 z) - 72 y^9 (-1 + z)^2 (149 + 87 z) +  y^8 (28080 - 43945 z - 17990 z^2 + 55603 z^3 - 21640 z^4) +  y (-1 + z)^2 z^2 (144 - 3573 z\\
  &+ 10219 z^2 - 12168 z^3 + 4234 z^4 +  864 z^5) - y^7 (39888 - 94511 z + 2851 z^2 + 137595 z^3 -  104887 z^4 + 19064 z^5) + y^6 (32832\\
  &- 125478 z + 90408 z^2 +  113163 z^3 - 158918 z^4 + 46681 z^5 + 1096 z^6) -  y^2 (-1 + z)^2 z (360 - 2430 z - 10125 z^2 + 18134 z^3\\
  &- 1460 z^4 -  13140 z^5 + 4320 z^6) + y^5 (-15336 + 94486 z - 141904 z^2 +  19649 z^3 + 66045 z^4 + 13557 z^5 - 53601 z^6 + 17104 z^7)\\
  &+  y^4 (3600 - 38177 z + 94480 z^2 - 70947 z^3 + 25180 z^4 - 94718 z^5 +  142546 z^6 - 75040 z^7 + 13184 z^8) +  y^3 (-288 + 7227 z\\
  &- 28965 z^2 + 25131 z^3 + 4151 z^4 + 35390 z^5 -  108748 z^6 + 97128 z^7 - 33618 z^8 + 2592 z^9)) +  48 (-1 + y)^2 y (-1 + z)^2 z (y\\
  &+ z)^4 (16 y^4 + 4 y^3 (-11 + 6 z) +  6 y^2 (8 - 9 z + 2 z^2) - 3 (-2 + 4 z - 3 z^2 + z^3) +  y (-26 + 48 z - 33 z^2 + 12 z^3)) H(0, z)\\
  &+ 144 (-1 + y)^2 y (-1 + z)^2  z (y + z)^4 (2 + 16 y^4 - 4 z + 3 z^2 - z^3 + 2 y^3 (-21 + 8 z) +  6 y^2 (7 - 6 z + 2 z^2) + y (-18\\
  &+ 24 z - 15 z^2 + 4 z^3)) H(1, z) +  48 (-1 + y)^2 y (-1 + z)^2 z (y + z)^4 (2 + 16 y^4 - 4 z + 3 z^2 - z^3 +  2 y^3 (-21 + 8 z)\\
  &+ 6 y^2 (7 - 6 z + 2 z^2) +  y (-18 + 24 z - 15 z^2 + 4 z^3)) H(2, y)) +  H(0, z)^2 ((y + z)^2 (1728 y^9 (-1 + z)^2 (-1 + 3 z)\\
  &-  216 (-1 + z)^6 z^3 (-1 + 4 z) + y^8 (9000 - 48470 z + 93492 z^2 -  78258 z^3 + 23552 z^4) + 4 y^7 (-4770 + 28897 z - 69529 z^2\\
  &+  79974 z^3 - 42772 z^4 + 8200 z^5) + y (-1 + z)^2 z^2  (144 - 3501 z + 19081 z^2 - 43239 z^3 + 50375 z^4 - 23976 z^5 +  1728 z^6)\\
  &+ y^6 (20808 - 148735 z + 427794 z^2 - 614728 z^3 +  447358 z^4 - 140145 z^5 + 9016 z^6) +  y^5 (-12024 + 107821 z - 367983 z^2\\
  &+ 619778 z^3 - 507386 z^4 +  132177 z^5 + 54601 z^6 - 26984 z^7) - y^2 (-1 + z)^2 z  (360 - 2934 z - 12651 z^2 + 74708 z^3 - 142537 z^4\\
  &+ 119942 z^5 -  36936 z^6 + 864 z^7) + y^4 (3312 - 41639 z + 174106 z^2 -  315241 z^3 + 188404 z^4 + 184665 z^5 - 356162 z^6\\
  &+ 199207 z^7 -  37336 z^8) + y^3 (-288 + 7155 z - 40995 z^2 + 60519 z^3 + 92277 z^4 -  437623 z^5 + 633399 z^6 - 444135 z^7 + 146323 z^8\\
  &- 16632 z^9)) +  48 (-1 + y)^2 y (-1 + z)^2 z (y + z)^4 (y^3 (-1 + 4 z) +  4 y (-1 + z)^2 (-1 + 4 z) + 3 y^2 (1 - 5 z + 4 z^2) +  2 (1\\
  &- 9 z + 21 z^2 - 21 z^3 + 8 z^4)) H(1, z) +  144 (-1 + y)^2 y (-1 + z)^2 z (y + z)^4 (y^3 (-1 + 4 z) +  4 y (-1 + z)^2 (-1 + 4 z)\\
  &+ 3 y^2 (1 - 5 z + 4 z^2) +  2 (1 - 9 z + 21 z^2 - 21 z^3 + 8 z^4)) H(2, y)) +  H(1, z)^2 (-1728 y^{12} (-1 + z)^2 + 216 (-1\\
  &+ z)^5 z^6 (-1 + 4 z) +  24 y^{11} (-1 + z)^2 (375 + 718 z) + y^{10} (-19080 - 28289 z + 250122 z^2 -  338949 z^3 + 136304 z^4) - y (-1\\
  &+ z)^2 z^5 (-8082 + 36428 z -  80967 z^2 + 91745 z^3 - 40992 z^4 + 1728 z^5) +  y^9 (20808 + 63133 z - 748547 z^2 + 1624467 z^3\\
  &- 1359901 z^4 +  400256 z^5) + y^2 (-1 + z)^2 z^4 (25944 - 171016 z + 438304 z^2 -  585767 z^3 + 364650 z^4 - 73560 z^5 + 864 z^6)\\
  &+  y^8 (-12024 - 69787 z + 1130000 z^2 - 3636494 z^3 + 4922656 z^4 -  3035579 z^5 + 701120 z^6) + y^7 (3312 + 52501 z - 1005911 z^2\\
  &+  4577790 z^3 - 9135618 z^4 + 9138769 z^5 - 4491307 z^6 + 860032 z^7) +  y^3 z^3 (32844 - 401168 z + 1810731 z^2 - 4338377 z^3\\
\end{aligned}
\end{equation*}
\begin{equation*}
\begin{aligned}
  &+ 6108582 z^4 -  5103134 z^5 + 2392815 z^6 - 537669 z^7 + 35376 z^8) +  y^4 z^2 (23424 - 374888 z + 2225864 z^2 - 6614805 z^3\\
  &+ 11116384 z^4 -  10963698 z^5 + 6184528 z^6 - 1787581 z^7 + 190880 z^8) +  y^5 z (6066 - 175600 z + 1559775 z^2 - 6176973 z^3\\
  &+ 12917254 z^4 -  15269476 z^5 + 10196989 z^6 - 3546779 z^7 + 488960 z^8) +  y^6 (-288 - 26312 z + 553704 z^2 - 3461777 z^3\\
  &+ 9763000 z^4 -  14401588 z^5 + 11527644 z^6 - 4728907 z^7 + 774416 z^8) -  96 (-1 + y)^2 y (-1 + z)^2 z (y + z)^4 (13 + 380 y^4 - 225 z\\
  &+ 747 z^2 -  915 z^3 + 380 z^4 + y^3 (-915 + 584 z) + 9 y^2 (83 - 117 z + 64 z^2) +  y (-225 + 654 z - 1053 z^2 + 584 z^3)) H(2, y)\\
  &-  96 (-1 + y)^2 y (-1 + z)^2 z (y + z)^4 (4 + 16 y^4 - 22 z + 45 z^2 -  43 z^3 + 16 z^4 + y^3 (-43 + 20 z) + 3 y^2 (15 - 17 z + 8 z^2)\\
  &+  y (-22 + 48 z - 51 z^2 + 20 z^3)) H(3, y)) -  2 (y + z)^2 (1728 y^{10} (-1 + z)^2 + 216 (-1 + z)^5 z^3 (-1 + 4 z) -  72 y^9 (-1\\
  &+ z)^2 (149 + 87 z) + y^8 (28080 - 43945 z - 17990 z^2 +  55603 z^3 - 21640 z^4) + y (-1 + z)^2 z^2 (144 - 3573 z + 10219 z^2\\
  &-  12168 z^3 + 4234 z^4 + 864 z^5) -  y^7 (39888 - 94511 z + 2851 z^2 + 137595 z^3 - 104887 z^4 + 19064 z^5) +  y^6 (32832 - 125478 z\\
  &+ 90408 z^2 + 113163 z^3 - 158918 z^4 + 46681 z^5 +  1096 z^6) - y^2 (-1 + z)^2 z (360 - 2430 z - 10125 z^2 + 18134 z^3 -  1460 z^4\\
  &- 13140 z^5 + 4320 z^6) +  y^5 (-15336 + 94486 z - 141904 z^2 + 19649 z^3 + 66045 z^4 + 13557 z^5 -  53601 z^6 + 17104 z^7) + y^4 (3600\\
  &- 38177 z + 94480 z^2 - 70947 z^3 +  25180 z^4 - 94718 z^5 + 142546 z^6 - 75040 z^7 + 13184 z^8) +  y^3 (-288 + 7227 z - 28965 z^2\\
  &+ 25131 z^3 + 4151 z^4 + 35390 z^5 -  108748 z^6 + 97128 z^7 - 33618 z^8 + 2592 z^9)) H(0, 0, y) -  2 (y + z)^2 (1728 y^9 (-1 + z)^2 (-1\\
  &+ 3 z) - 216 (-1 + z)^6 z^3  (-1 + 4 z) + y^8 (9000 - 48470 z + 93492 z^2 - 78258 z^3 + 23552 z^4) +  4 y^7 (-4770 + 28897 z - 69529 z^2\\
  &+ 79974 z^3 - 42772 z^4 + 8200 z^5) +  y (-1 + z)^2 z^2 (144 - 3501 z + 19081 z^2 - 43239 z^3 + 50375 z^4 -  23976 z^5 + 1728 z^6)\\
  &+ y^6 (20808 - 148735 z + 427794 z^2 -  614728 z^3 + 447358 z^4 - 140145 z^5 + 9016 z^6) +  y^5 (-12024 + 107821 z - 367983 z^2\\
  &+ 619778 z^3 - 507386 z^4 +  132177 z^5 + 54601 z^6 - 26984 z^7) - y^2 (-1 + z)^2 z  (360 - 2934 z - 12651 z^2 + 74708 z^3 - 142537 z^4\\
  &+ 119942 z^5 -  36936 z^6 + 864 z^7) + y^4 (3312 - 41639 z + 174106 z^2 - 315241 z^3 +  188404 z^4 + 184665 z^5 - 356162 z^6\\
  &+ 199207 z^7 - 37336 z^8) +  y^3 (-288 + 7155 z - 40995 z^2 + 60519 z^3 + 92277 z^4 - 437623 z^5 +  633399 z^6 - 444135 z^7 + 146323 z^8\\
  &- 16632 z^9)) H(0, 0, z) +  8 (-1 + y)^2 y (-1 + z) z (528 y^8 (-1 + z) -  3 (-1 + z)^2 z^4 (14 - 14 z + 3 z^2) + 6 y^7 (227 - 631 z\\
  &+ 404 z^2) +  6 y^6 (-223 + 1245 z - 1758 z^2 + 738 z^3) +  3 y^5 (182 - 2102 z + 5517 z^2 - 4865 z^3 + 1284 z^4) +  2 y^2 z^2 (-294\\
  &+ 1706 z - 2858 z^2 + 1581 z^3 + 225 z^4 - 354 z^5) +  y z^3 (-384 + 1354 z - 1582 z^2 + 531 z^3 + 261 z^4 - 180 z^5) -  4 y^3 z (96\\
  &- 1033 z + 2746 z^2 - 2667 z^3 + 666 z^4 + 180 z^5) +  y^4 (-42 + 2398 z - 11683 z^2 + 18420 z^3 - 10149 z^4 + 1128 z^5))  H(0, 1, z)\\
  &- 8 (-1 + y) y (-1 + z)^2 z (528 y^9 + 6 y^8 (-323 + 492 z) +  12 y^7 (237 - 866 z + 609 z^2) + 3 z^4 (-30 + 60 z - 49 z^2 + 19 z^3)\\
  &+  3 y^6 (-692 + 4784 z - 7993 z^2 + 3540 z^3) +  y z^3 (-144 + 1364 z - 1944 z^2 + 1374 z^3 - 405 z^4) +  y^5 (732 - 9572 z + 30498 z^2\\
  &- 31161 z^3 + 9960 z^4) +  y^2 z^2 (-204 + 3656 z - 8678 z^2 + 7902 z^3 - 3411 z^4 + 348 z^5) +  2 y^3 z (-72 + 2320 z - 8830 z^2\\
  &+ 11118 z^3 - 6117 z^4 + 1086 z^5) +  y^4 (-90 + 2804 z - 18275 z^2 + 34641 z^3 - 24864 z^4 + 6048 z^5))  H(0, 2, y) + 264 (-1\\
  &+ y)^2 y^2 (-1 + z)^2 z (y + z)^4  (16 y^3 + y^2 (-41 + 12 z) - 2 (7 - 6 z + 3 z^2) + 3 y (13 - 9 z + 4 z^2))  H(1, 0, y) - 8 (-1\\
  &+ y)^2 y (-1 + z) z  (-3 (-1 + z)^2 z^4 (30 - 30 z + 19 z^2) + y^7 (57 - 405 z + 348 z^2) +  3 y^6 (-49 + 392 z - 939 z^2 + 592 z^3)\\
  &+  3 y^5 (60 - 516 z + 1941 z^2 - 2857 z^3 + 1356 z^4) +  y z^3 (-144 + 956 z - 2180 z^2 + 2505 z^3 - 1581 z^4 + 444 z^5)\\
  &+  y^2 z^2 (-204 + 1868 z - 6197 z^2 + 9312 z^3 - 6951 z^4 + 2160 z^5) +  y^3 z (-144 + 1808 z - 8288 z^2 + 15567 z^3 - 13539 z^4\\
  &+ 4548 z^5) +  y^4 (-90 + 902 z - 5345 z^2 + 13260 z^3 - 14271 z^4 + 5472 z^5))  H(1, 0, z) + 2 (1728 y^{12} (-1 + z)^2 - 216 (-1\\
  &+ z)^5 z^6 (-1 + 4 z) -  24 y^{11} (-1 + z)^2 (375 + 718 z) + y^{10} (19080 + 28289 z - 250122 z^2 +  338949 z^3 - 136304 z^4) + y (-1\\
  &+ z)^2 z^5 (-8082 + 36428 z -  80967 z^2 + 91745 z^3 - 40992 z^4 + 1728 z^5) -  y^9 (20808 + 63133 z - 748547 z^2 + 1624467 z^3\\
  &- 1359901 z^4 +  400256 z^5) + y^8 (12024 + 69787 z - 1130000 z^2 + 3636494 z^3 -  4922656 z^4 + 3035579 z^5 - 701120 z^6) - y^2 (-1\\
  &+ z)^2 z^4  (25944 - 171016 z + 438304 z^2 - 585767 z^3 + 364650 z^4 - 73560 z^5 +  864 z^6) - y^7 (3312 + 52501 z - 1005911 z^2\\
  &+ 4577790 z^3 -  9135618 z^4 + 9138769 z^5 - 4491307 z^6 + 860032 z^7) +  y^6 (288 + 26312 z - 553704 z^2 + 3461777 z^3 - 9763000 z^4\\
  &+  14401588 z^5 - 11527644 z^6 + 4728907 z^7 - 774416 z^8) +  y^5 z (-6066 + 175600 z - 1559775 z^2 + 6176973 z^3 - 12917254 z^4\\
  &+  15269476 z^5 - 10196989 z^6 + 3546779 z^7 - 488960 z^8) +  y^4 z^2 (-23424 + 374888 z - 2225864 z^2 + 6614805 z^3 - 11116384 z^4\\
  &+  10963698 z^5 - 6184528 z^6 + 1787581 z^7 - 190880 z^8) +  y^3 z^3 (-32844 + 401168 z - 1810731 z^2 + 4338377 z^3 - 6108582 z^4\\
  &+  5103134 z^5 - 2392815 z^6 + 537669 z^7 - 35376 z^8)) H(1, 1, z) +  H(3, y) (96 (-1 + y)^2 y (-1 + z)^2 z (y + z)^4 (4 + 16 y^4 - 22 z\\
  &+  45 z^2 - 43 z^3 + 16 z^4 + y^3 (-43 + 20 z) +  3 y^2 (15 - 17 z + 8 z^2) + y (-22 + 48 z - 51 z^2 + 20 z^3))  H(0, 1, z) + 96 (-1\\
  &+ y)^2 y (-1 + z)^2 z (y + z)^4  (4 + 16 y^4 - 22 z + 45 z^2 - 43 z^3 + 16 z^4 + y^3 (-43 + 20 z) +  3 y^2 (15 - 17 z + 8 z^2) + y (-22\\
  &+ 48 z - 51 z^2 + 20 z^3))  H(1, 0, z) + 192 (-1 + y)^2 y (-1 + z)^2 z (y + z)^4  (4 + 16 y^4 - 22 z + 45 z^2 - 43 z^3 + 16 z^4\\
  &+ y^3 (-43 + 20 z) +  3 y^2 (15 - 17 z + 8 z^2) + y (-22 + 48 z - 51 z^2 + 20 z^3))  H(1, 1, z)) - 8 (-1 + y) y (-1 + z)^2 z  (528 y^9\\
  &+ 6 y^8 (-323 + 492 z) + 12 y^7 (237 - 866 z + 609 z^2) +  3 z^4 (-30 + 60 z - 49 z^2 + 19 z^3) +  3 y^6 (-692 + 4784 z - 7993 z^2\\
  &+ 3540 z^3) +  y z^3 (-144 + 1364 z - 1944 z^2 + 1374 z^3 - 405 z^4) +  y^5 (732 - 9572 z + 30498 z^2 - 31161 z^3 + 9960 z^4)\\
  &+  y^2 z^2 (-204 + 3656 z - 8678 z^2 + 7902 z^3 - 3411 z^4 + 348 z^5) +  2 y^3 z (-72 + 2320 z - 8830 z^2 + 11118 z^3 - 6117 z^4\\
  &+ 1086 z^5) +  y^4 (-90 + 2804 z - 18275 z^2 + 34641 z^3 - 24864 z^4 + 6048 z^5))  H(2, 0, y) + 2 (1728 y^{12} (-1 + z)^2 - 216 (-1\\
  &+ z)^5 z^6 (-1 + 4 z) -  24 y^{11} (-1 + z)^2 (375 + 718 z) + y^{10} (19080 + 28289 z - 250122 z^2 +  338949 z^3 - 136304 z^4) + y (-1\\
\end{aligned}
\end{equation*}
\begin{equation*}
\begin{aligned}
  &+ z)^2 z^5 (-8082 + 36428 z -  80967 z^2 + 91745 z^3 - 40992 z^4 + 1728 z^5) -  y^9 (20808 + 63133 z - 748547 z^2 + 1624467 z^3\\
  &- 1359901 z^4 +  400256 z^5) + y^8 (12024 + 69787 z - 1130000 z^2 + 3636494 z^3 -  4922656 z^4 + 3035579 z^5 - 701120 z^6) - y^2 (-1\\
  &+ z)^2 z^4  (25944 - 171016 z + 438304 z^2 - 585767 z^3 + 364650 z^4 - 73560 z^5 +  864 z^6) - y^7 (3312 + 52501 z - 1005911 z^2\\
  &+ 4577790 z^3 -  9135618 z^4 + 9138769 z^5 - 4491307 z^6 + 860032 z^7) +  y^6 (288 + 26312 z - 553704 z^2 + 3461777 z^3 - 9763000 z^4\\
  &+  14401588 z^5 - 11527644 z^6 + 4728907 z^7 - 774416 z^8) +  y^5 z (-6066 + 175600 z - 1559775 z^2 + 6176973 z^3 - 12917254 z^4\\
  &+  15269476 z^5 - 10196989 z^6 + 3546779 z^7 - 488960 z^8) +  y^4 z^2 (-23424 + 374888 z - 2225864 z^2 + 6614805 z^3 - 11116384 z^4\\
  &+  10963698 z^5 - 6184528 z^6 + 1787581 z^7 - 190880 z^8) +  y^3 z^3 (-32844 + 401168 z - 1810731 z^2 + 4338377 z^3 - 6108582 z^4\\
  &+  5103134 z^5 - 2392815 z^6 + 537669 z^7 - 35376 z^8)) H(2, 2, y) +  H(0, z) (-88 (-1 + y)^2 y (-1 + z) z (y + z)^4  (-5 (-1 + z)^2 (2\\
  &- 2 z + z^2) + 5 y^3 (1 - 5 z + 4 z^2) +  3 y^2 (-5 + 20 z - 24 z^2 + 8 z^3) + 4 y (5 - 20 z + 33 z^2 - 29 z^3 +  11 z^4)) + 96 (-1\\
  &+ y)^2 y (-1 + z)^2 z (y + z)^4  (1 + 8 y^4 - 9 z + 21 z^2 - 21 z^3 + 8 z^4 + y^3 (-21 + 8 z) +  3 y^2 (7 - 7 z + 4 z^2) + y (-9 + 18 z\\
  &- 21 z^2 + 8 z^3)) H(1, z)^2 -  264 (-1 + y)^2 y (-1 + z)^2 z (y + z)^4 (y^3 (-1 + 4 z) +  4 y (-1 + z)^2 (-1 + 4 z) + 3 y^2 (1 - 5 z\\
  &+ 4 z^2) +  2 (1 - 9 z + 21 z^2 - 21 z^3 + 8 z^4)) H(2, y) +  96 (-1 + y)^2 y (-1 + z)^2 z (y + z)^4 (8 y^4 + 2 y^3 (-11 + 6 z)\\
  &+  12 y^2 (2 - 3 z + 2 z^2) + y (-13 + 42 z - 57 z^2 + 24 z^3) +  3 (1 - 9 z + 21 z^2 - 21 z^3 + 8 z^4)) H(2, y)^2 +  H(1, z) (32 (-1\\
  &+ y)^2 y (-1 + z) z (-6 (-1 + z)^2 z^4 (1 - z + z^2) +  6 y^7 (1 - 10 z + 9 z^2) + 3 y^6 (-4 + 54 z - 155 z^2 + 104 z^3) +  3 y^5 (4\\
  &- 52 z + 279 z^2 - 497 z^3 + 262 z^4) +  y z^3 (30 + 8 z - 182 z^2 + 321 z^3 - 255 z^4 + 78 z^5) +  y^2 z^2 (48 + 38 z - 650 z^2\\
  &+ 1338 z^3 - 1185 z^4 + 408 z^5) +  2 y^3 z (15 + 28 z - 475 z^2 + 1158 z^3 - 1185 z^4 + 453 z^5) +  2 y^4 (-3 + 22 z - 301 z^2\\
  &+ 981 z^3 - 1260 z^4 + 552 z^5)) +  96 (-1 + y)^2 y (-1 + z)^2 z^2 (y + z)^4 (-14 + 39 z - 41 z^2 +  16 z^3 + 6 y^2 (-1 + 2 z) + 3 y (4\\
  &- 9 z + 4 z^2)) H(2, y) -  96 (-1 + y)^2 y (-1 + z)^2 z (y + z)^4 (4 + 16 y^4 - 22 z + 45 z^2 -  43 z^3 + 16 z^4 + y^3 (-43 + 20 z)\\
  &+ 3 y^2 (15 - 17 z + 8 z^2) +  y (-22 + 48 z - 51 z^2 + 20 z^3)) H(3, y)) -  96 (-1 + y)^2 y (-1 + z)^2 z (y + z)^4 (16 y^4 + 4 y^3 (-11\\
  &+ 6 z) +  6 y^2 (8 - 9 z + 2 z^2) - 3 (-2 + 4 z - 3 z^2 + z^3) +  y (-26 + 48 z - 33 z^2 + 12 z^3)) H(0, 0, y) -  96 (-1 + y)^2 y (-1\\
  &+ z)^2 z^2 (y + z)^4 (-14 + 39 z - 41 z^2 + 16 z^3 +  6 y^2 (-1 + 2 z) + 3 y (4 - 9 z + 4 z^2)) H(0, 0, z) -  96 (-1 + y)^2 y (-1\\
  &+ z)^2 z (y + z)^4 (2 + 16 y^4 - 4 z + 3 z^2 - z^3 +  2 y^3 (-21 + 8 z) + 6 y^2 (7 - 6 z + 2 z^2) +  y (-18 + 24 z - 15 z^2\\
  &+ 4 z^3)) H(0, 1, z) -  96 (-1 + y)^2 y (-1 + z)^2 z (-1 + y + z) (y + z)^4  (-4 + 18 z - 27 z^2 + 16 z^3 + y^2 (-2 + 8 z) + 2 y (2\\
  &- 7 z + 2 z^2))  H(0, 2, y) + 96 (-1 + y)^2 y (-1 + z)^2 z (y + z)^4  (2 + y^2 (3 - 9 z) - 4 z + 3 z^2 - z^3 + y^3 (-1 + 4 z) +  y (-4\\
  &+ 12 z - 9 z^2 + 4 z^3)) H(1, 0, z) - 96 (-1 + y)^2 y (-1 + z)^2  z (y + z)^4 (4 + 16 y^4 - 22 z + 45 z^2 - 43 z^3 + 16 z^4 +  y^3 (-43\\
  &+ 20 z) + 3 y^2 (15 - 17 z + 8 z^2) +  y (-22 + 48 z - 51 z^2 + 20 z^3)) H(1, 1, z) -  96 (-1 + y)^2 y (-1 + z)^2 z (-1 + y + z) (y\\
  &+ z)^4  (-4 + 18 z - 27 z^2 + 16 z^3 + y^2 (-2 + 8 z) + 2 y (2 - 7 z + 2 z^2))  H(2, 0, y) - 192 (-1 + y)^2 y (-1 + z)^2 z (y\\
  &+ z)^4  (8 y^4 + 2 y^3 (-11 + 6 z) + 12 y^2 (2 - 3 z + 2 z^2) +  y (-13 + 42 z - 57 z^2 + 24 z^3) + 3 (1 - 9 z + 21 z^2 - 21 z^3\\
  &+  8 z^4)) H(2, 2, y)) + 264 (-1 + y)^2 y (-1 + z)^2 z (y + z)^4  (4 + 16 y^4 - 22 z + 45 z^2 - 43 z^3 + 16 z^4 + y^3 (-43 + 20 z)\\
  &+  3 y^2 (15 - 17 z + 8 z^2) + y (-22 + 48 z - 51 z^2 + 20 z^3))  H(3, 2, y) + H(0, y) (-88 (-1 + y) y (-1 + z)^2 z (y + z)^4  (y^4 (-5\\
  &+ 44 z) + 4 y^3 (5 - 29 z + 6 z^2) +  5 (-2 + 4 z - 3 z^2 + z^3) - 5 y (-6 + 16 z - 12 z^2 + 5 z^3) +  y^2 (-35 + 132 z - 72 z^2\\
  &+ 20 z^3)) + 48 (-1 + y)^2 y (-1 + z)^2 z  (y + z)^4 (6 - 26 z + 48 z^2 - 44 z^3 + 16 z^4 + 3 y^3 (-1 + 4 z) +  3 y^2 (3 - 11 z + 4 z^2)\\
  &+ 6 y (-2 + 8 z - 9 z^2 + 4 z^3)) H(0, z)^2 +  96 (-1 + y)^2 y (-1 + z)^2 z (y + z)^4 (3 + 24 y^4 - 13 z + 24 z^2 -  22 z^3 + 8 z^4\\
  &+ 3 y^3 (-21 + 8 z) + 3 y^2 (21 - 19 z + 8 z^2) +  3 y (-9 + 14 z - 12 z^2 + 4 z^3)) H(1, z)^2 +  32 (-1 + y) y (-1 + z)^2 z (y^8 (-6\\
  &+ 78 z) +  3 y^7 (6 - 85 z + 136 z^2) + 6 z^4 (-1 + 2 z - 2 z^2 + z^3) +  3 y^6 (-8 + 107 z - 395 z^2 + 302 z^3) +  2 y z^3 (15 + 22 z\\
  &- 78 z^2 + 81 z^3 - 30 z^4) +  2 y^5 (9 - 91 z + 669 z^2 - 1185 z^3 + 552 z^4) +  y^2 z^2 (48 + 56 z - 602 z^2 + 837 z^3 - 465 z^4\\
  &+ 54 z^5) +  y^3 z (30 + 38 z - 950 z^2 + 1962 z^3 - 1491 z^4 + 312 z^5) +  y^4 (-6 + 8 z - 650 z^2 + 2316 z^3 - 2520 z^4\\
  &+ 786 z^5)) H(2, y) +  96 (-1 + y)^2 y (-1 + z)^2 z (y + z)^4 (1 + 8 y^4 - 9 z + 21 z^2 -  21 z^3 + 8 z^4 + y^3 (-21 + 8 z) + 3 y^2 (7\\
  &- 7 z + 4 z^2) +  y (-9 + 18 z - 21 z^2 + 8 z^3)) H(2, y)^2 +  H(0, z) (-264 (-1 + y)^2 y (-1 + z)^2 z (y + z)^4  (2 + y^2 (3 - 9 z)\\
  &- 4 z + 3 z^2 - z^3 + y^3 (-1 + 4 z) +  y (-4 + 12 z - 9 z^2 + 4 z^3)) + 96 (-1 + y)^2 y (-1 + z)^2 z  (-1 + y + z) (y + z)^4 (16 y^3\\
  &+ y^2 (-27 + 4 z) - 2 (2 - 2 z + z^2) +  2 y (9 - 7 z + 4 z^2)) H(1, z) + 96 (-1 + y)^2 y (-1 + z)^2 z  (-1 + y + z) (y + z)^4 (-4\\
  &+ 18 z - 27 z^2 + 16 z^3 + y^2 (-2 + 8 z) +  2 y (2 - 7 z + 2 z^2)) H(2, y)) +  H(1, z) (-264 (-1 + y)^2 y (-1 + z)^2 z (y + z)^4  (2\\
  &+ 16 y^4 - 4 z + 3 z^2 - z^3 + 2 y^3 (-21 + 8 z) +  6 y^2 (7 - 6 z + 2 z^2) + y (-18 + 24 z - 15 z^2 + 4 z^3)) +  96 (-1 + y)^2 y^2 (-1\\
  &+ z)^2 z (y + z)^4 (16 y^3 + y^2 (-41 + 12 z) -  2 (7 - 6 z + 3 z^2) + 3 y (13 - 9 z + 4 z^2)) H(2, y) -  96 (-1 + y)^2 y (-1\\
  &+ z)^2 z (y + z)^4 (4 + 16 y^4 - 22 z + 45 z^2 -  43 z^3 + 16 z^4 + y^3 (-43 + 20 z) + 3 y^2 (15 - 17 z + 8 z^2) +  y (-22 + 48 z\\
  &- 51 z^2 + 20 z^3)) H(3, y)) -  96 (-1 + y)^2 y^2 (-1 + z)^2 z (y + z)^4 (16 y^3 + y^2 (-41 + 12 z) -  2 (7 - 6 z + 3 z^2) + 3 y (13\\
  &- 9 z + 4 z^2)) H(0, 0, y) -  96 (-1 + y)^2 y (-1 + z)^2 z (y + z)^4 (6 - 26 z + 48 z^2 - 44 z^3 +  16 z^4 + 3 y^3 (-1 + 4 z) + 3 y^2 (3\\
  &- 11 z + 4 z^2) +  6 y (-2 + 8 z - 9 z^2 + 4 z^3)) H(0, 0, z) -  96 (-1 + y)^2 y (-1 + z)^2 z (-1 + y + z) (y + z)^4  (16 y^3 + y^2 (-27\\
  &+ 4 z) - 2 (2 - 2 z + z^2) + 2 y (9 - 7 z + 4 z^2))  H(0, 1, z) + 96 (-1 + y)^2 y (-1 + z)^2 z (y + z)^4  (2 + 16 y^4 - 4 z + 3 z^2\\
  &- z^3 + 2 y^3 (-21 + 8 z) +  6 y^2 (7 - 6 z + 2 z^2) + y (-18 + 24 z - 15 z^2 + 4 z^3)) H(0, 2, y) -  96 (-1 + y)^2 y^2 (-1 + z)^2 z (y\\
\end{aligned}
\end{equation*}
\begin{equation*}
\begin{aligned}
  &+ z)^4 (16 y^3 + y^2 (-41 + 12 z) -  2 (7 - 6 z + 3 z^2) + 3 y (13 - 9 z + 4 z^2)) H(1, 0, y) -  96 (-1 + y)^2 y (-1 + z)^2 z (-1 + y\\
  &+ z) (y + z)^4  (16 y^3 + y^2 (-27 + 4 z) - 2 (2 - 2 z + z^2) + 2 y (9 - 7 z + 4 z^2))  H(1, 0, z) - 192 (-1 + y)^2 y (-1 + z)^2 z (y\\
  &+ z)^4  (3 + 24 y^4 - 13 z + 24 z^2 - 22 z^3 + 8 z^4 + 3 y^3 (-21 + 8 z) +  3 y^2 (21 - 19 z + 8 z^2) + 3 y (-9 + 14 z - 12 z^2\\
  &+ 4 z^3))  H(1, 1, z) + 96 (-1 + y)^2 y (-1 + z)^2 z (y + z)^4  (2 + 16 y^4 - 4 z + 3 z^2 - z^3 + 2 y^3 (-21 + 8 z) +  6 y^2 (7 - 6 z\\
  &+ 2 z^2) + y (-18 + 24 z - 15 z^2 + 4 z^3)) H(2, 0, y) -  96 (-1 + y)^2 y (-1 + z)^2 z (y + z)^4 (4 + 16 y^4 - 22 z + 45 z^2 -  43 z^3\\
  &+ 16 z^4 + y^3 (-43 + 20 z) + 3 y^2 (15 - 17 z + 8 z^2) +  y (-22 + 48 z - 51 z^2 + 20 z^3)) H(2, 2, y) -  96 (-1 + y)^2 y (-1\\
  &+ z)^2 z (y + z)^4 (4 + 16 y^4 - 22 z + 45 z^2 -  43 z^3 + 16 z^4 + y^3 (-43 + 20 z) + 3 y^2 (15 - 17 z + 8 z^2) +  y (-22 + 48 z\\
  &- 51 z^2 + 20 z^3)) H(3, 2, y)) +  H(2, y) (-44 (-1 + y)^2 y (-1 + z)^2 z (y^7 (-9 + 84 z) +  3 y^6 (9 - 83 z + 144 z^2) - 9 z^4 (-2\\
  &+ 4 z - 3 z^2 + z^3) +  3 y^5 (-12 + 92 z - 373 z^2 + 324 z^3) +  y z^3 (-36 - 136 z + 276 z^2 - 249 z^3 + 84 z^4) +  y^2 z^2 (-60\\
  &- 292 z + 1053 z^2 - 1119 z^3 + 432 z^4) +  y^3 z (-36 - 292 z + 1608 z^2 - 2175 z^3 + 972 z^4) +  y^4 (18 - 136 z + 1053 z^2 - 2175 z^3\\
  &+ 1248 z^4)) -  96 (-1 + y)^2 y^2 (-1 + z)^2 z (y + z)^4 (16 y^3 + y^2 (-41 + 12 z) -  2 (7 - 6 z + 3 z^2) + 3 y (13 - 9 z\\
  &+ 4 z^2)) H(0, 0, y) -  288 (-1 + y)^2 y (-1 + z)^2 z (y + z)^4 (y^3 (-1 + 4 z) +  4 y (-1 + z)^2 (-1 + 4 z) + 3 y^2 (1 - 5 z + 4 z^2)\\
  &+  2 (1 - 9 z + 21 z^2 - 21 z^3 + 8 z^4)) H(0, 0, z) -  96 (-1 + y)^2 y (-1 + z)^2 z^2 (y + z)^4 (-14 + 39 z - 41 z^2 + 16 z^3\\
  &+  6 y^2 (-1 + 2 z) + 3 y (4 - 9 z + 4 z^2)) H(0, 1, z) +  96 (-1 + y)^2 y (-1 + z)^2 z (y + z)^4 (2 + 16 y^4 - 4 z + 3 z^2 - z^3\\
  &+  2 y^3 (-21 + 8 z) + 6 y^2 (7 - 6 z + 2 z^2) +  y (-18 + 24 z - 15 z^2 + 4 z^3)) H(0, 2, y) -  96 (-1 + y)^2 y^2 (-1 + z)^2 z (y\\
  &+ z)^4 (16 y^3 + y^2 (-41 + 12 z) -  2 (7 - 6 z + 3 z^2) + 3 y (13 - 9 z + 4 z^2)) H(1, 0, y) -  96 (-1 + y)^2 y (-1 + z)^2 z^2 (y\\
  &+ z)^4 (-14 + 39 z - 41 z^2 + 16 z^3 +  6 y^2 (-1 + 2 z) + 3 y (4 - 9 z + 4 z^2)) H(1, 0, z) +  192 (-1 + y)^2 y (-1 + z)^2 z (y\\
  &+ z)^4 (13 + 380 y^4 - 225 z + 747 z^2 -  915 z^3 + 380 z^4 + y^3 (-915 + 584 z) + 9 y^2 (83 - 117 z + 64 z^2) +  y (-225 + 654 z\\
  &- 1053 z^2 + 584 z^3)) H(1, 1, z) +  96 (-1 + y)^2 y (-1 + z)^2 z (y + z)^4 (2 + 16 y^4 - 4 z + 3 z^2 - z^3 +  2 y^3 (-21 + 8 z)\\
  &+ 6 y^2 (7 - 6 z + 2 z^2) +  y (-18 + 24 z - 15 z^2 + 4 z^3)) H(2, 0, y) -  96 (-1 + y)^2 y (-1 + z)^2 z (y + z)^4 (4 + 16 y^4 - 22 z\\
  &+ 45 z^2 -  43 z^3 + 16 z^4 + y^3 (-43 + 20 z) + 3 y^2 (15 - 17 z + 8 z^2) +  y (-22 + 48 z - 51 z^2 + 20 z^3)) H(2, 2, y) -  96 (-1\\
  &+ y)^2 y (-1 + z)^2 z (y + z)^4 (4 + 16 y^4 - 22 z + 45 z^2 -  43 z^3 + 16 z^4 + y^3 (-43 + 20 z) + 3 y^2 (15 - 17 z + 8 z^2) +  y (-22\\
  &+ 48 z - 51 z^2 + 20 z^3)) H(3, 2, y)) +  H(1, z) (-44 (-1 + y)^2 y (-1 + z)^2 z (y^7 (-9 + 84 z) +  3 y^6 (9 - 83 z + 144 z^2)\\
  &- 9 z^4 (-2 + 4 z - 3 z^2 + z^3) +  3 y^5 (-12 + 92 z - 373 z^2 + 324 z^3) +  y z^3 (-36 - 136 z + 276 z^2 - 249 z^3 + 84 z^4)\\
  &+  y^2 z^2 (-60 - 292 z + 1053 z^2 - 1119 z^3 + 432 z^4) +  y^3 z (-36 - 292 z + 1608 z^2 - 2175 z^3 + 972 z^4) +  y^4 (18 - 136 z\\
  &+ 1053 z^2 - 2175 z^3 + 1248 z^4)) -  96 (-1 + y)^2 y (-1 + z)^2 z (y + z)^4 (13 + 380 y^4 - 225 z + 747 z^2 -  915 z^3 + 380 z^4\\
  &+ y^3 (-915 + 584 z) + 9 y^2 (83 - 117 z + 64 z^2) +  y (-225 + 654 z - 1053 z^2 + 584 z^3)) H(2, y)^2 +  264 (-1 + y)^2 y (-1\\
  &+ z)^2 z (y + z)^4 (4 + 16 y^4 - 22 z + 45 z^2 -  43 z^3 + 16 z^4 + y^3 (-43 + 20 z) + 3 y^2 (15 - 17 z + 8 z^2) +  y (-22 + 48 z\\
  &- 51 z^2 + 20 z^3)) H(3, y) - 96 (-1 + y)^2 y (-1 + z)^2  z (y + z)^4 (4 + 16 y^4 - 22 z + 45 z^2 - 43 z^3 + 16 z^4 +  y^3 (-43 + 20 z)\\
  &+ 3 y^2 (15 - 17 z + 8 z^2) +  y (-22 + 48 z - 51 z^2 + 20 z^3)) H(2, y) H(3, y) -  288 (-1 + y)^2 y (-1 + z)^2 z (y + z)^4 (2 + 16 y^4\\
  &- 4 z + 3 z^2 - z^3 +  2 y^3 (-21 + 8 z) + 6 y^2 (7 - 6 z + 2 z^2) +  y (-18 + 24 z - 15 z^2 + 4 z^3)) H(0, 0, y) -  96 (-1 + y)^2 y (-1\\
  &+ z)^2 z^2 (y + z)^4 (-14 + 39 z - 41 z^2 + 16 z^3 +  6 y^2 (-1 + 2 z) + 3 y (4 - 9 z + 4 z^2)) H(0, 0, z) -  96 (-1 + y)^2 y (-1\\
  &+ z)^2 z (y + z)^4 (2 + 16 y^4 - 4 z + 3 z^2 - z^3 +  2 y^3 (-21 + 8 z) + 6 y^2 (7 - 6 z + 2 z^2) +  y (-18 + 24 z - 15 z^2\\
  &+ 4 z^3)) H(0, 1, z) -  96 (-1 + y)^2 y^2 (-1 + z)^2 z (y + z)^4 (16 y^3 + y^2 (-41 + 12 z) -  2 (7 - 6 z + 3 z^2) + 3 y (13 - 9 z\\
  &+ 4 z^2)) H(0, 2, y) +  96 (-1 + y)^2 y (-1 + z)^2 z (y + z)^4 (4 + 16 y^4 - 22 z + 45 z^2 -  43 z^3 + 16 z^4 + y^3 (-43 + 20 z)\\
  &+ 3 y^2 (15 - 17 z + 8 z^2) +  y (-22 + 48 z - 51 z^2 + 20 z^3)) H(0, 3, y) +  96 (-1 + y)^2 y (-1 + z)^2 z (y + z)^4 (2 + y^2 (3 - 9 z)\\
  &- 4 z + 3 z^2 -  z^3 + y^3 (-1 + 4 z) + y (-4 + 12 z - 9 z^2 + 4 z^3)) H(1, 0, z) -  96 (-1 + y)^2 y (-1 + z)^2 z (y + z)^4 (4 + 16 y^4\\
  &- 22 z + 45 z^2 -  43 z^3 + 16 z^4 + y^3 (-43 + 20 z) + 3 y^2 (15 - 17 z + 8 z^2) +  y (-22 + 48 z - 51 z^2 + 20 z^3)) H(1, 1, z)\\
  &-  96 (-1 + y)^2 y^2 (-1 + z)^2 z (y + z)^4 (16 y^3 + y^2 (-41 + 12 z) -  2 (7 - 6 z + 3 z^2) + 3 y (13 - 9 z + 4 z^2)) H(2, 0, y)\\
  &+  192 (-1 + y)^2 y (-1 + z)^2 z (y + z)^4 (13 + 380 y^4 - 225 z + 747 z^2 -  915 z^3 + 380 z^4 + y^3 (-915 + 584 z) + 9 y^2 (83 - 117 z\\
  &+ 64 z^2) +  y (-225 + 654 z - 1053 z^2 + 584 z^3)) H(2, 2, y) +  96 (-1 + y)^2 y (-1 + z)^2 z (y + z)^4 (4 + 16 y^4 - 22 z + 45 z^2\\
  &-  43 z^3 + 16 z^4 + y^3 (-43 + 20 z) + 3 y^2 (15 - 17 z + 8 z^2) +  y (-22 + 48 z - 51 z^2 + 20 z^3)) H(2, 3, y) +  96 (-1 + y)^2 y (-1\\
  &+ z)^2 z (y + z)^4 (4 + 16 y^4 - 22 z + 45 z^2 -  43 z^3 + 16 z^4 + y^3 (-43 + 20 z) + 3 y^2 (15 - 17 z + 8 z^2) +  y (-22 + 48 z\\
  &- 51 z^2 + 20 z^3)) H(3, 0, y) +  96 (-1 + y)^2 y (-1 + z)^2 z (y + z)^4 (4 + 16 y^4 - 22 z + 45 z^2 -  43 z^3 + 16 z^4 + y^3 (-43\\
  &+ 20 z) + 3 y^2 (15 - 17 z + 8 z^2) +  y (-22 + 48 z - 51 z^2 + 20 z^3)) H(3, 2, y)) +  48 (-1 + y)^2 y (-1 + z)^2 z (y + z)^4 (56\\
  &+ 596 y^4 + 34 z - 396 z^2 +  550 z^3 - 244 z^4 + y^3 (-1661 + 1420 z) +  3 y^2 (511 - 769 z + 204 z^2) - 4 y (131 - 237 z + 51 z^2\\
  &+ 38 z^3))  H(0, 0, 0, y) - 48 (-1 + y)^2 y (-1 + z)^2 z (y + z)^4  (-56 + 244 y^4 + 524 z - 1533 z^2 + 1661 z^3 - 596 z^4\\
  &+  2 y^3 (-275 + 76 z) + y^2 (396 + 204 z - 612 z^2) -  y (34 + 948 z - 2307 z^2 + 1420 z^3)) H(0, 0, 0, z) +  96 (-1 + y)^2 y (-1\\
  &+ z)^2 z (y + z)^4 (2 + 32 y^4 - 4 z + 3 z^2 - z^3 +  y^3 (-83 + 28 z) + 3 y^2 (27 - 21 z + 8 z^2) +  y (-32 + 36 z - 21 z^2\\
  &+ 4 z^3)) H(0, 0, 1, z) -  96 (-1 + y)^2 y (-1 + z)^2 z (y + z)^4 (32 y^4 + y^3 (-85 + 36 z) +  3 y^2 (29 - 27 z + 8 z^2) - 3 (-2 + 4 z\\
\end{aligned}
\end{equation*}
\begin{equation*}
\begin{aligned}
  &- 3 z^2 + z^3) +  y (-40 + 60 z - 39 z^2 + 12 z^3)) H(0, 0, 2, y) +  96 (-1 + y)^2 y^2 (-1 + z)^2 z (y + z)^4 (16 y^3 + y^2 (-41 + 12 z)\\
  &-  2 (7 - 6 z + 3 z^2) + 3 y (13 - 9 z + 4 z^2)) H(0, 1, 0, y) +  96 (-1 + y)^2 y (-1 + z)^2 z (y + z)^4 (-2 + 16 y^4 + 8 y^3 (-5 + z)\\
  &+  4 z - 3 z^2 + z^3 + 6 y^2 (6 - 3 z + 2 z^2) + y (-10 + 3 z^2 - 4 z^3))  H(0, 1, 0, z) + 96 (-1 + y)^2 y (-1 + z)^2 z (y\\
  &+ z)^4  (32 y^4 + y^3 (-85 + 36 z) + 3 y^2 (29 - 27 z + 8 z^2) -  3 (-2 + 4 z - 3 z^2 + z^3) + y (-40 + 60 z - 39 z^2\\
  &+ 12 z^3))  H(0, 1, 1, z) - 96 (-1 + y)^2 y (-1 + z)^2 z (y + z)^4  (32 y^4 + y^3 (-85 + 36 z) + 3 y^2 (29 - 27 z + 8 z^2) -  3 (-2\\
  &+ 4 z - 3 z^2 + z^3) + y (-40 + 60 z - 39 z^2 + 12 z^3))  H(0, 2, 0, y) - 96 (-1 + y)^2 y (-1 + z)^2 z (y + z)^4  (2 + 32 y^4 - 4 z\\
  &+ 3 z^2 - z^3 + y^3 (-83 + 28 z) +  3 y^2 (27 - 21 z + 8 z^2) + y (-32 + 36 z - 21 z^2 + 4 z^3))  H(0, 2, 2, y) + 96 (-1 + y)^2 y (-1\\
  &+ z)^2 z (y + z)^4  (4 + 16 y^4 - 22 z + 45 z^2 - 43 z^3 + 16 z^4 + y^3 (-43 + 20 z) +  3 y^2 (15 - 17 z + 8 z^2) + y (-22 + 48 z\\
  &- 51 z^2 + 20 z^3))  H(0, 3, 2, y) + 192 (-1 + y)^2 y^2 (-1 + z)^2 z (y + z)^4  (16 y^3 + y^2 (-41 + 12 z) - 2 (7 - 6 z + 3 z^2)\\
  &+ 3 y (13 - 9 z + 4 z^2))  H(1, 0, 0, y) - 288 (-1 + y)^2 y (-1 + z)^2 z (y + z)^4  (2 + y^2 (3 - 9 z) - 4 z + 3 z^2 - z^3 + y^3 (-1\\
  &+ 4 z) +  y (-4 + 12 z - 9 z^2 + 4 z^3)) H(1, 0, 0, z) +  96 (-1 + y)^2 y (-1 + z)^2 z (y + z)^4 (2 + 16 y^4 - 4 z + 3 z^2 - z^3\\
  &+  2 y^3 (-21 + 8 z) + 6 y^2 (7 - 6 z + 2 z^2) +  y (-18 + 24 z - 15 z^2 + 4 z^3)) H(1, 0, 1, z) +  96 (-1 + y)^2 y^2 (-1 + z)^2 z (y\\
  &+ z)^4 (16 y^3 + y^2 (-41 + 12 z) -  2 (7 - 6 z + 3 z^2) + 3 y (13 - 9 z + 4 z^2)) H(1, 0, 2, y) -  96 (-1 + y)^2 y (-1 + z)^2 z (y\\
  &+ z)^4 (2 + y^2 (3 - 9 z) - 4 z + 3 z^2 -  z^3 + y^3 (-1 + 4 z) + y (-4 + 12 z - 9 z^2 + 4 z^3)) H(1, 1, 0, z) +  96 (-1 + y)^2 y (-1\\
  &+ z)^2 z (y + z)^4 (30 + 776 y^4 - 472 z + 1539 z^2 -  1873 z^3 + 776 z^4 + y^3 (-1873 + 1188 z) +  3 y^2 (513 - 719 z + 392 z^2)\\
  &+ y (-472 + 1356 z - 2157 z^2 + 1188 z^3))  H(1, 1, 1, z) + 96 (-1 + y)^2 y^2 (-1 + z)^2 z (y + z)^4  (16 y^3 + y^2 (-41 + 12 z) - 2 (7\\
  &- 6 z + 3 z^2) + 3 y (13 - 9 z + 4 z^2))  H(1, 2, 0, y) - 96 (-1 + y)^2 y (-1 + z)^2 z (y + z)^4  (32 y^4 + y^3 (-85 + 36 z) + 3 y^2 (29\\
  &- 27 z + 8 z^2) -  3 (-2 + 4 z - 3 z^2 + z^3) + y (-40 + 60 z - 39 z^2 + 12 z^3))  H(2, 0, 0, y) - 96 (-1 + y)^2 y (-1 + z)^2 z (y\\
  &+ z)^4  (2 + 32 y^4 - 4 z + 3 z^2 - z^3 + y^3 (-83 + 28 z) +  3 y^2 (27 - 21 z + 8 z^2) + y (-32 + 36 z - 21 z^2\\
  &+ 4 z^3))  H(2, 0, 2, y) + 96 (-1 + y)^2 y^2 (-1 + z)^2 z (y + z)^4  (16 y^3 + y^2 (-41 + 12 z) - 2 (7 - 6 z + 3 z^2) + 3 y (13 - 9 z\\
  &+ 4 z^2))  H(2, 1, 0, y) - 96 (-1 + y)^2 y (-1 + z)^2 z (y + z)^4  (2 + 32 y^4 - 4 z + 3 z^2 - z^3 + y^3 (-83 + 28 z) +  3 y^2 (27\\
  &- 21 z + 8 z^2) + y (-32 + 36 z - 21 z^2 + 4 z^3))  H(2, 2, 0, y) + 96 (-1 + y)^2 y (-1 + z)^2 z (y + z)^4  (30 + 776 y^4 - 472 z\\
  &+ 1539 z^2 - 1873 z^3 + 776 z^4 +  y^3 (-1873 + 1188 z) + 3 y^2 (513 - 719 z + 392 z^2) +  y (-472 + 1356 z - 2157 z^2\\
  &+ 1188 z^3)) H(2, 2, 2, y) +  96 (-1 + y)^2 y (-1 + z)^2 z (y + z)^4 (4 + 16 y^4 - 22 z + 45 z^2 -  43 z^3 + 16 z^4 + y^3 (-43 + 20 z)\\
  &+ 3 y^2 (15 - 17 z + 8 z^2) +  y (-22 + 48 z - 51 z^2 + 20 z^3)) H(2, 3, 2, y) +  96 (-1 + y)^2 y (-1 + z)^2 z (y + z)^4 (4 + 16 y^4\\
  &- 22 z + 45 z^2 -  43 z^3 + 16 z^4 + y^3 (-43 + 20 z) + 3 y^2 (15 - 17 z + 8 z^2) +  y (-22 + 48 z - 51 z^2 + 20 z^3)) H(3, 0, 2, y)\\
  &+  96 (-1 + y)^2 y (-1 + z)^2 z (y + z)^4 (4 + 16 y^4 - 22 z + 45 z^2 -  43 z^3 + 16 z^4 + y^3 (-43 + 20 z) + 3 y^2 (15 - 17 z + 8 z^2)\\
  &+  y (-22 + 48 z - 51 z^2 + 20 z^3)) H(3, 2, 0, y) +  192 (-1 + y)^2 y (-1 + z)^2 z (y + z)^4 (4 + 16 y^4 - 22 z + 45 z^2 -  43 z^3\\
  &+ 16 z^4 + y^3 (-43 + 20 z) + 3 y^2 (15 - 17 z + 8 z^2) +  y (-22 + 48 z - 51 z^2 + 20 z^3)) H(3, 2, 2, y)\Bigg\}\Big/ \Big(192 (-1 + y)^2 y^2 (-1\\
  &+ z)^2 z^2 (-1 + y + z) (y + z)^4\Big);\\[10pt]
\mathcal{A}_{3;C_{F}^{2}}^{(2)} &=  
\Bigg\{-18 (-1 + y)^2 y (-1 + z)^2 z (y + z)^2 (2 + y^2 (3 - 9 z) - 4 z + 3 z^2 -  z^3 + y^3 (-1 + 4 z) + y (-4 + 12 z - 9 z^2 + 4 z^3))\\
  &-  12 (-1 + y)^2 y (-1 + z)^2 z (-1 + y + z) (y + z)^2  (-44 + 528 y^3 + 66 z - 99 z^2 + 64 z^3 + y^2 (-829 + 412 z) +  2 y (187 - 176 z\\
  &+ 58 z^2)) H(0, y)^3 - 12 (-1 + y)^2 y (-1 + z)^2 z  (-1 + y + z) (y + z)^2 (-44 + 64 y^3 + 374 z - 829 z^2 + 528 z^3 +  y^2 (-99\\
  &+ 116 z) + y (66 - 352 z + 412 z^2)) H(0, z)^3 -  12 (-1 + y)^2 y (-1 + z)^2 z (-1 + y + z) (y + z)^2  (4 + 72 y^3 + 46 z - 131 z^2\\
  &+ 72 z^3 + y^2 (-131 + 108 z) +  2 y (23 - 50 z + 54 z^2)) H(1, z)^3 -  2 (432 y^{10} (-1 + z)^2 - 54 (-1 + z)^5 z^4 (-1 + 4 z)\\
  &+  2 y^9 (-1 + z)^2 (-1125 + 1321 z) +  3 y^8 (1590 - 7130 z + 10312 z^2 - 5603 z^3 + 822 z^4) +  y^7 (-5202 + 29970 z - 47113 z^2\\
  &+ 16349 z^3 + 14834 z^4 - 8838 z^5) +  y (-1 + z)^2 z^3 (-1653 + 4845 z - 5811 z^2 + 4198 z^3 - 2002 z^4 +  432 z^5) + y^6 (3006\\
  &- 18611 z + 14103 z^2 + 65692 z^3 - 134393 z^4 +  90425 z^5 - 20168 z^6) - y^2 (-1 + z)^2 z^2  (792 - 14109 z + 32355 z^2 - 27545 z^3\\
  &+ 10016 z^4 - 1898 z^5 +  216 z^6) + y^5 (-828 + 945 z + 35948 z^2 - 178242 z^3 + 329525 z^4 -  282687 z^5 + 110081 z^6 - 14742 z^7)\\
  &+  y^4 (72 + 4209 z - 39045 z^2 + 174583 z^3 - 372816 z^4 + 403037 z^5 -  220595 z^6 + 54578 z^7 - 4050 z^8) +  y^3 z (-1401 + 13065 z\\
  &- 78255 z^2 + 213625 z^3 - 291633 z^4 +  209152 z^5 - 75802 z^6 + 11847 z^7 - 598 z^8)) H(2, y)^2 -  12 (-1 + y)^2 y (-1 + z)^2 z (-1\\
  &+ y + z) (y + z)^2  (4 + 72 y^3 + 46 z - 131 z^2 + 72 z^3 + y^2 (-131 + 108 z) +  2 y (23 - 50 z + 54 z^2)) H(2, y)^3\\
  &+  H(0, z)^2 (-108 (-1 + z)^6 z^3 (-1 + 4 z) - 4 y^9 (-1 + z)^2  (216 + 215 z) + y^8 (4500 - 8947 z - 426 z^2 + 9495 z^3 - 4692 z^4)\\
  &+  y (-1 + z)^3 z^2 (-72 + 2469 z - 8130 z^2 + 15633 z^3 - 12928 z^4 +  864 z^5) - y^7 (9540 - 30281 z + 21427 z^2 + 17363 z^3\\
  &- 25789 z^4 +  7740 z^5) - 2 y^6 (-5202 + 24062 z - 33217 z^2 + 1427 z^3 + 33635 z^4 -  28049 z^5 + 7274 z^6) - y^2 (-1 + z)^2 z (180\\
  &+ 114 z - 13297 z^2 +  48726 z^3 - 88199 z^4 + 76698 z^5 - 22076 z^6 + 432 z^7) -  4 y^5 (1503 - 9820 z + 19651 z^2 - 5048 z^3\\
  &- 33560 z^4 + 51210 z^5 -  31336 z^6 + 7400 z^7) + y^4 (1656 - 15965 z + 40998 z^2 + 3632 z^3 -  193426 z^4 + 388782 z^5 - 361522 z^6\\
  &+ 164423 z^7 - 28648 z^8) -  y^3 (144 - 2787 z + 8149 z^2 + 23010 z^3 - 166174 z^4 + 387958 z^5 -  474394 z^6 + 314077 z^7 - 100103 z^8\\
 \end{aligned}
 \end{equation*}
 \begin{equation*}
 \begin{aligned}
  &+ 10120 z^9) +  48 (-1 + y)^2 y (y^2 + 2 y (-1 + z) + 2 (-1 + z)^2) (-1 + z)^2 z  (-1 + y + z) (y + z)^2 (-1 + 4 z) H(1, z) +  48 (-1\\
  &+ y)^2 y (y^2 + 2 y (-1 + z) + 2 (-1 + z)^2) (-1 + z)^2 z  (-1 + y + z) (y + z)^2 (-1 + 4 z) H(2, y)) +  H(0, y)^2 (864 y^{10} (-1 + z)^2\\
  &+ 108 (-1 + z)^5 z^3 (-1 + 4 z) -  4 y^9 (-1 + z)^2 (1341 + 1234 z) + y^8 (14040 - 14991 z - 32938 z^2 +  54743 z^3 - 20800 z^4)\\
  &+ y^7 (-19944 + 36561 z + 52567 z^2 -  160807 z^3 + 117263 z^4 - 25640 z^5) - y (-1 + z)^2 z^2  (-72 + 2577 z - 8383 z^2 + 13377 z^3\\
  &- 10501 z^4 + 3020 z^5) -  4 y^6 (-4104 + 13505 z + 4189 z^2 - 53569 z^3 + 65725 z^4 - 30346 z^5 +  4627 z^6) + y^2 (-1 + z)^2 z (-180\\
  &- 366 z + 14249 z^2 - 35168 z^3 +  44729 z^4 - 26794 z^5 + 4744 z^6) -  2 y^5 (3834 - 21096 z + 12153 z^2 + 79661 z^3 - 164736 z^4\\
  &+ 132075 z^5 -  49685 z^6 + 7794 z^7) + y^4 (1800 - 16449 z + 20344 z^2 + 84562 z^3 -  275038 z^4 + 340574 z^5 - 219676 z^6 + 73813 z^7\\
  &- 9876 z^8) -  y^3 (144 - 2823 z + 4349 z^2 + 40704 z^3 - 163328 z^4 + 272002 z^5 -  250136 z^6 + 128747 z^7 - 31815 z^8 + 2156 z^9)\\
  &+  48 (-1 + y)^2 y (-1 + 4 y) (-1 + z)^2 z (-1 + y + z) (y + z)^2  (2 + 2 y^2 + 2 y (-2 + z) - 2 z + z^2) H(1, z) +  48 (-1 + y)^2 y (-1\\
  &+ 4 y) (-1 + z)^2 z (-1 + y + z) (y + z)^2  (2 + 2 y^2 + 2 y (-2 + z) - 2 z + z^2) H(2, y)) +  H(1, z)^2 (-2 (432 y^{10} (-1 + z)^2\\
  &- 54 (-1 + z)^5 z^4 (-1 + 4 z) +  2 y^9 (-1 + z)^2 (-1125 + 1321 z) + 3 y^8 (1590 - 7130 z + 10312 z^2 -  5603 z^3 + 822 z^4)\\
  &+ y^7 (-5202 + 29970 z - 47113 z^2 + 16349 z^3 +  14834 z^4 - 8838 z^5) + y (-1 + z)^2 z^3 (-1653 + 4845 z - 5811 z^2 +  4198 z^3\\
  &- 2002 z^4 + 432 z^5) + y^6 (3006 - 18611 z + 14103 z^2 +  65692 z^3 - 134393 z^4 + 90425 z^5 - 20168 z^6) -  y^2 (-1 + z)^2 z^2 (792\\
  &- 14109 z + 32355 z^2 - 27545 z^3 + 10016 z^4 -  1898 z^5 + 216 z^6) + y^5 (-828 + 945 z + 35948 z^2 - 178242 z^3 +  329525 z^4\\
  &- 282687 z^5 + 110081 z^6 - 14742 z^7) +  y^4 (72 + 4209 z - 39045 z^2 + 174583 z^3 - 372816 z^4 + 403037 z^5 -  220595 z^6 + 54578 z^7\\
  &- 4050 z^8) +  y^3 z (-1401 + 13065 z - 78255 z^2 + 213625 z^3 - 291633 z^4 +  209152 z^5 - 75802 z^6 + 11847 z^7 - 598 z^8)) -  60 (-1\\
  &+ y)^2 y (-1 + z)^2 z (-1 + y + z) (y + z)^2  (-4 + 56 y^3 + 50 z - 109 z^2 + 56 z^3 + y^2 (-109 + 84 z) +  y (50 - 92 z\\
  &+ 84 z^2)) H(2, y) - 96 (-1 + y)^2 y (-1 + z)^2 z  (-1 + y + z) (y + z)^2 (-4 + 8 y^3 + 14 z - 19 z^2 + 8 z^3 +  y^2 (-19 + 12 z)\\
  &+ 2 y (7 - 10 z + 6 z^2)) H(3, y)) -  2 (864 y^{10} (-1 + z)^2 + 108 (-1 + z)^5 z^3 (-1 + 4 z) -  4 y^9 (-1 + z)^2 (1341 + 1234 z)\\
  &+ y^8 (14040 - 14991 z - 32938 z^2 +  54743 z^3 - 20800 z^4) + y^7 (-19944 + 36561 z + 52567 z^2 -  160807 z^3 + 117263 z^4 - 25640 z^5)\\
  &- y (-1 + z)^2 z^2  (-72 + 2577 z - 8383 z^2 + 13377 z^3 - 10501 z^4 + 3020 z^5) -  4 y^6 (-4104 + 13505 z + 4189 z^2 - 53569 z^3\\
  &+ 65725 z^4 - 30346 z^5 +  4627 z^6) + y^2 (-1 + z)^2 z (-180 - 366 z + 14249 z^2 - 35168 z^3 +  44729 z^4 - 26794 z^5 + 4744 z^6)\\
  &-  2 y^5 (3834 - 21096 z + 12153 z^2 + 79661 z^3 - 164736 z^4 + 132075 z^5 -  49685 z^6 + 7794 z^7) + y^4 (1800 - 16449 z + 20344 z^2\\
  &+ 84562 z^3 -  275038 z^4 + 340574 z^5 - 219676 z^6 + 73813 z^7 - 9876 z^8) -  y^3 (144 - 2823 z + 4349 z^2 + 40704 z^3 - 163328 z^4\\
  &+ 272002 z^5 -  250136 z^6 + 128747 z^7 - 31815 z^8 + 2156 z^9)) H(0, 0, y) +  2 (108 (-1 + z)^6 z^3 (-1 + 4 z) + 4 y^9 (-1 + z)^2 (216\\
  &+ 215 z) +  y^8 (-4500 + 8947 z + 426 z^2 - 9495 z^3 + 4692 z^4) -  y (-1 + z)^3 z^2 (-72 + 2469 z - 8130 z^2 + 15633 z^3 - 12928 z^4\\
  &+  864 z^5) + y^7 (9540 - 30281 z + 21427 z^2 + 17363 z^3 - 25789 z^4 +  7740 z^5) + 2 y^6 (-5202 + 24062 z - 33217 z^2 + 1427 z^3\\
  &+ 33635 z^4 -  28049 z^5 + 7274 z^6) + y^2 (-1 + z)^2 z (180 + 114 z - 13297 z^2 +  48726 z^3 - 88199 z^4 + 76698 z^5 - 22076 z^6\\
  &+ 432 z^7) +  4 y^5 (1503 - 9820 z + 19651 z^2 - 5048 z^3 - 33560 z^4 + 51210 z^5 -  31336 z^6 + 7400 z^7) + y^4 (-1656 + 15965 z\\
  &- 40998 z^2 - 3632 z^3 +  193426 z^4 - 388782 z^5 + 361522 z^6 - 164423 z^7 + 28648 z^8) +  y^3 (144 - 2787 z + 8149 z^2 + 23010 z^3\\
  &- 166174 z^4 + 387958 z^5 -  474394 z^6 + 314077 z^7 - 100103 z^8 + 10120 z^9)) H(0, 0, z) -  48 (-1 + y)^2 y^2 z^2 (-1 + y + z) (y\\
  &+ z)^2 (-12 + 30 z - 26 z^2 + 8 z^3 +  3 y (4 - 9 z + 4 z^2)) H(0, 1, z) - 48 y^2 (-1 + z)^2 z^2 (-1 + y + z)  (y + z)^2 (8 y^3 + y (30\\
  &- 27 z) + 12 (-1 + z) + 2 y^2 (-13 + 6 z))  H(0, 2, y) - 48 (-1 + y)^2 y^2 z^2 (-1 + y + z) (y + z)^2  (-12 + 30 z - 26 z^2 + 8 z^3\\
  &+ 3 y (4 - 9 z + 4 z^2)) H(1, 0, z) +  4 (432 y^{10} (-1 + z)^2 - 54 (-1 + z)^5 z^4 (-1 + 4 z) +  2 y^9 (-1 + z)^2 (-1125 + 1321 z)\\
  &+  3 y^8 (1590 - 7130 z + 10312 z^2 - 5603 z^3 + 822 z^4) +  y^7 (-5202 + 29970 z - 47113 z^2 + 16349 z^3 + 14834 z^4 - 8838 z^5)\\
  &+  y (-1 + z)^2 z^3 (-1653 + 4845 z - 5811 z^2 + 4198 z^3 - 2002 z^4 +  432 z^5) + y^6 (3006 - 18611 z + 14103 z^2 + 65692 z^3\\
  &- 134393 z^4 +  90425 z^5 - 20168 z^6) - y^2 (-1 + z)^2 z^2  (792 - 14109 z + 32355 z^2 - 27545 z^3 + 10016 z^4 - 1898 z^5 +  216 z^6)\\
  &+ y^5 (-828 + 945 z + 35948 z^2 - 178242 z^3 + 329525 z^4 -  282687 z^5 + 110081 z^6 - 14742 z^7) +  y^4 (72 + 4209 z - 39045 z^2\\
  &+ 174583 z^3 - 372816 z^4 + 403037 z^5 -  220595 z^6 + 54578 z^7 - 4050 z^8) +  y^3 z (-1401 + 13065 z - 78255 z^2 + 213625 z^3\\
  &- 291633 z^4 +  209152 z^5 - 75802 z^6 + 11847 z^7 - 598 z^8)) H(1, 1, z) +  192 (-1 + y)^2 y (-1 + z)^2 z (-1 + y + z) (y + z)^2  (-4\\
  &+ 8 y^3 + 14 z - 19 z^2 + 8 z^3 + y^2 (-19 + 12 z) +  2 y (7 - 10 z + 6 z^2)) H(3, y) H(1, 1, z) +  H(0, y) (96 (-1 + y)^2 y (-1\\
  &+ 4 y) (-1 + z)^2 z (-1 + y + z) (y + z)^2  (2 + 2 y^2 + 2 y (-2 + z) - 2 z + z^2) H(1, z)^2 +  48 y^2 (-1 + z)^2 z^2 (-1 + y + z) (y\\
  &+ z)^2 (8 y^3 + y (30 - 27 z) +  12 (-1 + z) + 2 y^2 (-13 + 6 z)) H(2, y) + 96 (-1 + y)^2 y (-1 + 4 y)  (-1 + z)^2 z (-1 + y + z) (y\\
  &+ z)^2 (2 + 2 y^2 + 2 y (-2 + z) - 2 z +  z^2) H(1, z) H(2, y) - 192 (-1 + y)^2 y (-1 + 4 y) (-1 + z)^2 z  (-1 + y + z) (y + z)^2 (2\\
  &+ 2 y^2 + 2 y (-2 + z) - 2 z + z^2)  H(1, 1, z)) - 48 y^2 (-1 + z)^2 z^2 (-1 + y + z) (y + z)^2  (8 y^3 + y (30 - 27 z) + 12 (-1 + z)\\
  &+ 2 y^2 (-13 + 6 z)) H(2, 0, y) +  4 (432 y^{10} (-1 + z)^2 - 54 (-1 + z)^5 z^4 (-1 + 4 z) +  2 y^9 (-1 + z)^2 (-1125 + 1321 z)\\
  &+  3 y^8 (1590 - 7130 z + 10312 z^2 - 5603 z^3 + 822 z^4) +  y^7 (-5202 + 29970 z - 47113 z^2 + 16349 z^3 + 14834 z^4 - 8838 z^5)\\
  &+  y (-1 + z)^2 z^3 (-1653 + 4845 z - 5811 z^2 + 4198 z^3 - 2002 z^4 +  432 z^5) + y^6 (3006 - 18611 z + 14103 z^2 + 65692 z^3\\
  &- 134393 z^4 +  90425 z^5 - 20168 z^6) - y^2 (-1 + z)^2 z^2  (792 - 14109 z + 32355 z^2 - 27545 z^3 + 10016 z^4 - 1898 z^5 +  216 z^6)\\
  &+ y^5 (-828 + 945 z + 35948 z^2 - 178242 z^3 + 329525 z^4 -  282687 z^5 + 110081 z^6 - 14742 z^7) +  y^4 (72 + 4209 z - 39045 z^2\\
 \end{aligned}
 \end{equation*}
 \begin{equation*}
 \begin{aligned}
  &+ 174583 z^3 - 372816 z^4 + 403037 z^5 -  220595 z^6 + 54578 z^7 - 4050 z^8) +  y^3 z (-1401 + 13065 z - 78255 z^2 + 213625 z^3\\
  &- 291633 z^4 +  209152 z^5 - 75802 z^6 + 11847 z^7 - 598 z^8)) H(2, 2, y) +  H(0, z) (96 (-1 + y)^2 y (y^2 + 2 y (-1 + z) + 2 (-1\\
  &+ z)^2) (-1 + z)^2 z  (-1 + y + z) (y + z)^2 (-1 + 4 z) H(2, y)^2 +  H(1, z) (48 (-1 + y)^2 y^2 z^2 (-1 + y + z) (y + z)^2  (-12 + 30 z\\
  &- 26 z^2 + 8 z^3 + 3 y (4 - 9 z + 4 z^2)) +  96 (-1 + y)^2 y (y^2 + 2 y (-1 + z) + 2 (-1 + z)^2) (-1 + z)^2 z  (-1 + y + z) (y\\
  &+ z)^2 (-1 + 4 z) H(2, y)) -  192 (-1 + y)^2 y (y^2 + 2 y (-1 + z) + 2 (-1 + z)^2) (-1 + z)^2 z  (-1 + y + z) (y + z)^2 (-1\\
  &+ 4 z) H(2, 2, y)) +  H(2, y) (-96 (-1 + y)^2 y (-1 + 4 y) (-1 + z)^2 z (-1 + y + z) (y + z)^2  (2 + 2 y^2 + 2 y (-2 + z) - 2 z\\
  &+ z^2) H(0, 0, y) -  96 (-1 + y)^2 y (y^2 + 2 y (-1 + z) + 2 (-1 + z)^2) (-1 + z)^2 z  (-1 + y + z) (y + z)^2 (-1 + 4 z) H(0, 0, z)\\
  &-  96 (-1 + y)^2 y (y^2 + 2 y (-1 + z) + 2 (-1 + z)^2) (-1 + z)^2 z  (-1 + y + z) (y + z)^2 (-1 + 4 z) H(0, 1, z) +  96 (-1 + y)^2 y (-1\\
  &+ 4 y) (-1 + z)^2 z (-1 + y + z) (y + z)^2  (2 + 2 y^2 + 2 y (-2 + z) - 2 z + z^2) H(0, 2, y) -  96 (-1 + y)^2 y (-1 + 4 y) (-1\\
  &+ z)^2 z (-1 + y + z) (y + z)^2  (2 + 2 y^2 + 2 y (-2 + z) - 2 z + z^2) H(1, 0, y) -  96 (-1 + y)^2 y (y^2 + 2 y (-1 + z) + 2 (-1\\
  &+ z)^2) (-1 + z)^2 z  (-1 + y + z) (y + z)^2 (-1 + 4 z) H(1, 0, z) +  120 (-1 + y)^2 y (-1 + z)^2 z (-1 + y + z) (y + z)^2  (-4 + 56 y^3\\
  &+ 50 z - 109 z^2 + 56 z^3 + y^2 (-109 + 84 z) +  y (50 - 92 z + 84 z^2)) H(1, 1, z) + 96 (-1 + y)^2 y (-1 + 4 y)  (-1 + z)^2 z (-1 + y\\
  &+ z) (y + z)^2 (2 + 2 y^2 + 2 y (-2 + z) - 2 z +  z^2) H(2, 0, y) - 96 (-1 + y)^2 y (-1 + z)^2 z (-1 + y + z) (y + z)^2  (-4 + 8 y^3\\
  &+ 14 z - 19 z^2 + 8 z^3 + y^2 (-19 + 12 z) +  2 y (7 - 10 z + 6 z^2)) H(2, 2, y) - 96 (-1 + y)^2 y (-1 + z)^2 z  (-1 + y + z) (y\\
  &+ z)^2 (-4 + 8 y^3 + 14 z - 19 z^2 + 8 z^3 +  y^2 (-19 + 12 z) + 2 y (7 - 10 z + 6 z^2)) H(3, 2, y)) +  H(1, z) (-60 (-1 + y)^2 y (-1\\
  &+ z)^2 z (-1 + y + z) (y + z)^2  (-4 + 56 y^3 + 50 z - 109 z^2 + 56 z^3 + y^2 (-109 + 84 z) +  y (50 - 92 z + 84 z^2)) H(2, y)^2\\
  &- 96 (-1 + y)^2 y (-1 + z)^2 z  (-1 + y + z) (y + z)^2 (-4 + 8 y^3 + 14 z - 19 z^2 + 8 z^3 +  y^2 (-19 + 12 z) + 2 y (7 - 10 z\\
  &+ 6 z^2)) H(2, y) H(3, y) -  96 (-1 + y)^2 y (-1 + 4 y) (-1 + z)^2 z (-1 + y + z) (y + z)^2  (2 + 2 y^2 + 2 y (-2 + z) - 2 z\\
  &+ z^2) H(0, 0, y) -  96 (-1 + y)^2 y (y^2 + 2 y (-1 + z) + 2 (-1 + z)^2) (-1 + z)^2 z  (-1 + y + z) (y + z)^2 (-1 + 4 z) H(0, 0, z)\\
  &-  96 (-1 + y)^2 y (-1 + 4 y) (-1 + z)^2 z (-1 + y + z) (y + z)^2  (2 + 2 y^2 + 2 y (-2 + z) - 2 z + z^2) H(0, 1, z) -  96 (-1\\
  &+ y)^2 y (-1 + 4 y) (-1 + z)^2 z (-1 + y + z) (y + z)^2  (2 + 2 y^2 + 2 y (-2 + z) - 2 z + z^2) H(0, 2, y) -  96 (-1 + y)^2 y (-1\\
  &+ z)^2 z (-1 + y + z) (y + z)^2  (-4 + 8 y^3 + 14 z - 19 z^2 + 8 z^3 + y^2 (-19 + 12 z) +  2 y (7 - 10 z + 6 z^2)) H(1, 1, z) - 96 (-1\\
  &+ y)^2 y (-1 + 4 y)  (-1 + z)^2 z (-1 + y + z) (y + z)^2 (2 + 2 y^2 + 2 y (-2 + z) - 2 z +  z^2) H(2, 0, y) + 120 (-1 + y)^2 y (-1\\
  &+ z)^2 z (-1 + y + z) (y + z)^2  (-4 + 56 y^3 + 50 z - 109 z^2 + 56 z^3 + y^2 (-109 + 84 z) +  y (50 - 92 z + 84 z^2)) H(2, 2, y)\\
  &+ 96 (-1 + y)^2 y (-1 + z)^2 z  (-1 + y + z) (y + z)^2 (-4 + 8 y^3 + 14 z - 19 z^2 + 8 z^3 +  y^2 (-19 + 12 z) + 2 y (7 - 10 z\\
  &+ 6 z^2)) H(2, 3, y) +  96 (-1 + y)^2 y (-1 + z)^2 z (-1 + y + z) (y + z)^2  (-4 + 8 y^3 + 14 z - 19 z^2 + 8 z^3 + y^2 (-19 + 12 z)\\
  &+  2 y (7 - 10 z + 6 z^2)) H(3, 2, y)) + 72 (-1 + y)^2 y (-1 + z)^2 z  (-1 + y + z) (y + z)^2 (-44 + 528 y^3 + 66 z - 99 z^2 + 64 z^3\\
  &+  y^2 (-829 + 412 z) + 2 y (187 - 176 z + 58 z^2)) H(0, 0, 0, y) +  72 (-1 + y)^2 y (-1 + z)^2 z (-1 + y + z) (y + z)^2  (-44 + 64 y^3\\
  &+ 374 z - 829 z^2 + 528 z^3 + y^2 (-99 + 116 z) +  y (66 - 352 z + 412 z^2)) H(0, 0, 0, z) + 192 (-1 + y)^2 y (-1 + 4 y)  (-1\\
  &+ z)^2 z (-1 + y + z) (y + z)^2 (2 + 2 y^2 + 2 y (-2 + z) - 2 z + z^2)  H(0, 1, 1, z) - 192 (-1 + y)^2 y (-1 + 4 y) (-1 + z)^2 z (-1 + y\\
  &+ z)  (y + z)^2 (2 + 2 y^2 + 2 y (-2 + z) - 2 z + z^2) H(0, 2, 2, y) +  96 (-1 + y)^2 y (-1 + 4 y) (-1 + z)^2 z (-1 + y + z) (y\\
  &+ z)^2  (2 + 2 y^2 + 2 y (-2 + z) - 2 z + z^2) H(1, 0, 1, z) +  96 (-1 + y)^2 y (-1 + 4 y) (-1 + z)^2 z (-1 + y + z) (y + z)^2  (2\\
  &+ 2 y^2 + 2 y (-2 + z) - 2 z + z^2) H(1, 0, 2, y) +  72 (-1 + y)^2 y (-1 + z)^2 z (-1 + y + z) (y + z)^2  (-12 + 104 y^3 + 102 z\\
  &- 207 z^2 + 104 z^3 + 3 y^2 (-69 + 52 z) +  6 y (17 - 30 z + 26 z^2)) H(1, 1, 1, z) + 96 (-1 + y)^2 y (-1 + 4 y)  (-1 + z)^2 z (-1 + y\\
  &+ z) (y + z)^2 (2 + 2 y^2 + 2 y (-2 + z) - 2 z + z^2)  H(1, 2, 0, y) - 192 (-1 + y)^2 y (-1 + 4 y) (-1 + z)^2 z (-1 + y + z)  (y\\
  &+ z)^2 (2 + 2 y^2 + 2 y (-2 + z) - 2 z + z^2) H(2, 0, 2, y) +  96 (-1 + y)^2 y (-1 + 4 y) (-1 + z)^2 z (-1 + y + z) (y + z)^2  (2\\
  &+ 2 y^2 + 2 y (-2 + z) - 2 z + z^2) H(2, 1, 0, y) -  192 (-1 + y)^2 y (-1 + 4 y) (-1 + z)^2 z (-1 + y + z) (y + z)^2  (2 + 2 y^2\\
  &+ 2 y (-2 + z) - 2 z + z^2) H(2, 2, 0, y) +  72 (-1 + y)^2 y (-1 + z)^2 z (-1 + y + z) (y + z)^2  (-12 + 104 y^3 + 102 z - 207 z^2\\
  &+ 104 z^3 + 3 y^2 (-69 + 52 z) +  6 y (17 - 30 z + 26 z^2)) H(2, 2, 2, y) + 96 (-1 + y)^2 y (-1 + z)^2 z  (-1 + y + z) (y + z)^2 (-4\\
  &+ 8 y^3 + 14 z - 19 z^2 + 8 z^3 +  y^2 (-19 + 12 z) + 2 y (7 - 10 z + 6 z^2)) H(2, 3, 2, y) +  192 (-1 + y)^2 y (-1 + z)^2 z (-1 + y\\
  &+ z) (y + z)^2  (-4 + 8 y^3 + 14 z - 19 z^2 + 8 z^3 + y^2 (-19 + 12 z) +  2 y (7 - 10 z + 6 z^2)) H(3, 2, 2, y)\Bigg\}\Big/\Big(48 (-1 + y)^2 y^2 (-1\\
  &+ z)^2 z^2  (-1 + y + z) (y + z)^2\Big);\\[10pt]
\mathcal{A}_{3;n_{f}^{2}}^{(2)} &=
\Bigg\{4 (5 y^7 (-1 + 4 z) + y^6 (15 - 83 z + 80 z^2) -  5 z^4 (-2 + 4 z - 3 z^2 + z^3) + y^5 (-20 + 147 z - 273 z^2 + 140 z^3) +  y z^3 (40\\
  &- 154 z + 147 z^2 - 83 z^3 + 20 z^4) +  y^2 z^2 (60 - 254 z + 363 z^2 - 273 z^3 + 80 z^4) +  y^3 z (40 - 254 z + 462 z^2 - 439 z^3\\
  &+ 140 z^4) +  y^4 (10 - 154 z + 363 z^2 - 439 z^3 + 160 z^4)) -  9 (y + z)^4 (2 + y^2 (3 - 9 z) - 4 z + 3 z^2 - z^3 + y^3 (-1 + 4 z)\\
  &+  y (-4 + 12 z - 9 z^2 + 4 z^3)) H(0, y) -  9 (y + z)^4 (2 + y^2 (3 - 9 z) - 4 z + 3 z^2 - z^3 + y^3 (-1 + 4 z) +  y (-4 + 12 z - 9 z^2\\
  &+ 4 z^3)) H(0, z) +  36 y z (-5 y^3 + 3 y^4 + y z^2 + y^2 (6 + z - 6 z^2) +  z^2 (6 - 5 z + 3 z^2)) H(1, z) +  36 y z (-5 y^3 + 3 y^4\\
  &+ y z^2 + y^2 (6 + z - 6 z^2) +  z^2 (6 - 5 z + 3 z^2)) H(2, y)\Bigg\}\Big/\Big(216 y z (-1 + y + z) (y + z)^4\Big);\\
\end{aligned}
\end{equation*}
\begin{equation*}
\begin{aligned}
\mathcal{A}_{3;C_{A}C_{F}}^{(2)} &=
\Bigg\{-4 (-1 + y) y (-1 + z) z (y + z) ((-1 + z)^2 z (5450 - 5153 z + 2725 z^2) +  y^5 (2725 - 10952 z + 7336 z^2) + y^4 (-10603\\
  &+ 43407 z - 37764 z^2 +  4960 z^3) + y^3 (18481 - 86606 z + 117680 z^2 - 53624 z^3 + 4960 z^4) +  y (5450 - 38046 z + 86747 z^2\\
  &- 86606 z^3 + 43407 z^4 - 10952 z^5) +  y^2 (-16053 + 86747 z - 157946 z^2 + 117680 z^3 - 37764 z^4 +  7336 z^5)) + 36 (-1\\
  &+ y)^2 y (-1 + z)^2 z (y + z)^2  (272 + 3128 y^4 - 470 z + 417 z^2 - 419 z^3 + 200 z^4 +  y^3 (-8003 + 5572 z) + 3 y^2 (2355 - 3089 z\\
  &+ 936 z^2) +  y (-2462 + 4440 z - 2739 z^2 + 772 z^3)) H(0, y)^3 +  36 (-1 + y)^2 y (-1 + z)^2 z (y + z)^2 (272 + 200 y^4 - 2462 z\\
  &+ 7065 z^2 -  8003 z^3 + 3128 z^4 + y^3 (-419 + 772 z) +  3 y^2 (139 - 913 z + 936 z^2) + y (-470 + 4440 z - 9267 z^2\\
  &+ 5572 z^3))  H(0, z)^3 + 36 (-1 + y)^2 y (-1 + z)^2 z (y + z)^2  (60 + 1048 y^4 - 1190 z + 3387 z^2 - 3305 z^3 + 1048 z^4\\
  &+  y^3 (-3305 + 2652 z) + y^2 (3387 - 5793 z + 3624 z^2) +  y (-1190 + 3984 z - 5793 z^2 + 2652 z^3)) H(1, z)^3 +  9 (2592 y^{10} (-1\\
  &+ z)^2 - 324 (-1 + z)^5 z^4 (-1 + 4 z) +  4 y^9 (-1 + z)^2 (-3375 + 1232 z) +  y^8 (28620 - 84611 z + 64366 z^2 + 10459 z^3\\
  &- 18996 z^4) +  y (-1 + z)^2 z^3 (-10284 + 31286 z - 47769 z^2 + 47069 z^3 - 22936 z^4 +  2592 z^5) - y^7 (31212 - 112231 z\\
  &+ 31885 z^2 + 231313 z^3 -  265795 z^4 + 83616 z^5) + y^6 (18036 - 61763 z - 163792 z^2 +  837127 z^3 - 1162130 z^4 + 669294 z^5\\
  &- 136448 z^6) -  y^2 (-1 + z)^2 z^2 (8364 - 92610 z + 217312 z^2 - 239107 z^3 +  137650 z^4 - 33236 z^5 + 1296 z^6) -  y^5 (4968\\
  &+ 12031 z - 343845 z^2 + 1384321 z^3 - 2358789 z^4 +  1969504 z^5 - 787230 z^6 + 119040 z^7) +  y^4 (432 + 28202 z - 276976 z^2\\
  &+ 1189366 z^3 - 2507872 z^4 +  2799861 z^5 - 1679342 z^6 + 504259 z^7 - 58092 z^8) +  y^3 z (-8772 + 93570 z - 525300 z^2\\
  &+ 1423618 z^3 - 2064667 z^4 +  1697887 z^5 - 784219 z^6 + 182395 z^7 - 14512 z^8)) H(2, y)^2 +  36 (-1 + y)^2 y (-1 + z)^2 z (y\\
  &+ z)^2 (60 + 1048 y^4 - 1190 z + 3387 z^2 -  3305 z^3 + 1048 z^4 + y^3 (-3305 + 2652 z) +  y^2 (3387 - 5793 z + 3624 z^2) + y (-1190\\
  &+ 3984 z - 5793 z^2 +  2652 z^3)) H(2, y)^3 +  H(0, z)^2 (9 (324 (-1 + z)^6 z^3 (-1 + 4 z) + 16 y^9 (-1 + z)^2  (162 + 73 z)\\
  &+ y^8 (-13500 + 34093 z - 26250 z^2 + 5391 z^3 +  1180 z^4) + y^7 (28620 - 106481 z + 143631 z^2 - 77501 z^3 +  5003 z^4 + 6728 z^5)\\
  &- 2 y (-1 + z)^2 z^2 (108 - 3068 z + 15065 z^2 -  37089 z^3 + 46088 z^4 - 22256 z^5 + 1296 z^6) +  y^6 (-31212 + 163509 z\\
  &- 312112 z^2 + 240300 z^3 - 456 z^4 -  105981 z^5 + 44124 z^6) + 2 y^2 (-1 + z)^2 z  (270 - 1316 z - 14807 z^2 + 70951 z^3\\
  &- 142240 z^4 + 124888 z^5 -  36250 z^6 + 648 z^7) + y^5 (18036 - 132635 z + 331953 z^2 -  300344 z^3 - 147828 z^4 + 523715 z^5\\
  &- 390745 z^6 + 97848 z^7) +  y^4 (-4968 + 54838 z - 176524 z^2 + 144129 z^3 + 393770 z^4 -  1110625 z^5 + 1146516 z^6 - 538862 z^7\\
  &+ 92640 z^8) +  y^3 (432 - 9848 z + 42974 z^2 + 4423 z^3 - 410951 z^4 + 1156123 z^5 -  1515467 z^6 + 1025284 z^7 - 326466 z^8\\
  &+ 33496 z^9)) -  864 (-1 + y)^2 y (-1 + z)^2 z (y + z)^2 (y^3 (-1 + 4 z) +  4 y (-1 + z)^2 (-1 + 4 z) + 3 y^2 (1 - 5 z + 4 z^2)\\
  &+  2 (1 - 8 z + 18 z^2 - 17 z^3 + 6 z^4)) H(1, z) -  1728 (-1 + y)^2 y (-1 + z)^2 z (y + z)^2 (2 - 15 z + 33 z^2 - 30 z^3 +  10 z^4\\
  &+ y^3 (-1 + 4 z) + 4 y (-1 + z)^2 (-1 + 4 z) +  3 y^2 (1 - 5 z + 4 z^2)) H(2, y)) +  H(0, y)^2 (-9 (2592 y^{10} (-1 + z)^2 + 324 (-1\\
  &+ z)^5 z^3 (-1 + 4 z) -  4 y^9 (-1 + z)^2 (4023 + 4486 z) - 6 y^8 (-7020 + 5368 z + 21840 z^2 -  31731 z^3 + 11516 z^4) + y^7 (-59832\\
  &+ 90672 z + 223220 z^2 -  565474 z^3 + 397382 z^4 - 85968 z^5) - y (-1 + z)^2 z^2  (-216 + 6244 z - 21180 z^2 + 34761 z^3 - 27075 z^4\\
  &+ 7648 z^5) +  y^6 (49248 - 151456 z - 105440 z^2 + 735113 z^3 - 850650 z^4 +  378865 z^5 - 56004 z^6) + y^2 (-1 + z)^2 z (-540\\
  &+ 1876 z +  30168 z^2 - 80106 z^3 + 101421 z^4 - 58502 z^5 + 11408 z^6) -  y^5 (23004 - 128508 z + 71864 z^2 + 470215 z^3\\
  &- 932695 z^4 +  701645 z^5 - 235797 z^6 + 30272 z^7) +  y^4 (5400 - 53988 z + 88836 z^2 + 166115 z^3 - 638606 z^4 +  766830 z^5\\
  &- 456762 z^6 + 139069 z^7 - 16732 z^8) -  y^3 (432 - 9956 z + 29272 z^2 + 57505 z^3 - 334959 z^4 + 576238 z^5 -  518670 z^6\\
  &+ 256651 z^7 - 61569 z^8 + 5056 z^9)) -  432 (-1 + y)^2 y (-1 + 4 y) (-1 + z)^2 z (-1 + y + z) (y + z)^2  (2 + 2 y^2 + 2 y (-2 + z)\\
  &- 2 z + z^2) H(0, z) -  1728 (-1 + y)^2 y (-1 + z)^2 z (y + z)^2 (2 + 10 y^4 - 4 z + 3 z^2 -  z^3 + 2 y^3 (-15 + 8 z) + 3 y^2 (11\\
  &- 12 z + 4 z^2) +  y (-15 + 24 z - 15 z^2 + 4 z^3)) H(1, z) - 864 (-1 + y)^2 y (-1 + z)^2  z (y + z)^2 (2 + 12 y^4 - 4 z + 3 z^2\\
  &- z^3 + 2 y^3 (-17 + 8 z) +  12 y^2 (3 - 3 z + z^2) + y (-16 + 24 z - 15 z^2 + 4 z^3)) H(2, y)) +  H(1, z)^2 (9 (2592 y^{10} (-1 + z)^2\\
  &- 324 (-1 + z)^5 z^4 (-1 + 4 z) +  4 y^9 (-1 + z)^2 (-3375 + 1232 z) + y^8 (28620 - 84611 z + 64366 z^2 +  10459 z^3 - 18996 z^4)\\
  &+ y (-1 + z)^2 z^3 (-10284 + 31286 z -  47769 z^2 + 47069 z^3 - 22936 z^4 + 2592 z^5) -  y^7 (31212 - 112231 z + 31885 z^2\\
  &+ 231313 z^3 - 265795 z^4 +  83616 z^5) + y^6 (18036 - 61763 z - 163792 z^2 + 837127 z^3 -  1162130 z^4 + 669294 z^5 - 136448 z^6)\\
  &- y^2 (-1 + z)^2 z^2  (8364 - 92610 z + 217312 z^2 - 239107 z^3 + 137650 z^4 - 33236 z^5 +  1296 z^6) - y^5 (4968 + 12031 z\\
  &- 343845 z^2 + 1384321 z^3 -  2358789 z^4 + 1969504 z^5 - 787230 z^6 + 119040 z^7) +  y^4 (432 + 28202 z - 276976 z^2 + 1189366 z^3\\
  &- 2507872 z^4 +  2799861 z^5 - 1679342 z^6 + 504259 z^7 - 58092 z^8) +  y^3 z (-8772 + 93570 z - 525300 z^2 + 1423618 z^3\\
  &- 2064667 z^4 +  1697887 z^5 - 784219 z^6 + 182395 z^7 - 14512 z^8)) +  108 (-1 + y)^2 y (-1 + z)^2 z (y + z)^2 (124 + 1240 y^4\\
  &- 1510 z +  4011 z^2 - 3865 z^3 + 1240 z^4 + y^3 (-3865 + 2972 z) +  y^2 (4011 - 6609 z + 4008 z^2) + y (-1510 + 4752 z - 6609 z^2\\
  &+  2972 z^3)) H(2, y) + 1728 (-1 + y)^2 y (-1 + z)^2 z (y + z)^2  (4 + 12 y^4 - 20 z + 39 z^2 - 35 z^3 + 12 z^4 + 5 y^3 (-7 + 4 z)\\
  &+  3 y^2 (13 - 17 z + 8 z^2) + y (-20 + 48 z - 51 z^2 + 20 z^3))  H(3, y)) + 18 (2592 y^{10} (-1 + z)^2 + 324 (-1 + z)^5 z^3 (-1 + 4 z)\\
  &-  4 y^9 (-1 + z)^2 (4023 + 4486 z) - 6 y^8 (-7020 + 5368 z + 21840 z^2 -  31731 z^3 + 11516 z^4) + y^7 (-59832 + 90672 z\\
  &+ 223220 z^2 -  565474 z^3 + 397382 z^4 - 85968 z^5) - y (-1 + z)^2 z^2  (-216 + 6244 z - 21180 z^2 + 34761 z^3 - 27075 z^4\\
  &+ 7648 z^5) +  y^6 (49248 - 151456 z - 105440 z^2 + 735113 z^3 - 850650 z^4 +  378865 z^5 - 56004 z^6) + y^2 (-1 + z)^2 z (-540\\
  &+ 1876 z + 30168 z^2 -  80106 z^3 + 101421 z^4 - 58502 z^5 + 11408 z^6) -  y^5 (23004 - 128508 z + 71864 z^2 + 470215 z^3\\
\end{aligned}
\end{equation*}
\begin{equation*}
\begin{aligned}
  &- 932695 z^4 +  701645 z^5 - 235797 z^6 + 30272 z^7) +  y^4 (5400 - 53988 z + 88836 z^2 + 166115 z^3 - 638606 z^4 + 766830 z^5\\
  &-  456762 z^6 + 139069 z^7 - 16732 z^8) -  y^3 (432 - 9956 z + 29272 z^2 + 57505 z^3 - 334959 z^4 + 576238 z^5 -  518670 z^6\\
  &+ 256651 z^7 - 61569 z^8 + 5056 z^9)) H(0, 0, y) -  18 (324 (-1 + z)^6 z^3 (-1 + 4 z) + 16 y^9 (-1 + z)^2 (162 + 73 z) +  y^8 (-13500\\
  &+ 34093 z - 26250 z^2 + 5391 z^3 + 1180 z^4) +  y^7 (28620 - 106481 z + 143631 z^2 - 77501 z^3 + 5003 z^4 + 6728 z^5) -  2 y (-1\\
  &+ z)^2 z^2 (108 - 3068 z + 15065 z^2 - 37089 z^3 + 46088 z^4 -  22256 z^5 + 1296 z^6) + y^6 (-31212 + 163509 z - 312112 z^2\\
  &+  240300 z^3 - 456 z^4 - 105981 z^5 + 44124 z^6) +  2 y^2 (-1 + z)^2 z (270 - 1316 z - 14807 z^2 + 70951 z^3 - 142240 z^4\\
  &+  124888 z^5 - 36250 z^6 + 648 z^7) +  y^5 (18036 - 132635 z + 331953 z^2 - 300344 z^3 - 147828 z^4 +  523715 z^5 - 390745 z^6\\
  &+ 97848 z^7) +  y^4 (-4968 + 54838 z - 176524 z^2 + 144129 z^3 + 393770 z^4 -  1110625 z^5 + 1146516 z^6 - 538862 z^7 + 92640 z^8)\\
  &+  y^3 (432 - 9848 z + 42974 z^2 + 4423 z^3 - 410951 z^4 + 1156123 z^5 -  1515467 z^6 + 1025284 z^7 - 326466 z^8\\
  &+ 33496 z^9)) H(0, 0, z) -  216 (-1 + y)^2 y z (88 y^6 (-1 + z)^2 - (-1 + z)^3 z^2  (14 - 14 z + 3 z^2) + y^5 (-278 + 828 z - 816 z^2\\
  &+ 272 z^3) -  y (-1 + z)^2 z (-68 + 202 z - 114 z^2 - 61 z^3 + 60 z^4) +  2 y^4 (157 - 686 z + 1012 z^2 - 591 z^3 + 114 z^4)\\
  &-  y^3 (138 - 984 z + 2141 z^2 - 1802 z^3 + 433 z^4 + 68 z^5) +  y^2 (14 - 340 z + 1129 z^2 - 1351 z^3 + 427 z^4 + 293 z^5\\
  &- 172 z^6))  H(0, 1, z) + 216 y (-1 + z)^2 z (88 y^8 + y^7 (-470 + 456 z) +  2 y^6 (511 - 1064 z + 482 z^2) + z^2 (30 - 60 z + 49 z^2\\
  &- 19 z^3) +  y^5 (-1156 + 4028 z - 3857 z^2 + 1036 z^3) +  y z (20 - 342 z + 666 z^2 - 509 z^3 + 162 z^4) +  y^4 (716 - 3888 z\\
  &+ 6107 z^2 - 3483 z^3 + 564 z^4) +  y^2 (30 - 432 z + 1885 z^2 - 2545 z^3 + 1441 z^4 - 273 z^5) +  y^3 (-230 + 1944 z - 4787 z^2\\
  &+ 4380 z^3 - 1557 z^4 + 124 z^5))  H(0, 2, y) - 2376 (-1 + y)^2 y (-1 + 4 y) (-1 + z)^2 z (-1 + y + z)  (y + z)^2 (2 + 2 y^2\\
  &+ 2 y (-2 + z) - 2 z + z^2) H(1, 0, y) +  432 (-1 + y)^2 y z (-4 (-1 + z)^3 z^2 (1 - z + z^2) +  y^5 (-4 + 48 z - 87 z^2 + 40 z^3)\\
  &+ 4 y (-1 + z)^2 z  (-3 - 5 z + 27 z^2 - 31 z^3 + 13 z^4) +  y^4 (8 - 128 z + 407 z^2 - 465 z^3 + 172 z^4) +  y^3 (-8 + 124 z\\
  &- 552 z^2 + 1046 z^3 - 889 z^4 + 276 z^5) +  y^2 (4 - 28 z + 244 z^2 - 760 z^3 + 1079 z^4 - 735 z^5 + 196 z^6))  H(1, 0, z)\\
  &- 18 (2592 y^{10} (-1 + z)^2 - 324 (-1 + z)^5 z^4 (-1 + 4 z) +  4 y^9 (-1 + z)^2 (-3375 + 1232 z) +  y^8 (28620 - 84611 z + 64366 z^2\\
  &+ 10459 z^3 - 18996 z^4) +  y (-1 + z)^2 z^3 (-10284 + 31286 z - 47769 z^2 + 47069 z^3 - 22936 z^4 +  2592 z^5) - y^7 (31212\\
  &- 112231 z + 31885 z^2 + 231313 z^3 -  265795 z^4 + 83616 z^5) + y^6 (18036 - 61763 z - 163792 z^2 +  837127 z^3 - 1162130 z^4\\
  &+ 669294 z^5 - 136448 z^6) -  y^2 (-1 + z)^2 z^2 (8364 - 92610 z + 217312 z^2 - 239107 z^3 +  137650 z^4 - 33236 z^5 + 1296 z^6)\\
  &-  y^5 (4968 + 12031 z - 343845 z^2 + 1384321 z^3 - 2358789 z^4 +  1969504 z^5 - 787230 z^6 + 119040 z^7) +  y^4 (432 + 28202 z\\
  &- 276976 z^2 + 1189366 z^3 - 2507872 z^4 +  2799861 z^5 - 1679342 z^6 + 504259 z^7 - 58092 z^8) +  y^3 z (-8772 + 93570 z\\
  &- 525300 z^2 + 1423618 z^3 - 2064667 z^4 +  1697887 z^5 - 784219 z^6 + 182395 z^7 - 14512 z^8)) H(1, 1, z) +  H(3, y) (-864 (-1\\
  &+ y)^2 y (-1 + z)^2 z (-1 + y + z) (y + z)^2  (-4 + 8 y^3 + 14 z - 19 z^2 + 8 z^3 + y^2 (-19 + 12 z) +  2 y (7 - 10 z\\
  &+ 6 z^2)) H(0, 1, z) - 864 (-1 + y)^2 y (-1 + z)^2 z  (-1 + y + z) (y + z)^2 (-4 + 8 y^3 + 14 z - 19 z^2 + 8 z^3 +  y^2 (-19 + 12 z)\\
  &+ 2 y (7 - 10 z + 6 z^2)) H(1, 0, z) -  3456 (-1 + y)^2 y (-1 + z)^2 z (y + z)^2 (4 + 12 y^4 - 20 z + 39 z^2 -  35 z^3 + 12 z^4\\
  &+ 5 y^3 (-7 + 4 z) + 3 y^2 (13 - 17 z + 8 z^2) +  y (-20 + 48 z - 51 z^2 + 20 z^3)) H(1, 1, z)) +  216 y (-1 + z)^2 z (88 y^8\\
  &+ y^7 (-470 + 456 z) +  2 y^6 (511 - 1064 z + 482 z^2) + z^2 (30 - 60 z + 49 z^2 - 19 z^3) +  y^5 (-1156 + 4028 z - 3857 z^2\\
  &+ 1036 z^3) +  y z (20 - 342 z + 666 z^2 - 509 z^3 + 162 z^4) +  y^4 (716 - 3888 z + 6107 z^2 - 3483 z^3 + 564 z^4) +  y^2 (30\\
  &- 432 z + 1885 z^2 - 2545 z^3 + 1441 z^4 - 273 z^5) +  y^3 (-230 + 1944 z - 4787 z^2 + 4380 z^3 - 1557 z^4 + 124 z^5))  H(2, 0, y)\\
  &- 18 (2592 y^{10} (-1 + z)^2 - 324 (-1 + z)^5 z^4 (-1 + 4 z) +  4 y^9 (-1 + z)^2 (-3375 + 1232 z) +  y^8 (28620 - 84611 z + 64366 z^2\\
  &+ 10459 z^3 - 18996 z^4) +  y (-1 + z)^2 z^3 (-10284 + 31286 z - 47769 z^2 + 47069 z^3 - 22936 z^4 +  2592 z^5) - y^7 (31212\\
  &- 112231 z + 31885 z^2 + 231313 z^3 -  265795 z^4 + 83616 z^5) + y^6 (18036 - 61763 z - 163792 z^2 +  837127 z^3 - 1162130 z^4\\
  &+ 669294 z^5 - 136448 z^6) -  y^2 (-1 + z)^2 z^2 (8364 - 92610 z + 217312 z^2 - 239107 z^3 +  137650 z^4 - 33236 z^5 + 1296 z^6)\\
  &-  y^5 (4968 + 12031 z - 343845 z^2 + 1384321 z^3 - 2358789 z^4 +  1969504 z^5 - 787230 z^6 + 119040 z^7) +  y^4 (432 + 28202 z\\
  &- 276976 z^2 + 1189366 z^3 - 2507872 z^4 +  2799861 z^5 - 1679342 z^6 + 504259 z^7 - 58092 z^8) +  y^3 z (-8772 + 93570 z\\
  &- 525300 z^2 + 1423618 z^3 - 2064667 z^4 +  1697887 z^5 - 784219 z^6 + 182395 z^7 - 14512 z^8)) H(2, 2, y) +  H(0, z) (1188 (-1\\
  &+ y)^2 y^2 z^2 (-1 + y + z) (y + z)^2  (-12 + 30 z - 26 z^2 + 8 z^3 + 3 y (4 - 9 z + 4 z^2)) -  864 (-1 + y)^2 y (-1 + z)^2 z (y\\
  &+ z)^2 (4 y^4 + (-1 + z)^3 (-1 + 4 z) +  y^3 (-13 + 8 z) + 3 y^2 (5 - 7 z + 4 z^2) +  y (-7 + 18 z - 21 z^2 + 8 z^3)) H(1, z)^2\\
  &+  2376 (-1 + y)^2 y (y^2 + 2 y (-1 + z) + 2 (-1 + z)^2) (-1 + z)^2 z  (-1 + y + z) (y + z)^2 (-1 + 4 z) H(2, y) - 864 (-1\\
  &+ y)^2 y (-1 + z)^2  z (y + z)^2 (5 + 4 y^4 - 39 z + 87 z^2 - 81 z^3 + 28 z^4 +  y^3 (-15 + 16 z) + 3 y^2 (7 - 17 z + 12 z^2)\\
  &+  y (-15 + 66 z - 93 z^2 + 40 z^3)) H(2, y)^2 +  H(1, z) (-432 (-1 + y)^2 y z (-4 (-1 + z)^3 z^2 (1 - z + z^2) +  y^5 (-4 + 48 z\\
  &- 87 z^2 + 40 z^3) + 4 y (-1 + z)^2 z  (-3 - 5 z + 27 z^2 - 31 z^3 + 13 z^4) +  y^4 (8 - 128 z + 407 z^2 - 465 z^3 + 172 z^4)\\
  &+  y^3 (-8 + 124 z - 552 z^2 + 1046 z^3 - 889 z^4 + 276 z^5) +  y^2 (4 - 28 z + 244 z^2 - 760 z^3 + 1079 z^4 - 735 z^5 + 196 z^6))\\
  &-  864 (-1 + y)^2 y (-1 + z)^2 z (y + z)^2 (2 - 28 z + 69 z^2 - 67 z^3 +  24 z^4 + y^3 (-1 + 4 z) + 3 y^2 (1 - 7 z + 8 z^2) +  y (-4\\
  &+ 36 z - 63 z^2 + 28 z^3)) H(2, y) + 864 (-1 + y)^2 y  (-1 + z)^2 z (-1 + y + z) (y + z)^2 (-4 + 8 y^3 + 14 z - 19 z^2 +  8 z^3\\
  &+ y^2 (-19 + 12 z) + 2 y (7 - 10 z + 6 z^2)) H(3, y)) +  864 (-1 + y)^2 y (-1 + 4 y) (-1 + z)^2 z (-1 + y + z) (y + z)^2  (2 + 2 y^2\\
\end{aligned}
\end{equation*}
\begin{equation*}
\begin{aligned}
  &+ 2 y (-2 + z) - 2 z + z^2) H(0, 0, y) +  864 (-1 + y)^2 y (y^2 + 2 y (-1 + z) + 2 (-1 + z)^2) (-1 + z)^2 z  (-1 + y + z) (y\\
  &+ z)^2 (-1 + 4 z) H(0, 0, z) +  864 (-1 + y)^2 y (-1 + 4 y) (-1 + z)^2 z (-1 + y + z) (y + z)^2  (2 + 2 y^2 + 2 y (-2 + z) - 2 z\\
  &+ z^2) H(0, 1, z) +  864 (-1 + y)^2 y (-1 + z)^2 z (y + z)^2 (4 - 18 z + 33 z^2 - 27 z^3 +  8 z^4 + y^3 (-2 + 8 z) + 6 y^2 (1 - 4 z\\
  &+ 2 z^2) +  y (-8 + 36 z - 45 z^2 + 20 z^3)) H(0, 2, y) +  864 (-1 + y)^2 y (-1 + z)^2 z (-1 + y + z) (y + z)^2  (-4 + 8 y^3 + 14 z\\
  &- 19 z^2 + 8 z^3 + y^2 (-19 + 12 z) +  2 y (7 - 10 z + 6 z^2)) H(1, 1, z) + 864 (-1 + y)^2 y (-1 + z)^2 z  (y + z)^2 (4 - 18 z\\
  &+ 33 z^2 - 27 z^3 + 8 z^4 + y^3 (-2 + 8 z) +  6 y^2 (1 - 4 z + 2 z^2) + y (-8 + 36 z - 45 z^2 + 20 z^3)) H(2, 0, y) +  1728 (-1\\
  &+ y)^2 y (-1 + z)^2 z (y + z)^2 (5 + 4 y^4 - 39 z + 87 z^2 -  81 z^3 + 28 z^4 + y^3 (-15 + 16 z) + 3 y^2 (7 - 17 z + 12 z^2)\\
  &+  y (-15 + 66 z - 93 z^2 + 40 z^3)) H(2, 2, y)) -  2376 (-1 + y)^2 y (-1 + z)^2 z (-1 + y + z) (y + z)^2  (-4 + 8 y^3 + 14 z\\
  &- 19 z^2 + 8 z^3 + y^2 (-19 + 12 z) +  2 y (7 - 10 z + 6 z^2)) H(3, 2, y) +  H(0, y) (1188 y^2 (-1 + z)^2 z^2 (-1 + y + z) (y\\
  &+ z)^2  (8 y^3 + y (30 - 27 z) + 12 (-1 + z) + 2 y^2 (-13 + 6 z)) -  432 (-1 + y)^2 y (y^2 + 2 y (-1 + z) + 2 (-1 + z)^2) (-1\\
  &+ z)^2 z  (-1 + y + z) (y + z)^2 (-1 + 4 z) H(0, z)^2 -  864 (-1 + y)^2 y (-1 + z)^2 z (y + z)^2 (5 + 28 y^4 - 15 z + 21 z^2\\
  &-  15 z^3 + 4 z^4 + y^3 (-81 + 40 z) + y^2 (87 - 93 z + 36 z^2) +  y (-39 + 66 z - 51 z^2 + 16 z^3)) H(1, z)^2 -  432 y (-1\\
  &+ z)^2 z (y^7 (-4 + 52 z) + 4 y^6 (4 - 57 z + 49 z^2) -  4 z^2 (-1 + 2 z - 2 z^2 + z^3) + y^5 (-28 + 408 z - 735 z^2 +  276 z^3)\\
  &+ 4 y z (-3 - 7 z + 31 z^2 - 32 z^3 + 12 z^4) +  y^4 (28 - 360 z + 1079 z^2 - 889 z^3 + 172 z^4) +  y^2 (4 + 4 z + 244 z^2 - 552 z^3\\
  &+ 407 z^4 - 87 z^5) +  y^3 (-16 + 136 z - 760 z^2 + 1046 z^3 - 465 z^4 + 40 z^5)) H(2, y) -  864 (-1 + y)^2 y (-1 + z)^2 z (y\\
  &+ z)^2 (4 y^4 + (-1 + z)^3 (-1 + 4 z) +  y^3 (-13 + 8 z) + 3 y^2 (5 - 7 z + 4 z^2) +  y (-7 + 18 z - 21 z^2 + 8 z^3)) H(2, y)^2\\
  &+  H(0, z) (-864 (-1 + y)^2 y (-1 + z)^2 z (y + z)^2  (4 + 8 y^4 - 8 z + 6 z^2 - 2 z^3 + y^3 (-27 + 20 z) +  3 y^2 (11 - 15 z\\
  &+ 4 z^2) + 2 y (-9 + 18 z - 12 z^2 + 4 z^3))  H(1, z) - 864 (-1 + y)^2 y (-1 + z)^2 z (y + z)^2  (4 - 18 z + 33 z^2 - 27 z^3 + 8 z^4\\
  &+ y^3 (-2 + 8 z) +  6 y^2 (1 - 4 z + 2 z^2) + y (-8 + 36 z - 45 z^2 + 20 z^3)) H(2, y)) +  H(1, z) (2376 (-1 + y)^2 y (-1 + 4 y) (-1\\
  &+ z)^2 z (-1 + y + z) (y + z)^2  (2 + 2 y^2 + 2 y (-2 + z) - 2 z + z^2) - 864 (-1 + y)^2 y (-1 + z)^2 z  (y + z)^2 (2 + 24 y^4 - 4 z\\
  &+ 3 z^2 - z^3 + y^3 (-67 + 28 z) +  3 y^2 (23 - 21 z + 8 z^2) + y (-28 + 36 z - 21 z^2 + 4 z^3))  H(2, y) + 864 (-1 + y)^2 y (-1\\
  &+ z)^2 z (-1 + y + z) (y + z)^2  (-4 + 8 y^3 + 14 z - 19 z^2 + 8 z^3 + y^2 (-19 + 12 z) +  2 y (7 - 10 z + 6 z^2)) H(3, y)) + 864 (-1\\
  &+ y)^2 y (-1 + 4 y)  (-1 + z)^2 z (-1 + y + z) (y + z)^2 (2 + 2 y^2 + 2 y (-2 + z) - 2 z +  z^2) H(0, 0, y) + 864 (-1 + y)^2 y (y^2\\
  &+ 2 y (-1 + z) + 2 (-1 + z)^2)  (-1 + z)^2 z (-1 + y + z) (y + z)^2 (-1 + 4 z) H(0, 0, z) +  864 (-1 + y)^2 y (-1 + z)^2 z (y\\
  &+ z)^2 (4 + 8 y^4 - 8 z + 6 z^2 -  2 z^3 + y^3 (-27 + 20 z) + 3 y^2 (11 - 15 z + 4 z^2) +  2 y (-9 + 18 z - 12 z^2\\
  &+ 4 z^3)) H(0, 1, z) -  864 (-1 + y)^2 y (-1 + 4 y) (-1 + z)^2 z (-1 + y + z) (y + z)^2  (2 + 2 y^2 + 2 y (-2 + z) - 2 z\\
  &+ z^2) H(0, 2, y) +  864 (-1 + y)^2 y (-1 + 4 y) (-1 + z)^2 z (-1 + y + z) (y + z)^2  (2 + 2 y^2 + 2 y (-2 + z) - 2 z\\
  &+ z^2) H(1, 0, y) +  864 (-1 + y)^2 y (-1 + z)^2 z (y + z)^2 (4 + 8 y^4 - 8 z + 6 z^2 -  2 z^3 + y^3 (-27 + 20 z) + 3 y^2 (11 - 15 z\\
  &+ 4 z^2) +  2 y (-9 + 18 z - 12 z^2 + 4 z^3)) H(1, 0, z) +  1728 (-1 + y)^2 y (-1 + z)^2 z (y + z)^2 (5 + 28 y^4 - 15 z + 21 z^2\\
  &-  15 z^3 + 4 z^4 + y^3 (-81 + 40 z) + y^2 (87 - 93 z + 36 z^2) +  y (-39 + 66 z - 51 z^2 + 16 z^3)) H(1, 1, z) -  864 (-1\\
  &+ y)^2 y (-1 + 4 y) (-1 + z)^2 z (-1 + y + z) (y + z)^2  (2 + 2 y^2 + 2 y (-2 + z) - 2 z + z^2) H(2, 0, y) +  864 (-1 + y)^2 y (-1\\
  &+ z)^2 z (-1 + y + z) (y + z)^2  (-4 + 8 y^3 + 14 z - 19 z^2 + 8 z^3 + y^2 (-19 + 12 z) +  2 y (7 - 10 z + 6 z^2)) H(2, 2, y)\\
  &+ 864 (-1 + y)^2 y (-1 + z)^2 z  (-1 + y + z) (y + z)^2 (-4 + 8 y^3 + 14 z - 19 z^2 + 8 z^3 +  y^2 (-19 + 12 z) + 2 y (7 - 10 z\\
  &+ 6 z^2)) H(3, 2, y)) +  H(1, z) (1188 (-1 + y)^2 y (-1 + z)^2 z (-1 + y + z)  (y^4 (-3 + 20 z) - 3 z^2 (2 - 2 z + z^2) + y^3 (6\\
  &- 50 z + 60 z^2) +  2 y z (4 + 17 z - 25 z^2 + 10 z^3) +  y^2 (-6 + 34 z - 94 z^2 + 60 z^3)) + 108 (-1 + y)^2 y (-1 + z)^2 z  (y\\
  &+ z)^2 (124 + 1240 y^4 - 1510 z + 4011 z^2 - 3865 z^3 + 1240 z^4 +  y^3 (-3865 + 2972 z) + y^2 (4011 - 6609 z + 4008 z^2) +  y (-1510\\
  &+ 4752 z - 6609 z^2 + 2972 z^3)) H(2, y)^2 -  2376 (-1 + y)^2 y (-1 + z)^2 z (-1 + y + z) (y + z)^2  (-4 + 8 y^3 + 14 z - 19 z^2\\
  &+ 8 z^3 + y^2 (-19 + 12 z) +  2 y (7 - 10 z + 6 z^2)) H(3, y) + 1728 (-1 + y)^2 y (-1 + z)^2 z  (y + z)^2 (4 + 12 y^4 - 20 z + 39 z^2\\
  &- 35 z^3 + 12 z^4 +  5 y^3 (-7 + 4 z) + 3 y^2 (13 - 17 z + 8 z^2) +  y (-20 + 48 z - 51 z^2 + 20 z^3)) H(2, y) H(3, y) +  3456 (-1\\
  &+ y)^2 y (-1 + z)^2 z (y + z)^2 (2 + 10 y^4 - 4 z + 3 z^2 -  z^3 + 2 y^3 (-15 + 8 z) + 3 y^2 (11 - 12 z + 4 z^2) +  y (-15 + 24 z\\
  &- 15 z^2 + 4 z^3)) H(0, 0, y) +  864 (-1 + y)^2 y (-1 + z)^2 z (y + z)^2 (2 - 28 z + 69 z^2 - 67 z^3 +  24 z^4 + y^3 (-1 + 4 z)\\
  &+ 3 y^2 (1 - 7 z + 8 z^2) +  y (-4 + 36 z - 63 z^2 + 28 z^3)) H(0, 0, z) +  1728 (-1 + y)^2 y (-1 + z)^2 z (y + z)^2 (2 + 12 y^4\\
  &- 4 z + 3 z^2 -  z^3 + 2 y^3 (-17 + 8 z) + 12 y^2 (3 - 3 z + z^2) +  y (-16 + 24 z - 15 z^2 + 4 z^3)) H(0, 1, z) +  864 (-1\\
  &+ y)^2 y (-1 + z)^2 z (y + z)^2 (2 + 24 y^4 - 4 z + 3 z^2 - z^3 +  y^3 (-67 + 28 z) + 3 y^2 (23 - 21 z + 8 z^2) +  y (-28 + 36 z\\
  &- 21 z^2 + 4 z^3)) H(0, 2, y) -  864 (-1 + y)^2 y (-1 + z)^2 z (-1 + y + z) (y + z)^2  (-4 + 8 y^3 + 14 z - 19 z^2 + 8 z^3 + y^2 (-19\\
  &+ 12 z) +  2 y (7 - 10 z + 6 z^2)) H(0, 3, y) - 864 (-1 + y)^2 y (-1 + z)^2 z  (y + z)^2 (2 + y^2 (3 - 9 z) - 4 z + 3 z^2 - z^3\\
  &+ y^3 (-1 + 4 z) +  y (-4 + 12 z - 9 z^2 + 4 z^3)) H(1, 0, z) +  1728 (-1 + y)^2 y (-1 + z)^2 z (y + z)^2 (4 + 12 y^4 - 20 z + 39 z^2\\
  &-  35 z^3 + 12 z^4 + 5 y^3 (-7 + 4 z) + 3 y^2 (13 - 17 z + 8 z^2) +  y (-20 + 48 z - 51 z^2 + 20 z^3)) H(1, 1, z) +  864 (-1\\
  &+ y)^2 y (-1 + z)^2 z (y + z)^2 (2 + 24 y^4 - 4 z + 3 z^2 - z^3 +  y^3 (-67 + 28 z) + 3 y^2 (23 - 21 z + 8 z^2) +  y (-28 + 36 z\\
  &- 21 z^2 + 4 z^3)) H(2, 0, y) -  216 (-1 + y)^2 y (-1 + z)^2 z (y + z)^2 (124 + 1240 y^4 - 1510 z +  4011 z^2 - 3865 z^3 + 1240 z^4\\
\end{aligned}
\end{equation*}
\begin{equation*}
\begin{aligned}
  &+ y^3 (-3865 + 2972 z) +  y^2 (4011 - 6609 z + 4008 z^2) + y (-1510 + 4752 z - 6609 z^2 +  2972 z^3)) H(2, 2, y) - 1728 (-1\\
  &+ y)^2 y (-1 + z)^2 z (y + z)^2  (4 + 12 y^4 - 20 z + 39 z^2 - 35 z^3 + 12 z^4 + 5 y^3 (-7 + 4 z) +  3 y^2 (13 - 17 z + 8 z^2)\\
  &+ y (-20 + 48 z - 51 z^2 + 20 z^3))  H(2, 3, y) - 864 (-1 + y)^2 y (-1 + z)^2 z (-1 + y + z) (y + z)^2  (-4 + 8 y^3 + 14 z - 19 z^2\\
  &+ 8 z^3 + y^2 (-19 + 12 z) +  2 y (7 - 10 z + 6 z^2)) H(3, 0, y) - 1728 (-1 + y)^2 y (-1 + z)^2 z  (y + z)^2 (4 + 12 y^4 - 20 z\\
  &+ 39 z^2 - 35 z^3 + 12 z^4 +  5 y^3 (-7 + 4 z) + 3 y^2 (13 - 17 z + 8 z^2) +  y (-20 + 48 z - 51 z^2 + 20 z^3)) H(3, 2, y))\\
  &+  H(2, y) (1188 (-1 + y)^2 y (-1 + z)^2 z (-1 + y + z)  (y^4 (-3 + 20 z) - 3 z^2 (2 - 2 z + z^2) + y^3 (6 - 50 z + 60 z^2)\\
  &+  2 y z (4 + 17 z - 25 z^2 + 10 z^3) +  y^2 (-6 + 34 z - 94 z^2 + 60 z^3)) + 864 (-1 + y)^2 y (-1 + z)^2 z  (y + z)^2 (2 + 24 y^4\\
  &- 4 z + 3 z^2 - z^3 + y^3 (-67 + 28 z) +  3 y^2 (23 - 21 z + 8 z^2) + y (-28 + 36 z - 21 z^2 + 4 z^3))  H(0, 0, y) + 3456 (-1\\
  &+ y)^2 y (-1 + z)^2 z (y + z)^2  (2 - 15 z + 33 z^2 - 30 z^3 + 10 z^4 + y^3 (-1 + 4 z) +  4 y (-1 + z)^2 (-1 + 4 z) + 3 y^2 (1 - 5 z\\
  &+ 4 z^2)) H(0, 0, z) +  864 (-1 + y)^2 y (-1 + z)^2 z (y + z)^2 (2 - 28 z + 69 z^2 - 67 z^3 +  24 z^4 + y^3 (-1 + 4 z) + 3 y^2 (1\\
  &- 7 z + 8 z^2) +  y (-4 + 36 z - 63 z^2 + 28 z^3)) H(0, 1, z) -  1728 (-1 + y)^2 y (-1 + z)^2 z (y + z)^2 (2 + 12 y^4 - 4 z + 3 z^2\\
  &-  z^3 + 2 y^3 (-17 + 8 z) + 12 y^2 (3 - 3 z + z^2) +  y (-16 + 24 z - 15 z^2 + 4 z^3)) H(0, 2, y) +  864 (-1 + y)^2 y (-1\\
  &+ z)^2 z (y + z)^2 (2 + 24 y^4 - 4 z + 3 z^2 - z^3 +  y^3 (-67 + 28 z) + 3 y^2 (23 - 21 z + 8 z^2) +  y (-28 + 36 z - 21 z^2\\
  &+ 4 z^3)) H(1, 0, y) +  864 (-1 + y)^2 y (-1 + z)^2 z (y + z)^2 (2 - 28 z + 69 z^2 - 67 z^3 +  24 z^4 + y^3 (-1 + 4 z) + 3 y^2 (1\\
  &- 7 z + 8 z^2) +  y (-4 + 36 z - 63 z^2 + 28 z^3)) H(1, 0, z) -  216 (-1 + y)^2 y (-1 + z)^2 z (y + z)^2 (124 + 1240 y^4 - 1510 z\\
  &+  4011 z^2 - 3865 z^3 + 1240 z^4 + y^3 (-3865 + 2972 z) +  y^2 (4011 - 6609 z + 4008 z^2) + y (-1510 + 4752 z - 6609 z^2\\
  &+  2972 z^3)) H(1, 1, z) - 1728 (-1 + y)^2 y (-1 + z)^2 z (y + z)^2  (2 + 12 y^4 - 4 z + 3 z^2 - z^3 + 2 y^3 (-17 + 8 z) +  12 y^2 (3\\
  &- 3 z + z^2) + y (-16 + 24 z - 15 z^2 + 4 z^3)) H(2, 0, y) +  1728 (-1 + y)^2 y (-1 + z)^2 z (y + z)^2 (4 + 12 y^4 - 20 z + 39 z^2\\
  &-  35 z^3 + 12 z^4 + 5 y^3 (-7 + 4 z) + 3 y^2 (13 - 17 z + 8 z^2) +  y (-20 + 48 z - 51 z^2 + 20 z^3)) H(2, 2, y) +  1728 (-1\\
  &+ y)^2 y (-1 + z)^2 z (y + z)^2 (4 + 12 y^4 - 20 z + 39 z^2 -  35 z^3 + 12 z^4 + 5 y^3 (-7 + 4 z) + 3 y^2 (13 - 17 z + 8 z^2)\\
  &+  y (-20 + 48 z - 51 z^2 + 20 z^3)) H(3, 2, y)) -  216 (-1 + y)^2 y (-1 + z)^2 z (y + z)^2 (296 + 3224 y^4 - 518 z + 453 z^2\\
  &-  431 z^3 + 200 z^4 + y^3 (-8315 + 5764 z) +  3 y^2 (2475 - 3233 z + 984 z^2) + y (-2630 + 4728 z - 2919 z^2\\
  &+  820 z^3)) H(0, 0, 0, y) - 216 (-1 + y)^2 y (-1 + z)^2 z (y + z)^2  (296 + 200 y^4 - 2630 z + 7425 z^2 - 8315 z^3 + 3224 z^4\\
  &+  y^3 (-431 + 820 z) + 3 y^2 (151 - 973 z + 984 z^2) +  y (-518 + 4728 z - 9699 z^2 + 5764 z^3)) H(0, 0, 0, z) -  864 (-1\\
  &+ y)^2 y (-1 + z)^2 z (y + z)^2 (2 + 16 y^4 - 4 z + 3 z^2 - z^3 +  y^3 (-51 + 28 z) + 3 y^2 (19 - 21 z + 8 z^2) +  y (-24 + 36 z\\
  &- 21 z^2 + 4 z^3)) H(0, 0, 1, z) +  864 (-1 + y)^2 y (-1 + z)^2 z (y + z)^2 (16 y^4 + y^3 (-53 + 36 z) +  3 y^2 (21 - 27 z + 8 z^2)\\
  &- 3 (-2 + 4 z - 3 z^2 + z^3) +  y (-32 + 60 z - 39 z^2 + 12 z^3)) H(0, 0, 2, y) -  864 (-1 + y)^2 y (-1 + 4 y) (-1 + z)^2 z (-1 + y\\
  &+ z) (y + z)^2  (2 + 2 y^2 + 2 y (-2 + z) - 2 z + z^2) H(0, 1, 0, y) -  864 (-1 + y)^2 y^2 (-1 + z)^2 z (y + z)^2 (8 y^3 + y^2 (-25\\
  &+ 12 z) -  2 (5 - 6 z + 3 z^2) + 3 y (9 - 9 z + 4 z^2)) H(0, 1, 0, z) -  864 (-1 + y)^2 y (-1 + z)^2 z (y + z)^2 (48 y^4 + y^3 (-137\\
  &+ 68 z) +  3 y^2 (49 - 51 z + 16 z^2) - 5 (-2 + 4 z - 3 z^2 + z^3) +  y (-68 + 108 z - 69 z^2 + 20 z^3)) H(0, 1, 1, z) +  864 (-1\\
  &+ y)^2 y (-1 + z)^2 z (y + z)^2 (16 y^4 + y^3 (-53 + 36 z) +  3 y^2 (21 - 27 z + 8 z^2) - 3 (-2 + 4 z - 3 z^2 + z^3) +  y (-32 + 60 z\\
  &- 39 z^2 + 12 z^3)) H(0, 2, 0, y) +  2592 (-1 + y)^2 y (-1 + z)^2 z (y + z)^2 (2 + 16 y^4 - 4 z + 3 z^2 - z^3 +  5 y^3 (-9 + 4 z)\\
  &+ y^2 (47 - 45 z + 16 z^2) +  y (-20 + 28 z - 17 z^2 + 4 z^3)) H(0, 2, 2, y) -  864 (-1 + y)^2 y (-1 + z)^2 z (-1 + y + z) (y\\
  &+ z)^2  (-4 + 8 y^3 + 14 z - 19 z^2 + 8 z^3 + y^2 (-19 + 12 z) +  2 y (7 - 10 z + 6 z^2)) H(0, 3, 2, y) - 1728 (-1 + y)^2 y (-1\\
  &+ 4 y)  (-1 + z)^2 z (-1 + y + z) (y + z)^2 (2 + 2 y^2 + 2 y (-2 + z) - 2 z + z^2)  H(1, 0, 0, y) + 864 (-1 + y)^2 y (-1 + z)^2 z (y\\
  &+ z)^2  (2 + y^2 (3 - 9 z) - 4 z + 3 z^2 - z^3 + y^3 (-1 + 4 z) +  y (-4 + 12 z - 9 z^2 + 4 z^3)) H(1, 0, 0, z) -  1728 (-1\\
  &+ y)^2 y (-1 + z)^2 z (y + z)^2 (2 + 12 y^4 - 4 z + 3 z^2 - z^3 +  2 y^3 (-17 + 8 z) + 12 y^2 (3 - 3 z + z^2) +  y (-16 + 24 z\\
  &- 15 z^2 + 4 z^3)) H(1, 0, 1, z) -  864 (-1 + y)^2 y (-1 + z)^2 z (y + z)^2 (2 + 24 y^4 - 4 z + 3 z^2 - z^3 +  y^3 (-67 + 28 z)\\
  &+ 3 y^2 (23 - 21 z + 8 z^2) +  y (-28 + 36 z - 21 z^2 + 4 z^3)) H(1, 0, 2, y) +  864 (-1 + y)^2 y (-1 + z)^2 z (y + z)^2 (2 + y^2 (3\\
  &- 9 z) - 4 z + 3 z^2 -  z^3 + y^3 (-1 + 4 z) + y (-4 + 12 z - 9 z^2 + 4 z^3)) H(1, 1, 0, z) -  216 (-1 + y)^2 y (-1 + z)^2 z (y\\
  &+ z)^2 (156 + 1336 y^4 - 1670 z +  4323 z^2 - 4145 z^3 + 1336 z^4 + y^3 (-4145 + 3132 z) +  y^2 (4323 - 7017 z + 4200 z^2) + y (-1670\\
  &+ 5136 z - 7017 z^2 +  3132 z^3)) H(1, 1, 1, z) - 864 (-1 + y)^2 y (-1 + z)^2 z (y + z)^2  (2 + 24 y^4 - 4 z + 3 z^2 - z^3 + y^3 (-67\\
  &+ 28 z) +  3 y^2 (23 - 21 z + 8 z^2) + y (-28 + 36 z - 21 z^2 + 4 z^3))  H(1, 2, 0, y) + 864 (-1 + y)^2 y (-1 + z)^2 z (y\\
  &+ z)^2  (16 y^4 + y^3 (-53 + 36 z) + 3 y^2 (21 - 27 z + 8 z^2) -  3 (-2 + 4 z - 3 z^2 + z^3) + y (-32 + 60 z - 39 z^2\\
  &+ 12 z^3))  H(2, 0, 0, y) + 2592 (-1 + y)^2 y (-1 + z)^2 z (y + z)^2  (2 + 16 y^4 - 4 z + 3 z^2 - z^3 + 5 y^3 (-9 + 4 z) +  y^2 (47\\
  &- 45 z + 16 z^2) + y (-20 + 28 z - 17 z^2 + 4 z^3))  H(2, 0, 2, y) - 864 (-1 + y)^2 y (-1 + z)^2 z (y + z)^2  (2 + 24 y^4 - 4 z\\
  &+ 3 z^2 - z^3 + y^3 (-67 + 28 z) +  3 y^2 (23 - 21 z + 8 z^2) + y (-28 + 36 z - 21 z^2 + 4 z^3))  H(2, 1, 0, y) + 2592 (-1\\
  &+ y)^2 y (-1 + z)^2 z (y + z)^2  (2 + 16 y^4 - 4 z + 3 z^2 - z^3 + 5 y^3 (-9 + 4 z) +  y^2 (47 - 45 z + 16 z^2) + y (-20 + 28 z\\
  &- 17 z^2 + 4 z^3))  H(2, 2, 0, y) - 216 (-1 + y)^2 y (-1 + z)^2 z (y + z)^2  (156 + 1336 y^4 - 1670 z + 4323 z^2 - 4145 z^3\\
  &+ 1336 z^4 +  y^3 (-4145 + 3132 z) + y^2 (4323 - 7017 z + 4200 z^2) +  y (-1670 + 5136 z - 7017 z^2 + 3132 z^3)) H(2, 2, 2, y)\\
\end{aligned}
\end{equation*}
\begin{equation*}
\begin{aligned}
  &-  1728 (-1 + y)^2 y (-1 + z)^2 z (y + z)^2 (4 + 12 y^4 - 20 z + 39 z^2 -  35 z^3 + 12 z^4 + 5 y^3 (-7 + 4 z) + 3 y^2 (13 - 17 z\\
  &+ 8 z^2) +  y (-20 + 48 z - 51 z^2 + 20 z^3)) H(2, 3, 2, y) -  864 (-1 + y)^2 y (-1 + z)^2 z (-1 + y + z) (y + z)^2  (-4 + 8 y^3\\
  &+ 14 z - 19 z^2 + 8 z^3 + y^2 (-19 + 12 z) +  2 y (7 - 10 z + 6 z^2)) H(3, 0, 2, y) - 864 (-1 + y)^2 y (-1 + z)^2 z  (-1 + y + z) (y\\
  &+ z)^2 (-4 + 8 y^3 + 14 z - 19 z^2 + 8 z^3 +  y^2 (-19 + 12 z) + 2 y (7 - 10 z + 6 z^2)) H(3, 2, 0, y) -  3456 (-1 + y)^2 y (-1\\
  &+ z)^2 z (y + z)^2 (4 + 12 y^4 - 20 z + 39 z^2 -  35 z^3 + 12 z^4 + 5 y^3 (-7 + 4 z) + 3 y^2 (13 - 17 z + 8 z^2) +  y (-20 + 48 z\\
  &- 51 z^2 + 20 z^3)) H(3, 2, 2, y)\Bigg\}\Big/ \Big(864 (-1 + y)^2 y^2 (-1 + z)^2 z^2 (-1 + y + z) (y + z)^2\Big);\\[10pt]
\mathcal{A}_{3;C_{A}n_{f}}^{(2)} &=
\Bigg\{-8 (-1 + y) (-1 + z) (y + z) (103 y^6 (-1 + 4 z) +  6 y^5 (47 - 272 z + 170 z^2) + z^3 (206 - 385 z + 282 z^2 - 103 z^3) +  y^4 (-385\\
  &+ 2703 z - 4017 z^2 + 1000 z^3) +  3 y^2 z (206 - 1022 z + 1843 z^2 - 1339 z^3 + 340 z^4) +  y z^2 (618 - 2620 z + 2703 z^2 - 1632 z^3\\
  &+ 412 z^4) +  y^3 (206 - 2620 z + 5529 z^2 - 4976 z^3 + 1000 z^4)) +  9 (-1 + y) (-1 + z) (y + z)^4 (-56 + 72 y^4 + 529 z - 1569 z^2\\
  &+ 1834 z^3 -  744 z^4 - 4 y^3 (31 + 12 z) - 3 y^2 (5 - 182 z + 216 z^2) +  y (105 - 1020 z + 2160 z^2 - 1336 z^3)) H(0, y)^2 -  216 (-1\\
  &+ y) (-1 + z) z (-1 + y + z) (y + z)^4  (3 - 12 z + 10 z^2 + y (-3 + 6 z)) H(0, y)^3 -  9 (-1 + y) (-1 + z) (y + z)^4 (56 + 744 y^4\\
  &- 105 z + 15 z^2 + 124 z^3 -  72 z^4 + 2 y^3 (-917 + 668 z) + 3 y^2 (523 - 720 z + 216 z^2) +  y (-529 + 1020 z - 546 z^2\\
  &+ 48 z^3)) H(0, z)^2 -  216 (-1 + y) y (3 + 10 y^2 + 6 y (-2 + z) - 3 z) (-1 + z) (-1 + y + z)  (y + z)^4 H(0, z)^3 + 1080 (-1 + y) (-1\\
  &+ z) (-1 + y + z) (y + z)^4  (10 y^3 + 6 y^2 (-2 + z) + y (3 - 6 z + 6 z^2) + z (3 - 12 z + 10 z^2))  H(1, z)^3 - 72 (-1 + y) (-1\\
  &+ z) (358 y^8 + 14 y^7 (-63 + 152 z) +  y^6 (732 - 4821 z + 5640 z^2) + y^5 (-223 + 3798 z - 11757 z^2 +  9136 z^3) + z^4 (14 - 223 z\\
  &+ 732 z^2 - 882 z^3 + 358 z^4) +  y z^3 (128 - 1175 z + 3798 z^2 - 4821 z^3 + 2128 z^4) +  2 y^3 z (64 - 1109 z + 5394 z^2 - 8670 z^3\\
  &+ 4568 z^4) +  y^2 z^2 (84 - 2218 z + 8460 z^2 - 11757 z^3 + 5640 z^4) +  y^4 (14 - 1175 z + 8460 z^2 - 17340 z^3\\
  &+ 10532 z^4)) H(2, y)^2 +  1080 (-1 + y) (-1 + z) (-1 + y + z) (y + z)^4 (10 y^3 + 6 y^2 (-2 + z) +  y (3 - 6 z + 6 z^2) + z (3 - 12 z\\
  &+ 10 z^2)) H(2, y)^3 +  H(0, y) (18 (-1 + z) (y + z)^4 (y^4 (-31 + 220 z) +  y^3 (124 - 607 z + 96 z^2) + 31 (-2 + 4 z - 3 z^2 + z^3)\\
  &-  31 y (-6 + 16 z - 12 z^2 + 5 z^3) + y^2 (-217 + 759 z - 387 z^2 +  124 z^3)) + 216 (-1 + y) (-1 + z) (y + z)^4  (2 + y^2 (3 - 9 z)\\
  &- 4 z + 3 z^2 - z^3 + y^3 (-1 + 4 z) +  y (-4 + 12 z - 9 z^2 + 4 z^3)) H(0, z) + 216 (-1 + y) (-1 + z)  (y + z)^4 (2 + 16 y^4 - 4 z\\
  &+ 3 z^2 - z^3 + 2 y^3 (-21 + 8 z) +  6 y^2 (7 - 6 z + 2 z^2) + y (-18 + 24 z - 15 z^2 + 4 z^3)) H(1, z) +  288 (-1 + y) y (-1\\
  &+ z) z (-5 y^3 + 3 y^4 + y z^2 + y^2 (6 + z - 6 z^2) +  z^2 (6 - 5 z + 3 z^2)) H(2, y)) +  H(1, z)^2 (-72 (-1 + y) (-1 + z) (358 y^8\\
  &+ 14 y^7 (-63 + 152 z) +  y^6 (732 - 4821 z + 5640 z^2) + y^5 (-223 + 3798 z - 11757 z^2 +  9136 z^3) + z^4 (14 - 223 z + 732 z^2\\
  &- 882 z^3 + 358 z^4) +  y z^3 (128 - 1175 z + 3798 z^2 - 4821 z^3 + 2128 z^4) +  2 y^3 z (64 - 1109 z + 5394 z^2 - 8670 z^3 + 4568 z^4)\\
  &+  y^2 z^2 (84 - 2218 z + 8460 z^2 - 11757 z^3 + 5640 z^4) +  y^4 (14 - 1175 z + 8460 z^2 - 17340 z^3 + 10532 z^4)) +  3240 (-1 + y) (-1\\
  &+ z) (-1 + y + z) (y + z)^4 (10 y^3 + 6 y^2 (-2 + z) +  y (3 - 6 z + 6 z^2) + z (3 - 12 z + 10 z^2)) H(2, y)) +  H(0, z) (18 (-1 + y) (y\\
  &+ z)^4 (-31 (-1 + z)^2 (2 - 2 z + z^2) +  31 y^3 (1 - 5 z + 4 z^2) + y^2 (-93 + 372 z - 387 z^2 + 96 z^3) +  y (124 - 496 z + 759 z^2\\
  &- 607 z^3 + 220 z^4)) +  288 (-1 + y) y (-1 + z) z (-5 y^3 + 3 y^4 + y z^2 + y^2 (6 + z - 6 z^2) +  z^2 (6 - 5 z + 3 z^2)) H(1, z)\\
  &+ 216 (-1 + y) (-1 + z) (y + z)^4  (y^3 (-1 + 4 z) + 4 y (-1 + z)^2 (-1 + 4 z) + 3 y^2 (1 - 5 z + 4 z^2) +  2 (1 - 9 z + 21 z^2 - 21 z^3\\
  &+ 8 z^4)) H(2, y)) -  18 (-1 + y) (-1 + z) (y + z)^4 (-56 + 72 y^4 + 529 z - 1569 z^2 +  1834 z^3 - 744 z^4 - 4 y^3 (31 + 12 z)\\
  &- 3 y^2 (5 - 182 z + 216 z^2) +  y (105 - 1020 z + 2160 z^2 - 1336 z^3)) H(0, 0, y) +  18 (-1 + y) (-1 + z) (y + z)^4 (56 + 744 y^4\\
  &- 105 z + 15 z^2 + 124 z^3 -  72 z^4 + 2 y^3 (-917 + 668 z) + 3 y^2 (523 - 720 z + 216 z^2) +  y (-529 + 1020 z - 546 z^2\\
  &+ 48 z^3)) H(0, 0, z) -  72 (-1 + y) (-1 + z) (48 y^8 + 6 y^7 (-21 + 40 z) +  6 y^6 (21 - 102 z + 86 z^2) - 3 z^4 (-2 + 4 z - 3 z^2\\
  &+ z^3) +  3 y^5 (-18 + 196 z - 411 z^2 + 212 z^3) +  y z^3 (48 - 122 z + 120 z^2 - 57 z^3 + 12 z^4) +  2 y^2 z^2 (18 - 142 z + 234 z^2\\
  &- 153 z^3 + 42 z^4) +  4 y^3 z (12 - 92 z + 237 z^2 - 210 z^3 + 66 z^4) +  y^4 (6 - 248 z + 1053 z^2 - 1335 z^3 + 504 z^4)) H(0, 1, z)\\
  &+  72 (-1 + y) (-1 + z) (48 y^8 + 6 y^7 (-21 + 40 z) +  6 y^6 (21 - 102 z + 86 z^2) - 3 z^4 (-2 + 4 z - 3 z^2 + z^3) +  y z^4 (-82\\
  &+ 96 z - 57 z^2 + 12 z^3) +  4 y^3 z^2 (-94 + 249 z - 210 z^2 + 66 z^3) +  3 y^5 (-18 + 188 z - 411 z^2 + 212 z^3) +  2 y^2 z^2 (18\\
  &- 146 z + 234 z^2 - 153 z^3 + 42 z^4) +  y^4 (6 - 208 z + 1053 z^2 - 1335 z^3 + 504 z^4)) H(0, 2, y) -  216 (-1 + y) y (-1 + z) (y\\
  &+ z)^4 (16 y^3 + y^2 (-41 + 12 z) -  2 (7 - 6 z + 3 z^2) + 3 y (13 - 9 z + 4 z^2)) H(1, 0, y) +  72 (-1 + y) (-1 + z) (3 y^7 (-1 + 4 z)\\
  &+ y^6 (9 - 39 z + 48 z^2) -  3 z^4 (-2 + 4 z - 3 z^2 + z^3) + y z^4 (-40 + 60 z - 39 z^2 + 12 z^3) +  3 y^5 (-4 + 20 z - 51 z^2\\
  &+ 28 z^3) +  y^3 z^2 (-124 + 312 z - 285 z^2 + 84 z^3) +  y^2 z^2 (36 - 124 z + 207 z^2 - 153 z^3 + 48 z^4) +  y^4 (6 - 40 z + 207 z^2\\
  &- 285 z^3 + 96 z^4)) H(1, 0, z) +  144 (-1 + y) (-1 + z) (358 y^8 + 14 y^7 (-63 + 152 z) +  y^6 (732 - 4821 z + 5640 z^2) + y^5 (-223\\
  &+ 3798 z - 11757 z^2 +  9136 z^3) + z^4 (14 - 223 z + 732 z^2 - 882 z^3 + 358 z^4) +  y z^3 (128 - 1175 z + 3798 z^2 - 4821 z^3\\
  &+ 2128 z^4) +  2 y^3 z (64 - 1109 z + 5394 z^2 - 8670 z^3 + 4568 z^4) +  y^2 z^2 (84 - 2218 z + 8460 z^2 - 11757 z^3 + 5640 z^4)\\
  &+  y^4 (14 - 1175 z + 8460 z^2 - 17340 z^3 + 10532 z^4)) H(1, 1, z) +  H(2, y) (36 (-1 + y) (-1 + z) (y^7 (-9 + 84 z) +  3 y^6 (9 - 83 z\\
  &+ 144 z^2) - 9 z^4 (-2 + 4 z - 3 z^2 + z^3) +  3 y^5 (-12 + 70 z - 373 z^2 + 324 z^3) +  y z^3 (-168 - 26 z + 210 z^2 - 249 z^3\\
  &+ 84 z^4) +  y^2 z^2 (-60 - 314 z + 1053 z^2 - 1119 z^3 + 432 z^4) +  y^3 z (-168 - 314 z + 1740 z^2 - 2175 z^3 + 972 z^4) +  y^4 (18\\
\end{aligned}
\end{equation*}
\begin{equation*}
\begin{aligned}
  &- 26 z + 1053 z^2 - 2175 z^3 + 1248 z^4)) -  6480 (-1 + y) (-1 + z) (-1 + y + z) (y + z)^4 (10 y^3 + 6 y^2 (-2 + z) +  y (3 - 6 z\\
  &+ 6 z^2) + z (3 - 12 z + 10 z^2)) H(1, 1, z)) +  72 (-1 + y) (-1 + z) (48 y^8 + 6 y^7 (-21 + 40 z) +  6 y^6 (21 - 102 z + 86 z^2)\\
  &- 3 z^4 (-2 + 4 z - 3 z^2 + z^3) +  y z^4 (-82 + 96 z - 57 z^2 + 12 z^3) +  4 y^3 z^2 (-94 + 249 z - 210 z^2 + 66 z^3) +  3 y^5 (-18\\
  &+ 188 z - 411 z^2 + 212 z^3) +  2 y^2 z^2 (18 - 146 z + 234 z^2 - 153 z^3 + 42 z^4) +  y^4 (6 - 208 z + 1053 z^2 - 1335 z^3\\
  &+ 504 z^4)) H(2, 0, y) +  144 (-1 + y) (-1 + z) (358 y^8 + 14 y^7 (-63 + 152 z) +  y^6 (732 - 4821 z + 5640 z^2) + y^5 (-223 + 3798 z\\
  &- 11757 z^2 +  9136 z^3) + z^4 (14 - 223 z + 732 z^2 - 882 z^3 + 358 z^4) +  y z^3 (128 - 1175 z + 3798 z^2 - 4821 z^3 + 2128 z^4)\\
  &+  2 y^3 z (64 - 1109 z + 5394 z^2 - 8670 z^3 + 4568 z^4) +  y^2 z^2 (84 - 2218 z + 8460 z^2 - 11757 z^3 + 5640 z^4) +  y^4 (14 - 1175 z\\
  &+ 8460 z^2 - 17340 z^3 + 10532 z^4)) H(2, 2, y) +  H(1, z) (36 (-1 + y) (-1 + z) (y^7 (-9 + 84 z) +  3 y^6 (9 - 83 z + 144 z^2)\\
  &- 9 z^4 (-2 + 4 z - 3 z^2 + z^3) +  3 y^5 (-12 + 70 z - 373 z^2 + 324 z^3) +  y z^3 (-168 - 26 z + 210 z^2 - 249 z^3 + 84 z^4)\\
  &+  y^2 z^2 (-60 - 314 z + 1053 z^2 - 1119 z^3 + 432 z^4) +  y^3 z (-168 - 314 z + 1740 z^2 - 2175 z^3 + 972 z^4) +  y^4 (18 - 26 z\\
  &+ 1053 z^2 - 2175 z^3 + 1248 z^4)) +  3240 (-1 + y) (-1 + z) (-1 + y + z) (y + z)^4 (10 y^3 + 6 y^2 (-2 + z) +  y (3 - 6 z + 6 z^2)\\
  &+ z (3 - 12 z + 10 z^2)) H(2, y)^2 -  216 (-1 + y) (-1 + z) (y + z)^4 (4 + 16 y^4 - 22 z + 45 z^2 - 43 z^3 +  16 z^4 + y^3 (-43 + 20 z)\\
  &+ 3 y^2 (15 - 17 z + 8 z^2) +  y (-22 + 48 z - 51 z^2 + 20 z^3)) H(3, y) - 6480 (-1 + y) (-1 + z)  (-1 + y + z) (y + z)^4 (10 y^3\\
  &+ 6 y^2 (-2 + z) + y (3 - 6 z + 6 z^2) +  z (3 - 12 z + 10 z^2)) H(2, 2, y)) - 216 (-1 + y) (-1 + z) (y + z)^4  (4 + 16 y^4 - 22 z\\
  &+ 45 z^2 - 43 z^3 + 16 z^4 + y^3 (-43 + 20 z) +  3 y^2 (15 - 17 z + 8 z^2) + y (-22 + 48 z - 51 z^2 + 20 z^3))  H(3, 2, y) + 1296 (-1\\
  &+ y) (-1 + z) z (-1 + y + z) (y + z)^4  (3 - 12 z + 10 z^2 + y (-3 + 6 z)) H(0, 0, 0, y) +  1296 (-1 + y) y (3 + 10 y^2 + 6 y (-2 + z)\\
  &- 3 z) (-1 + z) (-1 + y + z)  (y + z)^4 H(0, 0, 0, z) - 6480 (-1 + y) (-1 + z) (-1 + y + z) (y + z)^4  (10 y^3 + 6 y^2 (-2 + z) + y (3\\
  &- 6 z + 6 z^2) + z (3 - 12 z + 10 z^2))  H(1, 1, 1, z) - 6480 (-1 + y) (-1 + z) (-1 + y + z) (y + z)^4  (10 y^3 + 6 y^2 (-2 + z) + y (3\\
  &- 6 z + 6 z^2) + z (3 - 12 z + 10 z^2))  H(2, 2, 2, y)\Bigg\}\Big/\Big(864 (-1 + y) y (-1 + z) z (-1 + y + z) (y + z)^4\Big);\\[10pt]
\mathcal{A}_{3;C_{F}n_{f}}^{(2)} &=
\Bigg\{(-1 + y) (-1 + z) ((-1 + z)^2 z^2 (490 - 463 z + 245 z^2) +  y^6 (245 - 982 z + 656 z^2) + y^5 (-953 + 4142 z - 4366 z^2\\
  &+ 1096 z^3) +  y^4 (1661 - 8739 z + 14467 z^2 - 8188 z^3 + 880 z^4) +  y z (980 - 4869 z + 9468 z^2 - 8739 z^3 + 4142 z^4 - 982 z^5)\\
  &+  y^3 (-1443 + 9468 z - 21992 z^2 + 21140 z^3 - 8188 z^4 + 1096 z^5) +  y^2 (490 - 4869 z + 15614 z^2 - 21992 z^3 + 14467 z^4\\
  &- 4366 z^5 +  656 z^6)) + 9 (-1 + y)^2 (-1 + 4 y) (-1 + z)^2 (y + z)^2  (-2 + 2 y^3 + 4 z - 3 z^2 + z^3 + y^2 (-6 + 4 z) + y (6 - 8 z\\
  &+ 3 z^2))  H(0, y)^2 - 9 (1 - y) (-1 + y) (y^3 + 3 y^2 (-1 + z) + 4 y (-1 + z)^2 +  2 (-1 + z)^3) (-1 + z)^2 (y + z)^2 (-1\\
  &+ 4 z) H(0, z)^2 +  63 (1 - y) (-1 + y) (-1 + z)^2 (y + z)^2 (4 + 8 y^4 - 18 z + 33 z^2 -  27 z^3 + 8 z^4 + y^3 (-27 + 20 z) + 3 y^2 (11\\
  &- 17 z + 8 z^2) +  y (-18 + 48 z - 51 z^2 + 20 z^3)) H(1, z)^2 +  H(0, y) (-27 y (-1 + z)^2 z (-1 + y + z) (y + z)^2  (8 y^3 + y (30\\
  &- 27 z) + 12 (-1 + z) + 2 y^2 (-13 + 6 z)) -  54 (-1 + y)^2 (-1 + 4 y) (-1 + z)^2 (-1 + y + z) (y + z)^2  (2 + 2 y^2 + 2 y (-2 + z)\\
  &- 2 z + z^2) H(1, z)) +  27 (1 - y) (-1 + y) (-1 + z)^2 (y^5 (-3 + 20 z) + y^4 (9 - 73 z + 80 z^2) -  3 z^2 (-2 + 4 z - 3 z^2 + z^3)\\
  &+ 6 y^3 (-2 + 15 z - 34 z^2 + 20 z^3) +  y z (-8 - 32 z + 90 z^2 - 73 z^3 + 20 z^4) +  2 y^2 (3 - 16 z + 81 z^2 - 102 z^3\\
  &+ 40 z^4)) H(2, y) +  63 (1 - y) (-1 + y) (-1 + z)^2 (y + z)^2 (4 + 8 y^4 - 18 z + 33 z^2 -  27 z^3 + 8 z^4 + y^3 (-27 + 20 z)\\
  &+ 3 y^2 (11 - 17 z + 8 z^2) +  y (-18 + 48 z - 51 z^2 + 20 z^3)) H(2, y)^2 +  H(0, z) (-27 (-1 + y)^2 y z (-1 + y + z) (y + z)^2  (-12\\
  &+ 30 z - 26 z^2 + 8 z^3 + 3 y (4 - 9 z + 4 z^2)) -  54 (-1 + y)^2 (y^2 + 2 y (-1 + z) + 2 (-1 + z)^2) (-1 + z)^2 (-1 + y + z)  (y\\
  &+ z)^2 (-1 + 4 z) H(2, y)) +  H(1, z) (27 (1 - y) (-1 + y) (-1 + z)^2 (-1 + y + z)  (y^4 (-3 + 20 z) - 3 z^2 (2 - 2 z + z^2) + y^3 (6\\
  &- 50 z + 60 z^2) +  2 y z (4 + 17 z - 25 z^2 + 10 z^3) +  y^2 (-6 + 34 z - 94 z^2 + 60 z^3)) + 54 (-1 + y)^2 (-1 + z)^2  (-1 + y + z) (y\\
  &+ z)^2 (-4 + 8 y^3 + 14 z - 19 z^2 + 8 z^3 +  y^2 (-19 + 12 z) + 2 y (7 - 10 z + 6 z^2)) H(3, y)) +  18 (1 - y) (1 - 5 y + 4 y^2) (-1\\
  &+ z)^2 (y + z)  (2 y^4 + 6 y^3 (-1 + z) + y^2 (6 - 14 z + 7 z^2) +  z (-2 + 4 z - 3 z^2 + z^3) + y (-2 + 10 z - 11 z^2\\
  &+ 4 z^3)) H(0, 0, y) +  18 (1 - y) (-1 + y) (-1 + z) (y + z) (1 - 5 z + 4 z^2)  (y^4 + 2 (-1 + z)^3 z + 2 y (-1 + z)^2 (-1 + 3 z)\\
  &+ y^3 (-3 + 4 z) +  y^2 (4 - 11 z + 7 z^2)) H(0, 0, z) + 54 (-1 + y)^2 (-1 + 4 y) (-1 + z)^2  (y + z)^2 (-2 + 2 y^3 + 4 z - 3 z^2 + z^3\\
  &+ y^2 (-6 + 4 z) +  y (6 - 8 z + 3 z^2)) H(0, 1, z) - 54 (-1 + y)^2 (-1 + 4 y) (-1 + z)^2  (y + z)^2 (-2 + 2 y^3 + 4 z - 3 z^2 + z^3\\
  &+ y^2 (-6 + 4 z) +  y (6 - 8 z + 3 z^2)) H(0, 2, y) + 54 (-1 + y)^2 (-1 + 4 y) (-1 + z)^2  (y + z)^2 (-2 + 2 y^3 + 4 z - 3 z^2 + z^3\\
  &+ y^2 (-6 + 4 z) +  y (6 - 8 z + 3 z^2)) H(1, 0, y) + 126 (-1 + y)^2 (-1 + z)^2 (y + z)^2  (4 + 8 y^4 - 18 z + 33 z^2 - 27 z^3 + 8 z^4\\
  &+ y^3 (-27 + 20 z) +  3 y^2 (11 - 17 z + 8 z^2) + y (-18 + 48 z - 51 z^2 + 20 z^3))  H(1, 1, z) - 54 (-1 + y)^2 (-1 + 4 y) (-1 + z)^2 (y\\
  &+ z)^2  (-2 + 2 y^3 + 4 z - 3 z^2 + z^3 + y^2 (-6 + 4 z) + y (6 - 8 z + 3 z^2))  H(2, 0, y) + 126 (-1 + y)^2 (-1 + z)^2 (y + z)^2  (4\\
  &+ 8 y^4 - 18 z + 33 z^2 - 27 z^3 + 8 z^4 + y^3 (-27 + 20 z) +  3 y^2 (11 - 17 z + 8 z^2) + y (-18 + 48 z - 51 z^2 + 20 z^3))  H(2, 2, y)\\
  &+ 54 (-1 + y)^2 (-1 + z)^2 (y + z)^2  (4 + 8 y^4 - 18 z + 33 z^2 - 27 z^3 + 8 z^4 + y^3 (-27 + 20 z) +  3 y^2 (11 - 17 z + 8 z^2)\\
  &+ y (-18 + 48 z - 51 z^2 + 20 z^3)) H(3, 2, y)\Bigg\}\Big/ \Big(108 (-1 + y)^2 y (-1 + z)^2 z (-1 + y + z) (y + z)^2\Big);\\[1pt]
\mathcal{A}_{4;C_{A}^{2}}^{(2)} &=   \frac{39}{20}~ \mathcal{A}_{0}; \qquad
\mathcal{A}_{4;C_{F}^{2}}^{(2)} =  -\frac{44}{5}~ \mathcal{A}_{0}; \qquad
\mathcal{A}_{4;n_{f}^{2}}^{(2)} = 0; \qquad
\mathcal{A}_{4;C_{A}C_{F}}^{(2)} =   \frac{93}{10}~ \mathcal{A}_{0}; \qquad
\mathcal{A}_{4;C_{A}n_{f}}^{(2)} = 0;\qquad
\mathcal{A}_{4;C_{F}n_{f}}^{(2)} = 0; \\
\end{aligned}
\end{equation*}
\begin{equation*}
\begin{aligned}
\mathcal{A}_{5;C_{A}^{2}}^{(2)} &= 
\Bigg\{6048 y^9 z + (-1 + z)^5 (814 - 814 z + 83 z^2) +  216 y^8 z (-177 + 74 z - 12 z^2 + 3 z^3) +  y (-1 + z)^4 (4884 - 4916 z - 10029 z^2\\
  &+ 20360 z^3 - 14040 z^4 +  6048 z^5) + y^7 (83 + 112808 z - 124552 z^2 + 79000 z^3 - 39863 z^4 +  8812 z^5) + y^2 (-1 + z)^3 (12293\\
  &- 2060 z - 65043 z^2 + 119980 z^3 -  76600 z^4 + 15984 z^5) - y^3 (-1 + z)^2 (-16695 - 44834 z + 223663 z^2 -  364320 z^3 + 263636 z^4\\
  &- 73816 z^5 + 2592 z^6) +  y^6 (-1229 - 199901 z + 397732 z^2 - 413860 z^3 + 281873 z^4 - 103639 z^5 +  14832 z^6) + y^5 (5714\\
  &+ 219568 z - 670767 z^2 + 965408 z^3 -  836204 z^4 + 417156 z^5 - 103639 z^6 + 8812 z^7) +  y^4 (-13040 - 131106 z + 629609 z^2\\
  &- 1215939 z^3 + 1324022 z^4 -  836204 z^5 + 281873 z^6 - 39863 z^7 + 648 z^8) +  36 (-1 + y)^4 (-1 + z)^4 (2 + y^2 (3 - 9 z) - 4 z\\
  &+ 3 z^2 - z^3 +  y^3 (-1 + 4 z) + y (-4 + 12 z - 9 z^2 + 4 z^3)) H(0, y) +  36 (-1 + y)^4 (-1 + z)^4 (2 + y^2 (3 - 9 z) - 4 z + 3 z^2\\
  &- z^3 +  y^3 (-1 + 4 z) + y (-4 + 12 z - 9 z^2 + 4 z^3)) H(0, z) -  36 (-1 + y)^4 (-1 + z)^4 (220 y^4 + y^3 (-443 + 324 z) +  3 y^2 (81\\
  &- 117 z + 32 z^2) + y (-8 + 24 z + 6 z^2 - 24 z^3) +  6 (-2 + 4 z - 3 z^2 + z^3)) H(1, y) + 36 (-1 + y)^4 (-1 + z)^4  (220 y^4\\
  &+ 4 y^3 (-111 + 82 z) + 6 y^2 (41 - 60 z + 16 z^2) +  5 (-2 + 4 z - 3 z^2 + z^3) - y (12 - 36 z + 3 z^2 + 20 z^3)) H(1, z) +  72 (-1\\
  &+ y)^4 (-1 + z)^4 (-11 + 110 y^4 + 6 z + 114 z^2 - 219 z^3 +  110 z^4 + y^3 (-219 + 152 z) + 3 y^2 (38 - 59 z + 32 z^2) +  y (6 + 30 z\\
  &- 177 z^2 + 152 z^3)) H(2, y)\Bigg\}\Big/ \Big(144 (-1 + y)^4 y (-1 + z)^4 z (-1 + y + z)\Big);\\[4pt]
\mathcal{A}_{5;C_{F}^{2}}^{(2)} &= 
\Bigg\{-432 y^9 z - 6 (-1 + z)^5 (31 - 30 z + 20 z^2) -  12 y^8 z (-261 + 147 z - 40 z^2 + 10 z^3) -  6 y (-1 + z)^4 (185 - 354 z + 224 z^2\\
  &+ 102 z^3 - 234 z^4 + 72 z^5) +  y^7 (-120 - 8820 z + 8139 z^2 - 1414 z^3 - 733 z^4 + 356 z^5) -  3 y^2 (-1 + z)^3 (960 - 2620 z\\
  &+ 2922 z^2 - 754 z^3 - 949 z^4 + 588 z^5) +  y^3 (-1 + z)^2 (-4260 + 13428 z - 23904 z^2 + 19624 z^3 - 6031 z^4 -  454 z^5 + 480 z^6)\\
  &+ y^6 (780 + 11256 z - 11571 z^2 - 4643 z^3 +  9499 z^4 - 4121 z^5 + 528 z^6) +  y^5 (-2286 - 2220 z - 5247 z^2 + 31232 z^3 - 36206 z^4\\
  &+ 18060 z^5 -  4121 z^6 + 356 z^7) + y^4 (3930 - 13818 z + 38097 z^2 - 69183 z^3 +  68534 z^4 - 36206 z^5 + 9499 z^6 - 733 z^7\\
  &- 120 z^8) -  12 (-1 + y)^4 (-1 + z)^4 (14 y^4 + 6 y^2 (-3 + z) z + y^3 (-23 + 28 z) +  y (13 - 18 z + 15 z^2 - 8 z^3) + 2 (-2 + 4 z\\
  &- 3 z^2 + z^3)) H(1, y) +  12 (-1 + y)^4 (-1 + z)^4 (2 + 14 y^4 - 4 z + 3 z^2 - z^3 +  y^3 (-26 + 40 z) + y^2 (9 - 45 z + 6 z^2) +  y (1\\
  &+ 18 z - 12 z^2 + 4 z^3)) H(1, z) + 12 (-1 + y)^4 (-1 + z)^4  (-2 + 14 y^4 + 9 z + 3 z^2 - 24 z^3 + 14 z^4 + 8 y^3 (-3 + 4 z)\\
  &+  3 y^2 (1 - 10 z + 4 z^2) + y (9 - 30 z^2 + 32 z^3)) H(2, y)\Bigg\}\Big/ \Big(6 (-1 + y)^4 y (-1 + z)^4 z (-1 + y + z)\Big);\\[4pt]
\mathcal{A}_{5;n_{f}^{2}}^{(2)} &= 0;\\[4pt]
\end{aligned}
\end{equation*}
\begin{equation*}
\begin{aligned}
\mathcal{A}_{5;C_{A}C_{F}}^{(2)} &=
   \Bigg\{3456 y^9 z + (-1 + z)^5 (2470 - 2290 z + 1613 z^2) +  72 y^8 z (-351 + 207 z - 64 z^2 + 16 z^3) +  y^7 (1613 + 64136 z - 47008 z^2\\
  &- 12968 z^3 + 21871 z^4 - 6908 z^5) +  y (-1 + z)^4 (14640 - 32276 z + 25053 z^2 - 2392 z^3 - 11448 z^4 +  3456 z^5) + y^2 (-1\\
  &+ z)^3 (37763 - 128708 z + 163755 z^2 - 83348 z^3 -  2296 z^4 + 14904 z^5) - y^3 (-1 + z)^2 (-55665 + 243490 z - 455855 z^2\\
  &+  399648 z^3 - 166612 z^4 + 22184 z^5 + 4608 z^6) -  y^6 (10355 + 47891 z + 31748 z^2 - 206372 z^3 + 191809 z^4 - 69095 z^5\\
  &+  7488 z^6) + y^5 (30050 - 97592 z + 392007 z^2 - 755056 z^3 + 679888 z^4 -  308028 z^5 + 69095 z^6 - 6908 z^7) +  y^4 (-51380\\
  &+ 292182 z - 867721 z^2 + 1421763 z^3 - 1305946 z^4 +  679888 z^5 - 191809 z^6 + 21871 z^7 + 1152 z^8) +  36 (-1 + y)^4 (-1\\
  &+ z)^4 (64 y^4 + 18 y^2 z (-5 + 2 z) +  4 y^3 (-27 + 34 z) + y (66 - 96 z + 81 z^2 - 44 z^3) +  11 (-2 + 4 z - 3 z^2 + z^3)) H(1, y)\\
  &- 36 (-1 + y)^4 (-1 + z)^4  (4 + 64 y^4 - 8 z + 6 z^2 - 2 z^3 + y^3 (-121 + 188 z) +  3 y^2 (13 - 69 z + 12 z^2) + 2 y (7 + 30 z\\
  &- 18 z^2 + 4 z^3)) H(1, z) -  72 (-1 + y)^4 (-1 + z)^4 (-9 + 32 y^4 + 29 z + 3 z^2 - 55 z^3 + 32 z^4 +  y^3 (-55 + 72 z) + y^2 (3 - 63 z\\
  &+ 36 z^2) +  y (29 - 18 z - 63 z^2 + 72 z^3)) H(2, y)\Bigg\}\Big/\Big(72 (-1 + y)^4 y (-1 + z)^4 z  (-1 + y + z)\Big);\\[4pt]
\mathcal{A}_{5;C_{A}n_{f}}^{(2)} &=
\Bigg\{-1296 y^9 z - 37 (-1 + z)^5 (2 - 2 z + z^2) -  18 y^8 z (-459 + 177 z - 8 z^2 + 2 z^3) -  y^7 (37 + 23464 z - 21866 z^2 + 9620 z^3\\
  &- 4555 z^4 + 1076 z^5) -  y (-1 + z)^4 (444 - 388 z - 1653 z^2 + 3376 z^3 - 3078 z^4 + 1296 z^5) -  y^2 (-1 + z)^3 (1147 + 878 z\\
  &- 10719 z^2 + 18518 z^3 - 12308 z^4 +  3186 z^5) + y^6 (259 + 38809 z - 65000 z^2 + 56576 z^3 - 38287 z^4 +  15203 z^5 - 2376 z^6)\\
  &+ y^3 (-1 + z)^2 (-1665 - 9214 z + 35399 z^2 -  55404 z^3 + 37768 z^4 - 9332 z^5 + 144 z^6) -  y^5 (814 + 40088 z - 106383 z^2\\
  &+ 140272 z^3 - 121660 z^4 + 62292 z^5 -  15203 z^6 + 1076 z^7) + y^4 (1480 + 24504 z - 100897 z^2 + 183975 z^3 -  196954 z^4\\
  &+ 121660 z^5 - 38287 z^6 + 4555 z^7 - 36 z^8) +  108 (-1 + y)^4 y (-1 + z)^4 (10 y^3 - 3 (-1 + z)^2 + 2 y^2 (-11 + 8 z) +  3 y (5 - 7 z\\
  &+ 2 z^2)) H(1, y) - 108 (-1 + y)^4 y (-1 + z)^4  (10 y^3 - 3 (-1 + z)^2 + 2 y^2 (-11 + 8 z) + 3 y (5 - 7 z + 2 z^2))  H(1, z) - 108 (-1\\
  &+ y)^4 (-1 + z)^4 (-1 + y + z)  (10 y^3 + 6 y^2 (-2 + z) + y (3 - 6 z + 6 z^2) + z (3 - 12 z + 10 z^2))  H(2, y)\Bigg\}\Big/\Big(72 (-1 + y)^4 y (-1\\
  &+ z)^4 z (-1 + y + z)\Big);\\[3pt]
\mathcal{A}_{5;C_{F}n_{f}}^{(2)} &=  -\frac{1}{9} ~ \mathcal{A}_{0};
\end{aligned}
\end{equation*}
\begin{equation*}
\begin{aligned}
\mathcal{A}_{6;C_{A}^{2}}^{(2)} &=
\Bigg\{8 (-1 + y) (-1 + z) (y + z) (-1008 y^{12} z + 4 (-1 + z)^4 z^5  (10 - z + 5 z^2) + 4 y^{11} (5 + 622 z - 51 z^2 - 1631 z^3 + 551 z^4)\\
  &-  y (-1 + z)^3 z^4 (200 - 96 z + 1182 z^2 - 2503 z^3 + 536 z^4 +  1008 z^5) + y^{10} (-84 + 1087 z - 12868 z^2 + 50961 z^3 - 50902 z^4\\
  &+  13822 z^5) - y^2 (-1 + z)^2 z^3 (-400 + 3298 z - 10431 z^2 +  24380 z^3 - 31260 z^4 + 13276 z^5 + 204 z^6) +  y^9 (176 - 9291 z\\
  &+ 57608 z^2 - 186114 z^3 + 279822 z^4 - 174455 z^5 +  38302 z^6) + y^8 (-264 + 11687 z - 100176 z^2 + 374314 z^3 -  739622 z^4\\
  &+ 736539 z^5 - 343702 z^6 + 61224 z^7) +  y^7 (276 - 6537 z + 90451 z^2 - 436329 z^3 + 1100987 z^4 - 1543684 z^5 +  1154814 z^6\\
  &- 427250 z^7 + 61224 z^8) +  y^3 z^2 (400 - 4252 z + 32055 z^2 - 129393 z^3 + 304882 z^4 -  436329 z^5 + 374314 z^6 - 186114 z^7\\
  &+ 50961 z^8 - 6524 z^9) +  2 y^6 (-82 + 1035 z - 24270 z^2 + 152441 z^3 - 484929 z^4 + 903470 z^5 -  973380 z^6 + 577407 z^7\\
  &- 171851 z^8 + 19151 z^9) +  y^4 z (200 - 4098 z + 32055 z^2 - 165612 z^3 + 515832 z^4 - 969858 z^5 +  1100987 z^6 - 739622 z^7\\
  &+ 279822 z^8 - 50902 z^9 + 2204 z^{10}) +  y^5 (40 - 696 z + 17427 z^2 - 129393 z^3 + 515832 z^4 - 1240356 z^5 +  1806940 z^6\\
  &- 1543684 z^7 + 736539 z^8 - 174455 z^9 + 13822 z^{10})) -  9 (-1 + y)^4 (-1 + z)^4 (y + z)^6 (496 y^4 + y^3 (-1129 + 1244 z)\\
  &+  3 y^2 (213 - 557 z + 284 z^2) + y (62 + 444 z - 870 z^2 + 416 z^3) +  4 (-17 - 46 z + 234 z^2 - 301 z^3 + 130 z^4)) H(0, y)^2\\
  &-  9 (-1 + y)^4 (-1 + z)^4 (y + z)^6 (-68 + 520 y^4 + 62 z + 639 z^2 -  1129 z^3 + 496 z^4 + 4 y^3 (-301 + 104 z) + y^2 (936 - 870 z\\
  &+ 852 z^2) +  y (-184 + 444 z - 1671 z^2 + 1244 z^3)) H(0, z)^2 +  9 (-1 + y)^4 (-1 + z)^4 (y + z)^6 (-104 + 992 y^4 + 246 z + 861 z^2\\
  &-  1995 z^3 + 992 z^4 + 5 y^3 (-399 + 436 z) +  3 y^2 (287 - 881 z + 440 z^2) + y (246 + 696 z - 2643 z^2 + 2180 z^3))  H(1, z)^2\\
  &+ H(1, y) (-48 (-1 + y)^3 (-1 + z)^4 (y + z)^6  (-98 + 160 y^5 + 247 z - 279 z^2 + 96 z^3 + y^4 (-697 + 232 z) +  y^3 (1221 - 871 z\\
  &+ 318 z^2) + y (503 - 856 z + 768 z^2 - 192 z^3) +  3 y^2 (-363 + 416 z - 270 z^2 + 32 z^3)) + 144 (-1 + y)^4 (-1 + z)^4  (y\\
  &+ z)^6 (84 y^4 + y^3 (-193 + 76 z) + 3 y^2 (51 - 39 z + 16 z^2) +  2 (-2 + 4 z - 3 z^2 + z^3) - 2 y (20 - 12 z + 3 z^2\\
  &+ 4 z^3))  H(1, z)) - 12 (672 y^{15} z - (-1 + z)^5 z^6 (-306 + 1094 z - 1417 z^2 +  640 z^3) - 8 y^{14} (80 + 211 z - 246 z^2 - 284 z^3\\
  &+ 71 z^4) -  3 y^{13} (-1539 + 3624 z - 2740 z^2 + 6256 z^3 - 6117 z^4 + 1188 z^5) +  y^{12} (-14579 + 71063 z - 127704 z^2 + 152312 z^3\\
  &- 144029 z^4 +  67593 z^5 - 10032 z^6) + y (-1 + z)^4 z^5 (-1644 + 10824 z -  23462 z^2 + 24135 z^3 - 10904 z^4 + 1000 z^5 + 672 z^6)\\
  &+  6 y^2 (-1 + z)^3 z^4 (947 - 7217 z + 22516 z^2 - 36737 z^3 + 33102 z^4 -  15206 z^5 + 2354 z^6 + 328 z^7) -  2 y^{11} (-13173 + 94377 z\\
  &- 256362 z^2 + 378172 z^3 - 374314 z^4 +  238281 z^5 - 76309 z^6 + 8656 z^7) -  2 y^{10} (14920 - 147049 z + 552045 z^2 - 1083308 z^3\\
  &+ 1307983 z^4 -  1034540 z^5 + 496204 z^6 - 121179 z^7 + 10892 z^8) +  y^9 (21725 - 293156 z + 1483434 z^2 - 3858986 z^3 + 5967465 z^4\\
  &-  5942620 z^5 + 3826296 z^6 - 1476766 z^7 + 296968 z^8 - 23016 z^9) +  2 y^3 (-1 + z)^2 z^3 (-3384 + 44062 z - 198605 z^2 + 462415 z^3\\
  &-  632873 z^4 + 523576 z^5 - 249468 z^6 + 60796 z^7 - 7112 z^8 +  1136 z^9) + y^8 (-9947 + 189503 z - 1308468 z^2 + 4503474 z^3\\
  &-  8996757 z^4 + 11383863 z^5 - 9480686 z^6 + 5112368 z^7 - 1673910 z^8 +  296968 z^9 - 21784 z^{10}) - 2 y^7 (-1312 + 38311 z\\
  &- 380649 z^2 +  1756308 z^3 - 4551510 z^4 + 7339866 z^5 - 7783882 z^6 + 5514200 z^7 -  2556184 z^8 + 738383 z^9 - 121179 z^{10}\\
  &+ 8656 z^{11}) +  y^5 z (-1644 + 60348 z - 580226 z^2 + 2711300 z^3 - 7384344 z^4 +  12777180 z^5 - 14679732 z^6 + 11383863 z^7\\
  &- 5942620 z^8 +  2069080 z^9 - 476562 z^{10} + 67593 z^{11} - 3564 z^{12}) +  y^4 z^2 (-5682 + 101660 z - 722026 z^2 + 2711300 z^3\\
  &- 6165396 z^4 +  9103020 z^5 - 8996757 z^6 + 5967465 z^7 - 2615966 z^8 + 748628 z^9 -  144029 z^{10} + 18351 z^{11} - 568 z^{12})\\
  &-  2 y^6 (153 - 8700 z + 141024 z^2 - 903687 z^3 + 3082698 z^4 -  6388590 z^5 + 8608206 z^6 - 7783882 z^7 + 4740343 z^8 - 1913148 z^9\\
  &+  496204 z^{10} - 76309 z^{11} + 5016 z^{12})) H(2, y) +  9 (-1 + y)^4 (-1 + z)^4 (y + z)^6 (-104 + 992 y^4 + 246 z + 861 z^2 -  1995 z^3\\
  &+ 992 z^4 + 5 y^3 (-399 + 436 z) +  3 y^2 (287 - 881 z + 440 z^2) + y (246 + 696 z - 2643 z^2 + 2180 z^3))  H(2, y)^2 + H(0, y) (-12 (-1\\
  &+ y)^3 (y + z)^6  (672 y^6 z + 49 (-1 + z)^5 (2 - 2 z + z^2) +  24 y^5 z (-93 + 74 z - 12 z^2 + 3 z^3) - y (-1 + z)^4  (-306 + 1096 z\\
  &- 1680 z^2 + 1205 z^3) + y^3 (-1 + z)^2  (304 - 3453 z + 4062 z^2 - 2983 z^3 + 984 z^4) +  y^2 (-1 + z)^3 (415 - 2620 z + 4200 z^2\\
  &- 3661 z^3 + 1156 z^4) +  y^4 (-97 + 3488 z - 6112 z^2 + 3784 z^3 - 1283 z^4 + 220 z^5)) -  432 (-1 + y)^4 (-1 + z)^4 (y + z)^6 (2\\
  &+ y^2 (3 - 9 z) - 4 z + 3 z^2 -  z^3 + y^3 (-1 + 4 z) + y (-4 + 12 z - 9 z^2 + 4 z^3)) H(0, z) +  144 (-1 + y)^4 (-1 + z)^4 (y\\
  &+ z)^6 (22 y^3 + 4 y^4 -  6 y^2 (10 - 9 z + 4 z^2) + y (40 - 60 z + 39 z^2 - 12 z^3) +  3 (-2 + 4 z - 3 z^2 + z^3)) H(1, z) + 144 (-1\\
  &+ y)^4 (-1 + z)^4  (y + z)^6 (2 + y^2 (3 - 9 z) - 4 z + 3 z^2 - z^3 + y^3 (-1 + 4 z) +  y (-4 + 12 z - 9 z^2 + 4 z^3)) H(2, y))\\
  &+  H(0, z) (-12 (-1 + z)^3 (y + z)^6 (-((-1 + z)^2 (98 - 110 z + 97 z^2)) +  y^7 (49 - 1205 z + 1156 z^2) + y^6 (-343 + 6500 z\\
  &- 7129 z^2 +  984 z^3) + y^5 (1078 - 15046 z + 18651 z^2 - 4951 z^3 + 220 z^4) +  y^3 (2205 - 15725 z + 24536 z^2 - 14560 z^3 + 3784 z^4\\
  &- 288 z^5) +  y^4 (-1960 + 19590 z - 27359 z^2 + 11012 z^3 - 1283 z^4 + 72 z^5) +  y^2 (-1519 + 7900 z - 13305 z^2 + 11272 z^3\\
  &- 6112 z^4 + 1776 z^5) +  y (588 - 2320 z + 3865 z^2 - 4061 z^3 + 3488 z^4 - 2232 z^5 +  672 z^6)) + 144 (-1 + y)^4 (-1 + z)^4 (y\\
  &+ z)^6  (88 y^4 + y^3 (-201 + 124 z) + 3 y^2 (49 - 57 z + 20 z^2) +  2 (-2 + 4 z - 3 z^2 + z^3) - 2 y (15 - 18 z + 6 z^2\\
  &+ 4 z^3)) H(1, y) +  144 (-1 + y)^4 (-1 + z)^4 (y + z)^6 (2 + y^2 (3 - 9 z) - 4 z + 3 z^2 -  z^3 + y^3 (-1 + 4 z) + y (-4 + 12 z - 9 z^2\\
  &+ 4 z^3)) H(1, z) -  288 (-1 + y)^4 (-1 + z)^4 (y + z)^6 (44 y^4 + 34 y^3 (-3 + 2 z) -  (-1 + z)^2 (-1 + 14 z + 2 z^2) + 3 y^2 (26\\
  &- 35 z + 14 z^2) -  y (21 - 48 z + 33 z^2 + 4 z^3)) H(2, y)) +  H(1, z) (-12 (672 y^{15} z - (-1 + z)^5 z^6 (86 - 134 z + 91 z^2)\\
  &-  8 y^{14} (80 + 211 z - 246 z^2 - 284 z^3 + 71 z^4) -  3 y^{13} (-1411 + 2984 z - 1460 z^2 + 4976 z^3 - 5477 z^4 + 1060 z^5)\\
  &+  y^{12} (-11927 + 54659 z - 85020 z^2 + 92588 z^3 - 96701 z^4 +  47481 z^5 - 6456 z^6) + y (-1 + z)^4 z^5 (708 - 1452 z + 1870 z^2\\
  &-  2253 z^3 + 2456 z^4 - 1560 z^5 + 672 z^6) -  2 y^{11} (-9295 + 64627 z - 158940 z^2 + 201252 z^3 - 181596 z^4 +  112239 z^5 - 30455 z^6\\
\end{aligned}
\end{equation*}
\begin{equation*}
\begin{aligned}
  &+ 1496 z^7) + 2 y^2 (-1 + z)^3 z^4  (-99 + 1107 z - 3248 z^2 + 8087 z^3 - 15336 z^4 + 17678 z^5 -  10838 z^6 + 2904 z^7) + 2 y^{10} (-8632\\
  &+ 86311 z - 304059 z^2 +  516984 z^3 - 511401 z^4 + 325416 z^5 - 105440 z^6 - 739 z^7 +  5592 z^8) - 2 y^3 (-1 + z)^2 z^3 (-536 - 852 z\\
  &+ 7455 z^2 - 903 z^3 -  46165 z^4 + 106706 z^5 - 114712 z^6 + 62368 z^7 - 14048 z^8 +  144 z^9) + 3 y^9 (3127 - 46340 z + 231586 z^2\\
  &- 539722 z^3 +  671983 z^4 - 471956 z^5 + 148964 z^6 + 36522 z^7 - 42364 z^8 +  8648 z^9) + y^8 (-2527 + 64339 z - 489624 z^2\\
  &+ 1601594 z^3 -  2644553 z^4 + 2271207 z^5 - 738570 z^6 - 439680 z^7 + 556890 z^8 -  212212 z^9 + 27760 z^{10}) + 2 y^7 (34 - 6481 z\\
  &+ 102519 z^2 -  502456 z^3 + 1134338 z^4 - 1263358 z^5 + 486190 z^6 + 442468 z^7 -  684158 z^8 + 375311 z^9 - 94127 z^{10} + 8712 z^{11})\\
  &+  y^4 z^2 (198 + 2720 z - 54406 z^2 + 215036 z^3 - 276280 z^4 -  222736 z^5 + 1183523 z^6 - 1711791 z^7 + 1308226 z^8 - 546976 z^9\\
  &+  107843 z^{10} - 5429 z^{11} + 72 z^{12}) +  y^5 z (708 + 2880 z - 75318 z^2 + 390360 z^3 - 862488 z^4 +  683692 z^5 + 668588 z^6\\
  &- 2089833 z^7 + 2146884 z^8 - 1135548 z^9 +  302226 z^{10} - 32683 z^{11} + 1204 z^{12}) +  2 y^6 (43 - 562 z - 22170 z^2 + 191307 z^3\\
  &- 617668 z^4 + 934096 z^5 -  486604 z^6 - 520444 z^7 + 1071647 z^8 - 795420 z^9 + 294048 z^{10} -  50805 z^{11} + 3204 z^{12})) - 144 (-1\\
  &+ y)^4 (-1 + z)^4 (y + z)^6  (100 y^4 + 2 y^3 (-127 + 60 z) + 6 y^2 (38 - 41 z + 24 z^2) +  z (-38 + 153 z - 203 z^2 + 88 z^3) +  y (-74\\
  &+ 144 z - 231 z^2 + 132 z^3)) H(2, y) -  144 (-1 + y)^4 (-1 + z)^4 (y + z)^6 (-4 + 36 y^4 - 4 z + 51 z^2 -  79 z^3 + 36 z^4 + y^3 (-79\\
  &+ 52 z) + y^2 (51 - 75 z + 48 z^2) +  y (-4 + 24 z - 75 z^2 + 52 z^3)) H(3, y)) +  18 (-1 + y)^4 (-1 + z)^4 (y + z)^6 (496 y^4\\
  &+ y^3 (-1129 + 1244 z) +  3 y^2 (213 - 557 z + 284 z^2) + y (62 + 444 z - 870 z^2 + 416 z^3) +  4 (-17 - 46 z + 234 z^2 - 301 z^3\\
  &+ 130 z^4)) H(0, 0, y) +  18 (-1 + y)^4 (-1 + z)^4 (y + z)^6 (-68 + 520 y^4 + 62 z + 639 z^2 -  1129 z^3 + 496 z^4 + 4 y^3 (-301\\
  &+ 104 z) + y^2 (936 - 870 z + 852 z^2) +  y (-184 + 444 z - 1671 z^2 + 1244 z^3)) H(0, 0, z) -  288 (-1 + y)^4 (-1 + z)^4 (y + z)^6 (2\\
  &+ 10 y^4 - 4 z + 3 z^2 - z^3 +  7 y^3 (-3 + 4 z) + 6 y^2 (2 - 6 z + z^2) +  y (-3 + 18 z - 12 z^2 + 4 z^3)) H(0, 1, z) +  144 (-1\\
  &+ y)^4 (-1 + z)^4 (y + z)^6 (-8 + 36 y^4 + 2 z + 27 z^2 - 37 z^3 +  16 z^4 + 11 y^3 (-7 + 4 z) + y^2 (45 - 51 z + 36 z^2) -  y (-4\\
  &+ 12 z + 3 z^2 + 4 z^3)) H(0, 2, y) -  144 (-1 + y)^4 y (-1 + z)^4 (y + z)^6 (4 y^3 + y^2 (19 + 12 z) +  4 (7 - 6 z + 3 z^2) - 3 y (17\\
  &- 9 z + 8 z^2)) H(1, 0, y) -  144 (-1 + y)^4 (-1 + z)^4 (-1 + y + z) (y + z)^6  (88 y^3 + 13 y^2 (-9 + 4 z) - 2 (2 - 2 z + z^2) + y (42\\
  &- 38 z + 8 z^2))  H(1, 0, z) + 288 (-1 + y)^4 y (-1 + z)^4 (y + z)^6  (5 + 2 y^3 + 6 z - 3 z^2 + 4 y^2 (-1 + 6 z) + 3 y (-1 - 9 z\\
  &+ 2 z^2))  H(1, 1, y) - 18 (-1 + y)^4 (-1 + z)^4 (y + z)^6  (-72 + 1792 y^4 + 182 z + 909 z^2 - 2011 z^3 + 992 z^4 +  y^3 (-4043\\
  &+ 3204 z) + 3 y^2 (911 - 1505 z + 664 z^2) +  y (-410 + 1560 z - 3123 z^2 + 2244 z^3)) H(1, 1, z) +  144 (-1 + y)^4 (-1 + z)^4 (y\\
  &+ z)^6 (84 y^4 + y^3 (-193 + 76 z) +  3 y^2 (51 - 39 z + 16 z^2) + 2 (-2 + 4 z - 3 z^2 + z^3) -  2 y (20 - 12 z + 3 z^2\\
  &+ 4 z^3)) H(1, 2, y) +  144 (-1 + y)^4 (-1 + z)^4 (y + z)^6 (-4 + 4 y^4 + 46 z - 159 z^2 +  205 z^3 - 88 z^4 + y^3 (21 + 4 z) + y^2 (-57\\
  &+ 75 z - 84 z^2) +  y (36 - 108 z + 219 z^2 - 140 z^3)) H(2, 0, y) -  144 (-1 + y)^4 (-1 + z)^4 (y + z)^6 (20 y^4 + y^3 (-49 + 60 z)\\
  &+  3 y^2 (11 - 35 z + 24 z^2) + 4 z (-10 + 27 z - 25 z^2 + 8 z^3) +  4 y (-1 + 18 z - 30 z^2 + 12 z^3)) H(2, 1, y) -  18 (-1 + y)^4 (-1\\
  &+ z)^4 (y + z)^6 (-104 + 1536 y^4 - 26 z + 1821 z^2 -  3227 z^3 + 1536 z^4 + y^3 (-3227 + 2756 z) +  3 y^2 (607 - 1217 z + 632 z^2)\\
  &+ y (-26 + 1272 z - 3651 z^2 + 2756 z^3))  H(2, 2, y) - 144 (-1 + y)^4 (-1 + z)^4 (y + z)^6  (-4 + 36 y^4 - 4 z + 51 z^2 - 79 z^3\\
  &+ 36 z^4 + y^3 (-79 + 52 z) +  y^2 (51 - 75 z + 48 z^2) + y (-4 + 24 z - 75 z^2 + 52 z^3)) H(3, 2, y)\Bigg\}\Big/ \Big(576 (-1 + y)^4 y (-1\\
  &+ z)^4 z (-1 + y + z) (y + z)^6\Big);\\[5pt]
\mathcal{A}_{6;C_{F}^{2}}^{(2)} &=
\Bigg\{4 (-1 + y) (-1 + z) (y + z) (288 y^{13} z - 3 (-1 + z)^5 z^5  (90 - 90 z + 43 z^2) + y^{12} (-129 - 1327 z + 1720 z^2 + 488 z^3\\
  &-  176 z^4) + y^{11} (915 + 342 z - 6679 z^2 + 3734 z^3 + 1168 z^4 -  56 z^5) + y (-1 + z)^4 z^4 (1350 - 4968 z + 5525 z^2 - 2086 z^3\\
  &-  175 z^4 + 288 z^5) + y^2 (-1 + z)^3 z^3 (-2700 + 15672 z - 38367 z^2 +  41471 z^3 - 16563 z^4 - 1519 z^5 + 1720 z^6) +  y^{10} (-2910\\
  &+ 11667 z - 6846 z^2 - 9291 z^3 + 9388 z^4 - 7000 z^5 +  3264 z^6) + y^9 (5340 - 38596 z + 84883 z^2 - 66628 z^3 + 1089 z^4 +  39320 z^5\\
  &- 36152 z^6 + 10744 z^7) + y^3 (-1 + z)^2 z^2  (2700 - 29448 z + 105101 z^2 - 188876 z^3 + 176750 z^4 - 72056 z^5 -  359 z^6 + 4710 z^7\\
  &+ 488 z^8) + y^8 (-6045 + 62541 z - 210950 z^2 +  320503 z^3 - 225267 z^4 + 21438 z^5 + 93980 z^6 - 69672 z^7 +  15200 z^8) + y^7 (4179\\
  &- 59394 z + 271749 z^2 - 614432 z^3 +  753510 z^4 - 469026 z^5 + 49426 z^6 + 123492 z^7 - 69672 z^8 +  10744 z^9) + y^6 (-1620 + 33497 z\\
  &- 206288 z^2 + 659603 z^3 -  1187244 z^4 + 1201898 z^5 - 610940 z^6 + 49426 z^7 + 93980 z^8 -  36152 z^9 + 3264 z^{10}) + y^4 z (1350\\
  &- 23772 z + 166697 z^2 -  565574 z^3 + 1068831 z^4 - 1187244 z^5 + 753510 z^6 - 225267 z^7 +  1089 z^8 + 9388 z^9 + 1168 z^{10}\\
  &- 176 z^{11}) +  y^5 (270 - 10368 z + 93483 z^2 - 428526 z^3 + 1068831 z^4 - 1510552 z^5 +  1201898 z^6 - 469026 z^7 + 21438 z^8\\
  &+ 39320 z^9 - 7000 z^{10} -  56 z^{11})) - 3 (-1 + y)^4 (-1 + z)^4 (-1 + y + z)^2 (y + z)^6  (46 + 704 y^3 - 112 z + 254 z^2 - 192 z^3\\
  &+ y^2 (-717 + 404 z) -  2 y (-21 + 11 z + 88 z^2)) H(0, y)^2 + 3 (-1 + y)^4 (-1 + z)^4  (-1 + y + z)^2 (y + z)^6 (-46 + 192 y^3 - 42 z\\
  &+ 717 z^2 - 704 z^3 +  2 y^2 (-127 + 88 z) + y (112 + 22 z - 404 z^2)) H(0, z)^2 +  9 (-1 + y)^4 (-1 + z)^4 (-1 + y + z)^2 (y + z)^6  (4\\
  &+ 56 y^3 + 34 z - 101 z^2 + 56 z^3 + y^2 (-101 + 84 z) +  y (34 - 76 z + 84 z^2)) H(1, z)^2 +  H(1, y) (-8 (-1 + y)^2 (-1 + z)^4 (-1 + y\\
  &+ z) (y + z)^6  (150 + 136 y^6 - 406 z + 411 z^2 - 155 z^3 + y^5 (-761 + 460 z) +  y^4 (1819 - 2033 z + 456 z^2) + y^2 (1859 - 3607 z\\
  &+ 2367 z^2 -  465 z^3) + y^3 (-2401 + 3730 z - 1665 z^2 + 164 z^3) +  y (-802 + 1856 z - 1569 z^2 + 474 z^3)) - 96 (-1 + y)^4 y (-1\\
  &+ z)^4  (-1 + y + z)^2 (y + z)^6 (3 + 14 y^2 - 3 z + 3 y (-5 + 2 z)) H(1, z)) +  H(0, y) (8 (-1 + y)^2 y (-1 + y + z)^2 (y\\
\end{aligned}
\end{equation*}
\begin{equation*}
\begin{aligned}
  &+ z)^6  (144 y^5 z + 12 (-1 + z)^3 z (7 - 6 z + z^2) +  4 y^4 z (-153 + 111 z - 40 z^2 + 10 z^3) - 3 y (-1 + z)^2  (-4 - 162 z + 220 z^2\\
  &- 108 z^3 + 11 z^4) +  y^3 (12 + 1112 z - 1955 z^2 + 1610 z^3 - 799 z^4 + 164 z^5) +  2 y^2 (-12 - 511 z + 1395 z^2 - 1516 z^3 + 831 z^4\\
  &- 189 z^5 +  2 z^6)) + 144 (-1 + y)^4 y (-1 + z)^4 (-1 + y + z)^2 (y + z)^6  (4 + 12 y^2 - 4 z + y (-15 + 8 z)) H(1, z)) +  8 (-1 + y\\
  &+ z) (144 y^{15} z + (-1 + z)^5 z^6 (-132 + 340 z - 341 z^2 +  136 z^3) + 4 y^{14} (34 - 397 z + 567 z^2 - 176 z^3 + 44 z^4) +  y^{13} (-1021\\
  &+ 8884 z - 22137 z^2 + 20586 z^3 - 8560 z^4 + 1816 z^5) +  y (-1 + z)^4 z^5 (468 - 3576 z + 7726 z^2 - 7779 z^3 + 3972 z^4 -  1012 z^5\\
  &+ 144 z^6) + y^{12} (3405 - 30315 z + 100809 z^2 - 152893 z^3 +  116054 z^4 - 46660 z^5 + 8448 z^6) + 3 y^2 (-1 + z)^3 z^4  (-748 + 5012 z\\
  &- 15992 z^2 + 27958 z^3 - 27830 z^4 + 16002 z^5 -  5111 z^6 + 756 z^7) + y^{11} (-6602 + 66866 z - 275775 z^2 + 575010 z^3 -  652809 z^4\\
  &+ 420138 z^5 - 149708 z^6 + 23168 z^7) +  y^{10} (8150 - 98054 z + 493695 z^2 - 1321709 z^3 + 2031803 z^4 -  1859997 z^5 + 1019924 z^6\\
  &- 313380 z^7 + 41296 z^8) -  y^3 (-1 + z)^2 z^3 (-1416 + 28792 z - 145538 z^2 + 374050 z^3 -  574998 z^4 + 551544 z^5 - 328166 z^6\\
  &+ 113833 z^7 - 19178 z^8 +  704 z^9) + y^9 (-6561 + 96216 z - 598074 z^2 + 2006252 z^3 -  3969623 z^4 + 4828260 z^5 - 3670950 z^6\\
  &+ 1713820 z^7 - 448924 z^8 +  49872 z^9) + y^8 (3361 - 62011 z + 494076 z^2 - 2075590 z^3 +  5157019 z^4 - 8018097 z^5 + 8000348 z^6\\
  &- 5125514 z^7 + 2032884 z^8 -  448924 z^9 + 41296 z^{10}) + 2 y^7 (-500 + 12419 z - 137577 z^2 +  734318 z^3 - 2271456 z^4 + 4429489 z^5\\
  &- 5635698 z^6 + 4719742 z^7 -  2562757 z^8 + 856910 z^9 - 156690 z^{10} + 11584 z^{11}) +  y^4 z^2 (2244 - 31624 z + 242798 z^2\\
  &- 1036852 z^3 + 2692286 z^4 -  4542912 z^5 + 5157019 z^6 - 3969623 z^7 + 2031803 z^8 - 652809 z^9 +  116054 z^{10} - 8560 z^{11} + 176 z^{12})\\
  &+  y^5 z (468 - 21768 z + 204538 z^2 - 1036852 z^3 + 3244920 z^4 -  6575600 z^5 + 8858978 z^6 - 8018097 z^7 + 4828260 z^8 - 1859997 z^9\\
  &+  420138 z^{10} - 46660 z^{11} + 1816 z^{12}) +  2 y^6 (66 - 2724 z + 49908 z^2 - 346959 z^3 + 1346143 z^4 - 3287800 z^5 +  5273177 z^6\\
  &- 5635698 z^7 + 4000174 z^8 - 1835475 z^9 + 509962 z^{10} -  74854 z^{11} + 4224 z^{12})) H(2, y) + 9 (-1 + y)^4 (-1 + z)^4  (-1 + y + z)^2 (y\\
  &+ z)^6 (4 + 56 y^3 + 34 z - 101 z^2 + 56 z^3 +  y^2 (-101 + 84 z) + y (34 - 76 z + 84 z^2)) H(2, y)^2 +  H(0, z) (8 (-1 + z)^2 z (-1 + y\\
  &+ z)^2 (y + z)^6  (12 (-1 + z)^2 z + y^6 (12 - 33 z + 4 z^2) +  2 y^5 (-54 + 195 z - 189 z^2 + 82 z^3) +  y^4 (336 - 1341 z + 1662 z^2\\
  &- 799 z^3 + 40 z^4) -  2 y^3 (240 - 1065 z + 1516 z^2 - 805 z^3 + 80 z^4) +  y^2 (324 - 1620 z + 2790 z^2 - 1955 z^3 + 444 z^4)\\
  &+  2 y (-42 + 231 z - 511 z^2 + 556 z^3 - 306 z^4 + 72 z^5)) +  48 (-1 + y)^4 y (-1 + z)^4 (-1 + y + z)^2 (y + z)^6  (6 + 8 y^2 - 6 z\\
  &+ 3 y (-5 + 4 z)) H(1, y) - 48 (-1 + y)^4 (-1 + z)^4  (-1 + y + z)^2 (y + z)^6 (8 y^3 + 3 y^2 (-5 + 4 z) +  6 y (1 + z - 4 z^2) - 3 z (4\\
  &- 15 z + 12 z^2)) H(2, y)) +  H(1, z) (8 (-1 + y + z) (144 y^{15} z + 6 (-1 + z)^5 z^6 (3 - 2 z + 2 z^2) +  4 y^{14} (34 - 397 z + 567 z^2\\
  &- 176 z^3 + 44 z^4) +  y^{13} (-866 + 8100 z - 20569 z^2 + 19018 z^3 - 7767 z^4 + 1652 z^5) +  2 y (-1 + z)^4 z^5 (-216 + 221 z + 222 z^2\\
  &- 456 z^3 + 370 z^4 -  234 z^5 + 72 z^6) + y^{12} (2374 - 23858 z + 83917 z^2 - 129245 z^3 +  97321 z^4 - 38669 z^5 + 7008 z^6) + y^2 (-1\\
  &+ z)^3 z^4  (6 - 780 z - 2588 z^2 + 13184 z^3 - 18942 z^4 + 14294 z^5 - 6479 z^6 +  1452 z^7) + y^{11} (-3622 + 43744 z - 199491 z^2\\
  &+ 436042 z^3 -  501379 z^4 + 321120 z^5 - 113638 z^6 + 17512 z^7) +  y^{10} (3290 - 50918 z + 300333 z^2 - 878721 z^3 + 1407853 z^4\\
  &-  1304275 z^5 + 713376 z^6 - 217490 z^7 + 28280 z^8) -  y^3 (-1 + z)^2 z^3 (1584 - 2338 z - 14788 z^2 + 72720 z^3 -  149820 z^4\\
  &+ 171732 z^5 - 117066 z^6 + 45939 z^7 - 8478 z^8 +  160 z^9) + 2 y^9 (-863 + 18146 z - 145902 z^2 + 567991 z^3 -  1216836 z^4\\
  &+ 1534256 z^5 - 1178239 z^6 + 547769 z^7 - 141466 z^8 +  15288 z^9) + 2 y^8 (213 - 6663 z + 89049 z^2 - 480126 z^3 +  1360821 z^4\\
  &- 2265678 z^5 + 2330148 z^6 - 1502619 z^7 + 590049 z^8 -  126886 z^9 + 11116 z^{10}) + 2 y^7 (3 + 76 z - 31201 z^2 + 259264 z^3\\
  &-  985699 z^4 + 2156109 z^5 - 2918981 z^6 + 2511740 z^7 - 1366047 z^8 +  447153 z^9 - 77957 z^{10} + 5324 z^{11}) +  y^4 z^2 (-6 + 5026 z\\
  &+ 3058 z^2 - 165362 z^3 + 688552 z^4 -  1444642 z^5 + 1850424 z^6 - 1531142 z^7 + 816755 z^8 - 264717 z^9 +  44437 z^{10} - 2423 z^{11}\\
  &+ 40 z^{12}) +  y^5 z (-432 - 66 z + 18020 z^2 - 196646 z^3 + 926280 z^4 -  2346594 z^5 + 3574134 z^6 - 3448086 z^7 + 2127944 z^8\\
  &- 811205 z^9 +  172740 z^{10} - 16485 z^{11} + 540 z^{12}) +  2 y^6 (-9 + 845 z + 4861 z^2 - 81055 z^3 + 437656 z^4 - 1299435 z^5\\
  &+  2341652 z^6 - 2665679 z^7 + 1940166 z^8 - 884963 z^9 + 236428 z^{10} -  31919 z^{11} + 1596 z^{12})) + 48 (-1 + y)^4 (-1 + z)^4 (-1 + y\\
  &+ z)^2  (y + z)^6 (-4 + 28 y^3 + 18 z - 21 z^2 + 8 z^3 + 4 y^2 (-8 + 5 z) +  2 y (5 - 10 z + 2 z^2)) H(2, y) - 144 (-1 + y)^4 (-1\\
  &+ z)^4  (-1 + y + z)^2 (y + z)^6 (12 y^3 + y^2 (-15 + 8 z) +  y (4 - 8 z + 8 z^2) + z (4 - 15 z + 12 z^2)) H(3, y)) +  6 (-1 + y)^4 (-1\\
  &+ z)^4 (-1 + y + z)^2 (y + z)^6  (46 + 704 y^3 - 112 z + 254 z^2 - 192 z^3 + y^2 (-717 + 404 z) -  2 y (-21 + 11 z + 88 z^2)) H(0, 0, y)\\
  &- 6 (-1 + y)^4 (-1 + z)^4  (-1 + y + z)^2 (y + z)^6 (-46 + 192 y^3 - 42 z + 717 z^2 - 704 z^3 +  2 y^2 (-127 + 88 z) + y (112 + 22 z\\
  &- 404 z^2)) H(0, 0, z) -  96 (-1 + y)^4 (-1 + 4 y) (-1 + z)^4 (-1 + y + z)^2 (y + z)^6  (2 + 2 y^2 + 2 y (-2 + z) - 2 z\\
  &+ z^2) H(0, 1, y) -  144 (-1 + y)^4 y (-1 + z)^4 (-1 + y + z)^2 (y + z)^6  (4 + 12 y^2 - 4 z + y (-15 + 8 z)) H(0, 1, z) +  48 (-1\\
  &+ y)^4 (-1 + z)^4 (-1 + y + z)^2 (y + z)^6  (52 y^3 + y^2 (-81 + 40 z) - 2 (2 - 2 z + z^2) + 4 y (9 - 8 z + 2 z^2))  H(0, 2, y)\\
  &- 144 (-1 + y)^4 y (-1 + z)^4 (-1 + y + z)^2 (y + z)^6  (4 + 12 y^2 - 4 z + y (-15 + 8 z)) H(1, 0, y) -  48 (-1 + y)^4 y (-1 + z)^4 (-1\\
  &+ y + z)^2 (y + z)^6  (6 + 8 y^2 - 6 z + 3 y (-5 + 4 z)) H(1, 0, z) +  48 (-1 + y)^4 (-1 + z)^4 (-1 + y + z)^2 (y + z)^6  (20 y^3\\
  &+ y^2 (-9 + 8 z) + 2 (2 - 2 z + z^2) - 4 y (3 - 2 z + 2 z^2))  H(1, 1, y) + 6 (-1 + y)^4 (-1 + z)^4 (-1 + y + z)^2 (y + z)^6  (56 y^3\\
  &+ y^2 (63 - 156 z) - 18 y (3 - 10 z + 14 z^2) -  3 (4 + 34 z - 101 z^2 + 56 z^3)) H(1, 1, z) -  96 (-1 + y)^4 y (-1 + z)^4 (-1 + y\\
  &+ z)^2 (y + z)^6  (3 + 14 y^2 - 3 z + 3 y (-5 + 2 z)) H(1, 2, y) +  48 (-1 + y)^4 (-1 + z)^4 (-1 + y + z)^2 (y + z)^6  (36 y^3\\
  &+ 3 y^2 (-15 + 8 z) + z (-6 + 15 z - 8 z^2) -  6 y (-2 + z + 2 z^2)) H(2, 0, y) - 48 (-1 + y)^4 (-1 + z)^4  (-1 + y + z)^2 (y\\
  &+ z)^6 (36 y^3 + 16 y (-1 + z)^2 +  4 (-1 + z)^2 (-1 + 4 z) + y^2 (-47 + 32 z)) H(2, 1, y) +  6 (-1 + y)^4 (-1 + z)^4 (-1 + y + z)^2 (y\\
\end{aligned}
\end{equation*}
\begin{equation*}
\begin{aligned}
  &+ z)^6  (-76 + 184 y^3 + 170 z - 241 z^2 + 184 z^3 + y^2 (-241 + 36 z) +  2 y (85 - 94 z + 18 z^2)) H(2, 2, y) - 144 (-1 + y)^4 (-1\\
  &+ z)^4  (-1 + y + z)^2 (y + z)^6 (12 y^3 + y^2 (-15 + 8 z) + y (4 - 8 z + 8 z^2) +  z (4 - 15 z + 12 z^2)) H(3, 2, y)\Bigg\}\Big/\Big(48 (-1\\
  &+ y)^4 y (-1 + z)^4 z  (-1 + y + z)^2 (y + z)^6\Big);\\[5pt]
\mathcal{A}_{6;n_{f}^{2}}^{(2)} &= \frac{1}{18} ~ \mathcal{A}_{0} ;\\[5pt]
\mathcal{A}_{6;C_{A}C_{F}}^{(2)} &=
\Bigg\{-2 (-1 + y) (-1 + z) (y + z) (1152 y^{13} z - 9 (-1 + z)^5 z^5  (90 - 94 z + 41 z^2) + 3 y^{12} (-123 - 2483 z + 4229 z^2\\
  &- 1283 z^3 +  428 z^4) + 3 y (-1 + z)^4 z^4 (1350 - 5784 z + 5943 z^2 - 1088 z^3 -  947 z^4 + 384 z^5) + 3 y^{11} (897 + 5004 z\\
  &- 25436 z^2 + 30556 z^3 -  16597 z^4 + 4808 z^5) + 3 y^2 (-1 + z)^3 z^3  (-2700 + 13740 z - 36361 z^2 + 35633 z^3 - 2149 z^4\\
  &- 12749 z^5 +  4229 z^6) + y^{10} (-8730 + 9231 z + 146355 z^2 - 412685 z^3 +  451131 z^4 - 253782 z^5 + 61568 z^6) +  2 y^9 (8100\\
  &- 47868 z - 594 z^2 + 355100 z^3 - 742645 z^4 + 700596 z^5 -  341189 z^6 + 68500 z^7) - y^3 (-1 + z)^2 z^2  (-8100 + 99768 z\\
  &- 310125 z^2 + 455122 z^3 - 249510 z^4 - 144438 z^5 +  240896 z^6 - 83970 z^7 + 3849 z^8) +  y^8 (-18405 + 190647 z - 410874 z^2\\
  &- 280262 z^3 + 2163621 z^4 -  3406317 z^5 + 2683478 z^6 - 1092120 z^7 + 177144 z^8) +  y^7 (12699 - 194892 z + 695613 z^2\\
  &- 809704 z^3 - 883105 z^4 +  3818004 z^5 - 4987759 z^6 + 3306568 z^7 - 1092120 z^8 + 137000 z^9) +  y^6 (-4896 + 111537 z\\
  &- 565908 z^2 + 1469879 z^3 - 1595836 z^4 -  916967 z^5 + 4424978 z^6 - 4987759 z^7 + 2683478 z^8 - 682378 z^9 +  61568 z^{10})\\
  &+ y^4 z (4050 - 65520 z + 517761 z^2 - 1698158 z^3 +  2639853 z^4 - 1595836 z^5 - 883105 z^6 + 2163621 z^7 - 1485290 z^8\\
  &+  451131 z^9 - 49791 z^{10} + 1284 z^{11}) +  y^5 (810 - 33552 z + 257043 z^2 - 1175140 z^3 + 2639853 z^4 -  2346720 z^5 - 916967 z^6\\
  &+ 3818004 z^7 - 3406317 z^8 + 1401192 z^9 -  253782 z^{10} + 14424 z^{11})) + 9 (-1 + y)^4 (-1 + z)^4 (-1 + y + z)  (y + z)^6 (-27\\
  &+ 612 y^4 + 7 z - 18 z^2 + 130 z^3 - 92 z^4 +  6 y^3 (-219 + 206 z) + 6 y^2 (127 - 251 z + 78 z^2) -  3 y (11 - 102 z + 26 z^2\\
  &+ 52 z^3)) H(0, y)^2 -  9 (-1 + y)^4 (-1 + z)^4 (-1 + y + z) (y + z)^6  (27 + 92 y^4 + 33 z - 762 z^2 + 1314 z^3 - 612 z^4\\
  &+ 26 y^3 (-5 + 6 z) +  y^2 (18 + 78 z - 468 z^2) - y (7 + 306 z - 1506 z^2 + 1236 z^3))  H(0, z)^2 - 9 (-1 + y)^4 (-1 + z)^4 (-1\\
  &+ y + z) (y + z)^6  (-58 + 80 y^4 + 136 z + 27 z^2 - 185 z^3 + 80 z^4 + 5 y^3 (-37 + 68 z) +  3 y^2 (9 - 139 z + 112 z^2) + y (136\\
  &- 12 z - 417 z^2 + 340 z^3))  H(1, z)^2 + H(0, y) (-12 (-1 + y)^2 y (-1 + y + z) (y + z)^6  (192 y^6 z + 4 y^5 z (-255 + 207 z\\
  &- 64 z^2 + 16 z^3) -  6 (-1 + z)^4 (1 + 18 z - 47 z^2 + 32 z^3) +  3 y (-1 + z)^3 (3 + 6 z + 242 z^2 - 344 z^3 + 134 z^4)\\
  &-  y^2 (-1 + z)^2 (-63 - 1120 z + 692 z^2 + 644 z^3 - 824 z^4 + 208 z^5) +  y^4 (27 + 2224 z - 3883 z^2 + 2826 z^3 - 1250 z^4\\
  &+ 248 z^5) -  y^3 (75 + 2315 z - 5945 z^2 + 5649 z^3 - 2530 z^4 + 376 z^5 +  60 z^6)) - 72 (-1 + y)^4 (-1 + z)^4 (-1 + y + z) (y\\
  &+ z)^6  (56 y^4 - 3 y^2 (-7 + 9 z) + y^3 (-91 + 60 z) +  y (20 - 36 z + 27 z^2 - 12 z^3) + 3 (-2 + 4 z - 3 z^2 + z^3))  H(1, z))\\
  &+ H(1, y) (12 (-1 + y)^3 (-1 + z)^4 (-1 + y + z) (y + z)^6  (-406 + 388 y^5 + 1020 z - 1029 z^2 + 383 z^3 + y^4 (-1877 + 1092 z)\\
  &+  y^3 (3592 - 3873 z + 1116 z^2) + y (1814 - 3714 z + 3018 z^2 -  859 z^3) + y^2 (-3511 + 5475 z - 3111 z^2 + 476 z^3)) +  72 (-1\\
  &+ y)^4 (-1 + z)^4 (-1 + y + z) (y + z)^6  (2 + 24 y^4 - 4 z + 3 z^2 - z^3 + 4 y^3 (-7 + 10 z) -  6 y^2 (2 + 3 z + 2 z^2) + y (14\\
  &- 3 z^2 + 4 z^3)) H(1, z)) -  6 (-1 + y + z) (384 y^{15} z + (-1 + z)^5 z^6 (-726 + 1882 z - 2105 z^2 +  776 z^3) + 8 y^{14} (97\\
  &- 739 z + 1077 z^2 - 452 z^3 + 113 z^4) +  y^{13} (-5985 + 40168 z - 92560 z^2 + 89128 z^3 - 40267 z^4 + 8364 z^5) +  y (-1\\
  &+ z)^4 z^5 (2964 - 19836 z + 41602 z^2 - 41559 z^3 + 20360 z^4 -  4376 z^5 + 384 z^6) + y^{12} (20167 - 150791 z + 451180 z^2\\
  &-  659196 z^3 + 502581 z^4 - 202437 z^5 + 35424 z^6) +  2 y^2 (-1 + z)^3 z^4 (-6765 + 42879 z - 127932 z^2 + 212727 z^3\\
  &-  204354 z^4 + 112598 z^5 - 33356 z^6 + 4308 z^7) +  2 y^{11} (-19473 + 173943 z - 646524 z^2 + 1262820 z^3 - 1387944 z^4\\
  &+  876471 z^5 - 304141 z^6 + 45232 z^7) +  2 y^{10} (23690 - 260707 z + 1196939 z^2 - 2976262 z^3 + 4350991 z^4 -  3849912 z^5\\
  &+ 2048644 z^6 - 608211 z^7 + 77132 z^8) -  2 y^3 (-1 + z)^2 z^3 (-4252 + 81656 z - 390703 z^2 + 946461 z^3 -  1380495 z^4\\
  &+ 1270200 z^5 - 730084 z^6 + 245894 z^7 - 40948 z^8 +  1808 z^9) + y^9 (-37381 + 518516 z - 2983546 z^2 + 9301958 z^3\\
  &-  17356961 z^4 + 20206452 z^5 - 14838192 z^6 + 6708682 z^7 -  1702288 z^8 + 183528 z^9) + y^8 (18775 - 338839 z + 2538420 z^2\\
  &-  9955302 z^3 + 23235465 z^4 - 34337223 z^5 + 32931038 z^6 -  20432684 z^7 + 7885302 z^8 - 1702288 z^9 + 154264 z^{10})\\
  &+  2 y^7 (-2756 + 69365 z - 731925 z^2 + 3664120 z^3 - 10625532 z^4 +  19591238 z^5 - 23845782 z^6 + 19305676 z^7 - 10216342 z^8\\
  &+  3354341 z^9 - 608211 z^{10} + 45232 z^{11}) +  y^4 z^2 (13530 - 180320 z + 1309202 z^2 - 5314260 z^3 + 13155096 z^4 -  21251064 z^5\\
  &+ 23235465 z^6 - 17356961 z^7 + 8701982 z^8 -  2775888 z^9 + 502581 z^{10} - 40267 z^{11} + 904 z^{12}) +  y^5 z (2964 - 126348 z\\
  &+ 1116534 z^2 - 5314260 z^3 + 15703800 z^4 -  30293056 z^5 + 39182476 z^6 - 34337223 z^7 + 20206452 z^8 -  7699824 z^9\\
  &+ 1752942 z^{10} - 202437 z^{11} + 8364 z^{12}) +  2 y^6 (363 - 15846 z + 276864 z^2 - 1809523 z^3 + 6577548 z^4 -  15146528 z^5\\
  &+ 23154650 z^6 - 23845782 z^7 + 16465519 z^8 -  7419096 z^9 + 2048644 z^{10} - 304141 z^{11} + 17712 z^{12})) H(2, y) -  9 (-1 + y)^4 (-1\\
  &+ z)^4 (-1 + y + z) (y + z)^6  (-58 + 80 y^4 + 136 z + 27 z^2 - 185 z^3 + 80 z^4 + 5 y^3 (-37 + 68 z) +  3 y^2 (9 - 139 z\\
  &+ 112 z^2) + y (136 - 12 z - 417 z^2 + 340 z^3))  H(2, y)^2 + H(0, z) (12 (-1 + z)^2 z (-1 + y + z) (y + z)^6  (-3 (-1 + z)^3 (2\\
  &+ 9 z) + 2 y^7 (96 - 201 z + 104 z^2) +  2 y^6 (-525 + 1119 z - 620 z^2 + 30 z^3) -  4 y^5 (-597 + 1257 z - 625 z^2 - 94 z^3\\
\end{aligned}
\end{equation*}
\begin{equation*}
\begin{aligned}
  &+ 62 z^4) -  y (-1 + z)^2 (-84 - 159 z + 760 z^2 - 636 z^3 + 192 z^4) +  y^2 (-678 + 699 z + 2869 z^2 - 5945 z^3 + 3883 z^4\\
  &- 828 z^5) -  2 y^4 (1443 - 2829 z + 710 z^2 + 1265 z^3 - 625 z^4 + 32 z^5) +  y^3 (1944 - 3165 z - 1860 z^2 + 5649 z^3 - 2826 z^4\\
  &+ 256 z^5)) -  72 (-1 + y)^4 (-1 + z)^4 (-1 + y + z) (y + z)^6  (-2 + 32 y^4 + 4 z - 3 z^2 + z^3 + y^3 (-93 + 68 z) +  3 y^2 (29\\
  &- 39 z + 16 z^2) - y (24 - 36 z + 15 z^2 + 4 z^3)) H(1, y) +  288 (-1 + y)^4 (-1 + z)^4 (-1 + y + z) (y + z)^6  (1 + 8 y^4 - 4 z\\
  &- 6 z^2 + 23 z^3 - 14 z^4 + 4 y^3 (-6 + 5 z) +  12 y^2 (2 - 3 z + z^2) + y (-9 + 18 z + 3 z^2 - 16 z^3)) H(2, y))\\
  &+  H(1, z) (-6 (-1 + y + z) (384 y^{15} z + (-1 + z)^5 z^6  (86 - 122 z + 97 z^2) + 8 y^{14} (97 - 739 z + 1077 z^2 - 452 z^3\\
  &+  113 z^4) + y^{13} (-5219 + 36152 z - 84156 z^2 + 80352 z^3 -  35693 z^4 + 7412 z^5) + y (-1 + z)^4 z^5 (-1908 + 2348 z - 286 z^2\\
  &-  1305 z^3 + 1608 z^4 - 1272 z^5 + 384 z^6) +  y^{12} (15045 - 117921 z + 362964 z^2 - 532604 z^3 + 400231 z^4 -  158267 z^5\\
  &+ 27480 z^6) + 2 y^2 (-1 + z)^3 z^4  (-675 - 195 z - 1902 z^2 + 12975 z^3 - 20676 z^4 + 17018 z^5 -  8414 z^6 + 1980 z^7)\\
  &+ 2 y^{11} (-12039 + 115335 z - 450060 z^2 +  899424 z^3 - 987042 z^4 + 612507 z^5 - 208045 z^6 + 30304 z^7) +  2 y^{10} (11498\\
  &- 141313 z + 701873 z^2 - 1830608 z^3 + 2724693 z^4 -  2395960 z^5 + 1248128 z^6 - 359939 z^7 + 43932 z^8) -  2 y^3 (-1\\
  &+ z)^2 z^3 (3868 - 2404 z - 34723 z^2 + 117005 z^3 -  200991 z^4 + 214578 z^5 - 145420 z^6 + 58736 z^7 - 11368 z^8 +  256 z^9)\\
  &+ y^9 (-12895 + 213812 z - 1415134 z^2 + 4813534 z^3 -  9392031 z^4 + 11059916 z^5 - 8020592 z^6 + 3525722 z^7 - 858756 z^8\\
  &+  87192 z^9) + y^8 (3685 - 89137 z + 912288 z^2 - 4186350 z^3 +  10584771 z^4 - 16190181 z^5 + 15575102 z^6 - 9479600 z^7\\
  &+  3525138 z^8 - 718540 z^9 + 59752 z^{10}) +  2 y^7 (-112 + 5193 z - 179561 z^2 + 1190012 z^3 - 3904772 z^4 +  7681554 z^5\\
  &- 9620890 z^6 + 7790928 z^7 - 4023410 z^8 + 1256653 z^9 -  210091 z^{10} + 13920 z^{11}) + y^4 z^2 (1350 + 16400 z + 21642 z^2\\
  &-  611924 z^3 + 2286652 z^4 - 4384532 z^5 + 5218711 z^6 - 4097291 z^7 +  2122526 z^8 - 683628 z^9 + 116791 z^{10} - 6825 z^{11}\\
  &+ 128 z^{12}) +  y^5 z (-1908 - 10644 z + 125622 z^2 - 840744 z^3 + 3311496 z^4 -  7624456 z^5 + 10838140 z^6 - 9884853 z^7\\
  &+ 5831340 z^8 -  2153460 z^9 + 452070 z^{10} - 43743 z^{11} + 1524 z^{12}) +  2 y^6 (-43 + 3050 z + 40106 z^2 - 411629 z^3 + 1786766 z^4\\
  &-  4630398 z^5 + 7634856 z^6 - 8152400 z^7 + 5636279 z^8 - 2461502 z^9 +  635000 z^{10} - 83925 z^{11} + 4224 z^{12})) - 288 (-1\\
  &+ y)^4 (-1 + z)^4  (-1 + y + z) (y + z)^6 (4 y^3 + 4 y^4 + 3 (-1 + z) z^2 -  3 y^2 (6 - 7 z + 6 z^2) - 2 y (-5 + 9 z - 12 z^2\\
  &+ 6 z^3)) H(2, y) +  432 (-1 + y)^4 (-1 + z)^4 (-1 + y + z)^2 (y + z)^6  (12 y^3 + y^2 (-15 + 8 z) + y (4 - 8 z + 8 z^2) + z (4\\
  &- 15 z + 12 z^2))  H(3, y)) - 18 (-1 + y)^4 (-1 + z)^4 (-1 + y + z) (y + z)^6  (-27 + 612 y^4 + 7 z - 18 z^2 + 130 z^3 - 92 z^4\\
  &+ 6 y^3 (-219 + 206 z) +  6 y^2 (127 - 251 z + 78 z^2) - 3 y (11 - 102 z + 26 z^2 + 52 z^3))  H(0, 0, y) + 18 (-1 + y)^4 (-1\\
  &+ z)^4 (-1 + y + z) (y + z)^6  (27 + 92 y^4 + 33 z - 762 z^2 + 1314 z^3 - 612 z^4 + 26 y^3 (-5 + 6 z) +  y^2 (18 + 78 z - 468 z^2)\\
  &- y (7 + 306 z - 1506 z^2 + 1236 z^3))  H(0, 0, z) + 144 (-1 + y)^4 y (-1 + z)^4 (-1 + y + z) (y + z)^6  (16 y^3 + y^2 (-41 + 12 z)\\
  &- 2 (7 - 6 z + 3 z^2) + 3 y (13 - 9 z + 4 z^2))  H(0, 1, y) + 72 (-1 + y)^4 (-1 + z)^4 (-1 + y + z)^2 (y + z)^6  (2 + 64 y^3 - 2 z\\
  &+ z^2 + 8 y^2 (-9 + 5 z) - 2 y (-6 + 7 z + 2 z^2))  H(0, 1, z) - 72 (-1 + y)^4 (-1 + z)^4 (-1 + y + z) (y + z)^6  (104 y^4 + 2 (-1\\
  &+ z)^3 (-1 + 4 z) + y^3 (-245 + 148 z) +  3 y^2 (65 - 77 z + 28 z^2) + 8 y (-7 + 12 z - 9 z^2 + 2 z^3))  H(0, 2, y) + 72 (-1\\
  &+ y)^4 (-1 + z)^4 (-1 + y + z) (y + z)^6  (56 y^4 - 3 y^2 (-7 + 9 z) + y^3 (-91 + 60 z) +  y (20 - 36 z + 27 z^2 - 12 z^3) + 3 (-2\\
  &+ 4 z - 3 z^2 + z^3))  H(1, 0, y) + 72 (-1 + y)^4 (-1 + z)^4 (-1 + y + z) (y + z)^6  (-2 + 32 y^4 + 4 z - 3 z^2 + z^3 + y^3 (-93\\
  &+ 68 z) +  3 y^2 (29 - 39 z + 16 z^2) - y (24 - 36 z + 15 z^2 + 4 z^3)) H(1, 0, z) -  216 (-1 + y)^4 y (-1 + z)^4 (-1 + y + z) (y\\
  &+ z)^6  (6 + 8 y^3 + 4 z - 2 z^2 + y^2 (-13 + 28 z) + y (-1 - 27 z + 4 z^2))  H(1, 1, y) + 18 (-1 + y)^4 (-1 + z)^4 (-1 + y + z) (y\\
  &+ z)^6  (-50 + 16 y^4 + 120 z + 39 z^2 - 189 z^3 + 80 z^4 + y^3 (-253 + 356 z) +  y^2 (327 - 741 z + 528 z^2) + y (-40 + 228 z\\
  &- 549 z^2 + 356 z^3))  H(1, 1, z) + 72 (-1 + y)^4 (-1 + z)^4 (-1 + y + z) (y + z)^6  (2 + 24 y^4 - 4 z + 3 z^2 - z^3 + 4 y^3 (-7\\
  &+ 10 z) -  6 y^2 (2 + 3 z + 2 z^2) + y (14 - 3 z^2 + 4 z^3)) H(1, 2, y) -  288 (-1 + y)^4 (-1 + z)^4 (-1 + y + z) (y + z)^6  (-1\\
  &+ 14 y^4 + 9 z - 24 z^2 + 24 z^3 - 8 z^4 + y^3 (-23 + 16 z) -  3 y^2 (-2 + z + 4 z^2) + y (4 - 18 z + 36 z^2 - 20 z^3)) H(2, 0, y)\\
  &+  144 (-1 + y)^4 (-1 + z)^4 (-1 + y + z) (y + z)^6  (4 + 32 y^4 - 38 z + 87 z^2 - 77 z^3 + 24 z^4 + y^3 (-75 + 68 z) +  3 y^2 (19\\
  &- 39 z + 20 z^2) + y (-18 + 84 z - 111 z^2 + 44 z^3))  H(2, 1, y) - 18 (-1 + y)^4 (-1 + z)^4 (-1 + y + z) (y + z)^6  (90 + 176 y^4\\
  &- 280 z + 381 z^2 - 367 z^3 + 176 z^4 + y^3 (-367 + 12 z) -  3 y^2 (-127 + 45 z + 48 z^2) + y (-280 + 396 z - 135 z^2\\
  &+ 12 z^3))  H(2, 2, y) + 432 (-1 + y)^4 (-1 + z)^4 (-1 + y + z)^2 (y + z)^6  (12 y^3 + y^2 (-15 + 8 z) + y (4 - 8 z + 8 z^2) + z (4\\
  &- 15 z + 12 z^2))  H(3, 2, y)\Bigg\}\Big/\Big(144 (-1 + y)^4 y (-1 + z)^4 z (-1 + y + z)^2 (y + z)^6\Big);\\[5pt]
\mathcal{A}_{6;C_{A}n_{f}}^{(2)} &=
\Bigg\{(-1 + y) (-1 + z) (y + z) (864 y^{12} z + 7 (-1 + z)^4 z^5 (2 - 2 z + z^2) +  y^{11} (7 - 3181 z + 2877 z^2 + 3053 z^3 - 1028 z^4)\\
  &+  y (-1 + z)^3 z^4 (-70 + 138 z + 447 z^2 - 697 z^3 - 589 z^4 + 864 z^5) -  2 y^{10} (21 - 1831 z + 3460 z^2 + 8127 z^3 - 12227 z^4\\
  &+ 3314 z^5) +  y^9 (112 - 93 z - 3293 z^2 + 54841 z^3 - 123339 z^4 + 85304 z^5 -  18716 z^6) + y^2 (-1 + z)^2 z^3 (140 - 770 z\\
  &- 1605 z^2 + 8180 z^3 -  8502 z^4 - 1166 z^5 + 2877 z^6) -  y^8 (168 + 2705 z - 24018 z^2 + 117638 z^3 - 307100 z^4 + 350637 z^5\\
  &-  170234 z^6 + 30204 z^7) + y^7 (147 + 1554 z - 26467 z^2 + 142908 z^3 -  438373 z^4 + 710452 z^5 - 567495 z^6 + 212662 z^7\\
  &- 30204 z^8) +  y^6 (-70 + 177 z + 10620 z^2 - 97217 z^3 + 373368 z^4 - 794957 z^5 +  925784 z^6 - 567495 z^7 + 170234 z^8 - 18716 z^9)\\
\end{aligned}
\end{equation*}
\begin{equation*}
\begin{aligned}
  &+  y^3 z^2 (140 + 376 z - 7503 z^2 + 37294 z^3 - 97217 z^4 + 142908 z^5 -  117638 z^6 + 54841 z^7 - 16254 z^8 + 3053 z^9) +  y^5 (14\\
  &- 348 z + 75 z^2 + 37294 z^3 - 197115 z^4 + 514818 z^5 -  794957 z^6 + 710452 z^7 - 350637 z^8 + 85304 z^9 - 6628 z^{10}) +  y^4 z (70\\
  &- 1050 z - 7503 z^2 + 62552 z^3 - 197115 z^4 + 373368 z^5 -  438373 z^6 + 307100 z^7 - 123339 z^8 + 24454 z^9 - 1028 z^{10})) +  12 (-1\\
  &+ y)^4 (y + z)^6 (72 y^5 z - (-1 + z)^5 (2 - 2 z + z^2) +  y^4 z (-171 + 177 z - 8 z^2 + 2 z^3) + y^3 (-1 + z)^2  (-1 + 150 z - 33 z^2\\
  &+ 12 z^3) + 3 y^2 (-1 + z)^3  (-1 + 25 z - 33 z^2 + 22 z^3) + y (-1 + z)^4 (-4 + 30 z - 81 z^2 +  64 z^3)) H(0, y) + 27 (-1 + y)^4 (-1\\
  &+ z)^4 z (-1 + y + z) (y + z)^6  (3 - 12 z + 10 z^2 + y (-3 + 6 z)) H(0, y)^2 +  27 (-1 + y)^4 y (3 + 10 y^2 + 6 y (-2 + z) - 3 z) (-1\\
  &+ z)^4 (-1 + y + z)  (y + z)^6 H(0, z)^2 - 108 (-1 + y)^4 (-1 + z)^4 (-1 + y + z) (y + z)^6  (10 y^3 + 6 y^2 (-2 + z) + y (3 - 6 z\\
  &+ 6 z^2) + z (3 - 12 z + 10 z^2))  H(1, z)^2 + H(1, y) (12 (-1 + y)^4 (-1 + z)^4 (y + z)^6  (22 y^4 + z (-1 + 3 z) + 2 y^3 (-29 + 8 z)\\
  &- 3 y (7 - 6 z + 3 z^2) +  3 y^2 (19 - 13 z + 6 z^2)) - 72 (-1 + y)^4 y  (3 + 10 y^2 + 6 y (-2 + z) - 3 z) (-1 + z)^4 (-1 + y + z) (y\\
  &+ z)^6  H(1, z)) + 6 (144 y^{15} z - (-1 + z)^5 z^6 (-2 + 44 z - 73 z^2 +  44 z^3) - 2 y^{14} (22 + 371 z - 477 z^2 - 80 z^3 + 20 z^4)\\
  &+  y^{13} (293 + 1168 z - 5070 z^2 + 2496 z^3 + 821 z^4 - 140 z^5) +  y (-1 + z)^4 z^5 (-24 + 332 z - 904 z^2 + 1005 z^3 - 360 z^4\\
  &- 166 z^5 +  144 z^6) + y^{12} (-849 + 873 z + 9576 z^2 - 16678 z^3 + 5423 z^4 +  311 z^5 + 192 z^6) + 6 y^2 (-1 + z)^3 z^4 (5 - 169 z\\
  &+ 700 z^2 -  1162 z^3 + 833 z^4 + 15 z^5 - 368 z^6 + 159 z^7) +  2 y^{11} (696 - 3138 z - 1425 z^2 + 18410 z^3 - 21716 z^4 + 10176 z^5\\
  &-  3835 z^6 + 976 z^7) + 2 y^{10} (-700 + 5626 z - 9744 z^2 - 13652 z^3 +  51737 z^4 - 54675 z^5 + 33289 z^6 - 13437 z^7 + 2420 z^8)\\
  &+  2 y^3 (-1 + z)^2 z^3 (-168 + 1335 z - 5622 z^2 + 12199 z^3 - 12840 z^4 +  3543 z^5 + 5796 z^6 - 5603 z^7 + 1408 z^8 + 80 z^9)\\
  &+  y^9 (869 - 11156 z + 40020 z^2 - 28260 z^3 - 100851 z^4 + 231456 z^5 -  227034 z^6 + 134608 z^7 - 45724 z^8 + 6360 z^9) +  y^8 (-313\\
  &+ 6709 z - 39528 z^2 + 82844 z^3 - 7079 z^4 - 224019 z^5 +  385714 z^6 - 333886 z^7 + 169290 z^8 - 45724 z^9 + 4840 z^{10}) +  2 y^7 (27\\
  &- 1188 z + 11322 z^2 - 42860 z^3 + 61782 z^4 + 15764 z^5 -  162079 z^6 + 229116 z^7 - 166943 z^8 + 67304 z^9 - 13437 z^{10} +  976 z^{11})\\
  &+ y^5 z (-24 + 1104 z - 16920 z^2 + 80478 z^3 - 172200 z^4 +  157568 z^5 + 31528 z^6 - 224019 z^7 + 231456 z^8 - 109350 z^9\\
  &+  20352 z^{10} + 311 z^{11} - 140 z^{12}) +  y^4 z^2 (-30 + 3342 z - 24582 z^2 + 80478 z^3 - 141088 z^4 + 123564 z^5 -  7079 z^6 - 100851 z^7\\
  &+ 103474 z^8 - 43432 z^9 + 5423 z^{10} + 821 z^{11} -  40 z^{12}) + 2 y^6 (-1 + 214 z - 3666 z^2 + 24778 z^3 - 70544 z^4 +  78784 z^5\\
  &+ 23768 z^6 - 162079 z^7 + 192857 z^8 - 113517 z^9 +  33289 z^{10} - 3835 z^{11} + 96 z^{12})) H(2, y) -  108 (-1 + y)^4 (-1 + z)^4 (-1 + y\\
  &+ z) (y + z)^6  (10 y^3 + 6 y^2 (-2 + z) + y (3 - 6 z + 6 z^2) + z (3 - 12 z + 10 z^2))  H(2, y)^2 + H(0, z) (12 (-1 + z)^4 (y + z)^6 (2\\
  &- 4 z + 3 z^2 - z^3 +  y^7 (-1 + 64 z) + y^6 (7 - 337 z + 66 z^2) +  y^5 (-22 + 738 z - 297 z^2 + 12 z^3) +  y^3 (-45 + 584 z - 591 z^2\\
  &+ 228 z^3 - 8 z^4) +  y^4 (40 - 866 z + 570 z^2 - 57 z^3 + 2 z^4) +  y^2 (31 - 225 z + 333 z^2 - 334 z^3 + 177 z^4) +  y (-12 + 46 z\\
  &- 84 z^2 + 152 z^3 - 171 z^4 + 72 z^5)) -  72 (-1 + y)^4 y (3 + 10 y^2 + 6 y (-2 + z) - 3 z) (-1 + z)^4 (-1 + y + z)  (y + z)^6 H(1, y)\\
  &+ 72 (-1 + y)^4 y (3 + 10 y^2 + 6 y (-2 + z) - 3 z)  (-1 + z)^4 (-1 + y + z) (y + z)^6 H(2, y)) +  H(1, z) (6 (144 y^{15} z + (-1\\
  &+ z)^5 z^6 (2 - 2 z + z^2) -  2 y^{14} (22 + 371 z - 477 z^2 - 80 z^3 + 20 z^4) +  y^{13} (293 + 1168 z - 5070 z^2 + 2496 z^3 + 821 z^4\\
  &- 140 z^5) +  y (-1 + z)^4 z^5 (-24 + 78 z - 16 z^2 - 225 z^3 + 400 z^4 - 342 z^5 +  144 z^6) + y^{12} (-843 + 831 z + 9720 z^2\\
  &- 16954 z^3 + 5717 z^4 +  149 z^5 + 228 z^6) + 2 y^2 (-1 + z)^3 z^4 (15 - 186 z + 305 z^2 +  667 z^3 - 2514 z^4 + 3319 z^5 - 2176 z^6\\
  &+ 609 z^7) +  2 y^{11} (683 - 3014 z - 1956 z^2 + 19678 z^3 - 23503 z^4 + 11652 z^5 -  4496 z^6 + 1100 z^7) + 2 y^{10} (-678 + 5313 z\\
  &- 8022 z^2 - 18773 z^3 +  60879 z^4 - 64750 z^5 + 39999 z^6 - 15912 z^7 + 2808 z^8) -  2 y^3 (-1 + z)^2 z^3 (168 - 900 z + 2040 z^2\\
  &+ 85 z^3 - 10206 z^4 +  22176 z^5 - 23052 z^6 + 12257 z^7 - 2704 z^8 + 8 z^9) +  y^9 (833 - 10292 z + 33702 z^2 - 4660 z^3 - 153421 z^4\\
  &+ 305336 z^5 -  293084 z^6 + 170960 z^7 - 56910 z^8 + 7824 z^9) +  y^8 (-299 + 6023 z - 32496 z^2 + 49134 z^3 + 86751 z^4 - 389079 z^5\\
  &+  574814 z^6 - 474284 z^7 + 234114 z^8 - 62510 z^9 + 6680 z^{10}) +  2 y^7 (26 - 1040 z + 8961 z^2 - 27672 z^3 + 8010 z^4 + 134028 z^5\\
  &-  332223 z^6 + 391604 z^7 - 268514 z^8 + 106908 z^9 - 22068 z^{10} +  1764 z^{11}) + y^4 z^2 (-30 + 2672 z - 15902 z^2 + 33612 z^3\\
  &+  1816 z^4 - 152124 z^5 + 346211 z^6 - 406001 z^7 + 279114 z^8 -  108142 z^9 + 19425 z^{10} - 655 z^{11} + 4 z^{12}) +  y^5 z (-24 + 822 z\\
  &- 11768 z^2 + 45164 z^3 - 41376 z^4 - 142688 z^5 +  485144 z^6 - 685563 z^7 + 546704 z^8 - 249740 z^9 + 58512 z^{10} -  5199 z^{11}\\
  &+ 156 z^{12}) + 2 y^6 (-1 + 187 z - 2784 z^2 + 16395 z^3 -  31152 z^4 - 31320 z^5 + 222248 z^6 - 400783 z^7 + 385277 z^8 -  215282 z^9\\
  &+ 66695 z^{10} - 9876 z^{11} + 540 z^{12})) +  72 (-1 + y)^4 (-1 + z)^4 (-1 + y + z) (y + z)^6  (10 y^3 + 6 y^2 (-2 + z) + y (3 - 6 z + 6 z^2)\\
  &+ z (3 - 12 z + 10 z^2))  H(2, y)) - 54 (-1 + y)^4 (-1 + z)^4 z (-1 + y + z) (y + z)^6  (3 - 12 z + 10 z^2 + y (-3 + 6 z)) H(0, 0, y)\\
  &-  54 (-1 + y)^4 y (3 + 10 y^2 + 6 y (-2 + z) - 3 z) (-1 + z)^4 (-1 + y + z)  (y + z)^6 H(0, 0, z) + 72 (-1 + y)^4 y (3 + 10 y^2\\
  &+ 6 y (-2 + z) - 3 z)  (-1 + z)^4 (-1 + y + z) (y + z)^6 H(1, 0, z) +  72 (-1 + y)^4 (-1 + z)^4 (-1 + y + z) (y + z)^6  (40 y^3\\
  &+ 24 y^2 (-2 + z) + 3 y (4 - 7 z + 6 z^2) +  3 z (3 - 12 z + 10 z^2)) H(1, 1, z) -  72 (-1 + y)^4 y (3 + 10 y^2 + 6 y (-2 + z)\\
  &- 3 z) (-1 + z)^4 (-1 + y + z)  (y + z)^6 H(1, 2, y) + 72 (-1 + y)^4 (-1 + z)^4 z (-1 + y + z) (y + z)^6  (3 - 12 z + 10 z^2 + y (-3\\
  &+ 6 z)) H(2, 0, y) +  288 (-1 + y)^4 (-1 + z)^4 (-1 + y + z) (y + z)^6  (10 y^3 + 6 y^2 (-2 + z) + y (3 - 6 z + 6 z^2) + z (3 - 12 z\\
  &+ 10 z^2))  H(2, 2, y)\Bigg\}\Big/\Big(144 (-1 + y)^4 y (-1 + z)^4 z (-1 + y + z) (y + z)^6\Big);\\[5pt]
\mathcal{A}_{6;C_{F}n_{f}}^{(2)} &=
\Bigg\{45 (-1 + z)^3 z^6 (2 - 2 z + z^2) + 3 y^{11} (15 - 6 z - 39 z^2 + 28 z^3) -  3 y (-1 + z)^2 z^5 (180 - 528 z + 506 z^2 - 167 z^3\\
  &+ 6 z^4) +  3 y^{10} (-75 + 179 z + 81 z^2 - 417 z^3 + 224 z^4) +  y^9 (495 - 2538 z + 2541 z^2 + 3020 z^3 - 5974 z^4 + 2432 z^5)\\
\end{aligned}
\end{equation*}
\begin{equation*}
\begin{aligned}
  &+  y^8 (-585 + 5121 z - 10623 z^2 + 3201 z^3 + 13112 z^4 - 15226 z^5 +  5024 z^6) - 3 y^2 z^4 (450 - 2184 z + 4776 z^2 - 5694 z^3\\
  &+ 3541 z^4 -  847 z^5 - 81 z^6 + 39 z^7) + 2 y^7 (180 - 2613 z + 8541 z^2 -  11754 z^3 + 2089 z^4 + 12163 z^5 - 11756 z^6 + 3180 z^7)\\
  &+  y^3 z^3 (-1800 + 13656 z - 34518 z^2 + 41110 z^3 - 23508 z^4 + 3201 z^5 +  3020 z^6 - 1251 z^7 + 84 z^8) + 2 y^4 z^2 (-675 + 6828 z\\
  &- 24831 z^2 +  39456 z^3 - 26784 z^4 + 2089 z^5 + 6556 z^6 - 2987 z^7 + 336 z^8) +  2 y^5 z (-270 + 3276 z - 17259 z^2 + 39456 z^3\\
  &- 38370 z^4 + 7389 z^5 +  12163 z^6 - 7613 z^7 + 1216 z^8) +  2 y^6 (-45 + 1332 z - 7164 z^2 + 20555 z^3 - 26784 z^4 + 7389 z^5\\
  &+  13973 z^6 - 11756 z^7 + 2512 z^8) - 12 (-1 + y)^2 (-1 + 4 y) (-1 + z)^2  (y + z)^6 (-2 + 2 y^3 + 4 z - 3 z^2 + z^3 + y^2 (-6 + 4 z)\\
  &+  y (6 - 8 z + 3 z^2)) H(1, y) + 3 (-1 + y)^2 (-1 + z)^2  (32 y^{10} + y^9 (-101 + 244 z) + 3 y^8 (37 - 241 z + 280 z^2) -  z^6 (-2 + 4 z\\
  &- 3 z^2 + z^3) + 2 y^7 (-22 + 351 z - 1125 z^2 +  856 z^3) + y z^5 (-36 + 4 z + 54 z^2 - 39 z^3 + 4 z^4) +  6 y^2 z^4 (69 - 146 z\\
  &+ 118 z^2 - 55 z^3 + 12 z^4) +  2 y^3 z^3 (-348 - 98 z + 825 z^2 - 659 z^3 + 200 z^4) +  6 y^5 z (-6 - 206 z + 527 z^2 - 726 z^3\\
  &+ 332 z^4) +  2 y^4 z^2 (207 - 198 z + 1263 z^2 - 1518 z^3 + 572 z^4) +  2 y^6 (1 - 98 z + 1110 z^2 - 1987 z^3 + 1132 z^4)) H(1, z)\\
  &+  3 (-1 + y)^2 (-1 + z)^2 (32 y^{10} + 5 y^9 (-21 + 52 z) +  3 y^8 (41 - 269 z + 328 z^2) + 6 y^7 (-10 + 145 z - 469 z^2 + 384 z^3)\\
  &+  z^6 (10 - 60 z + 123 z^2 - 105 z^3 + 32 z^4) +  6 y^2 z^4 (89 - 302 z + 516 z^2 - 469 z^3 + 164 z^4) +  y z^5 (12 - 348 z + 870 z^2\\
  &- 807 z^3 + 260 z^4) +  6 y^5 z (2 - 302 z + 927 z^2 - 1400 z^3 + 724 z^4) +  2 y^3 z^3 (-268 - 778 z + 2781 z^2 - 2961 z^3 + 1152 z^4)\\
  &+  2 y^4 z^2 (267 - 778 z + 3213 z^2 - 4200 z^3 + 1860 z^4) +  2 y^6 (5 - 174 z + 1548 z^2 - 2961 z^3 + 1860 z^4)) H(2, y)\Bigg\}\Big/ \Big(36 (-1\\
  &+ y)^2 y (-1 + z)^2 z (-1 + y + z) (y + z)^6\Big).\\
\end{aligned}
\end{equation*}
}